\documentclass[useAMS,usenatbib]{mn2e}
\usepackage{psfig}
\usepackage{graphicx}
\usepackage{amssymb}

\title[Lensing estimates of cluster density profiles]
{The effects of ellipticity and substructure on estimates of cluster density profiles based on lensing and kinematics}

\author[Meneghetti et al.]{
\parbox[t]{\textwidth}{
Massimo Meneghetti$^{1,3}$,
Matthias Bartelmann$^{1}$,
Adrian Jenkins$^{2}$,
Carlos Frenk$^2$.
}
\vspace*{6pt} \\ 
$^1$ITA, Universit\"at Heidelberg, Albert-\"Uberle-Str.~2, 69120 Heidelberg, Germany \\
$^2$Institute for Computational Cosmology, Physics Department, Durham, DH1 3LE, U.K.\\
$^3$INAF-Osservatorio Astronomico di Bologna, Via Ranzani 1, 40127 Bologna, Italy \\
\vspace{-0.5cm}
}

\date{Accepted ---. Received ---;in original form ---}

\pubyear{2005}
\newcommand{\dd}{{\rm d}}

\newcommand{\plotone}[1]
           {\centering \leavevmode \psfig{file=#1,width=\columnwidth,clip=}}
\newcommand{\plotside}[1]
           {\centering \leavevmode \psfig{file=#1,width=\textwidth,clip=}}
\newcommand{\plottwo}[2]
            {\hbox{\psfig{file=#1,width=4.0cm,clip=}
                             \hfil  \psfig{file=#2,width=4.0cm,clip=}}}
\newcommand{\plotthree}[3]
            {\centering \leavevmode \hbox{\psfig{file=#1,width=6.0cm,clip=}
                              \psfig{file=#2,width=6.0cm,clip=} 
                  \psfig{file=#3,width=6.0cm,clip=}}}
\newcommand{\plottwovert}[2]
            {\centering \leavevmode \psfig{file=#1,width=\columnwidth,clip=}
                             \hfil  \psfig{file=#2,width=\columnwidth,clip=}}

\newcommand{\bvec}[1]{\mbox{\boldmath $#1$}}

\begin{document}
\def\simlt{\lower.5ex\hbox{$\; \buildrel < \over \sim \;$}}
\def\simgt{\lower.5ex\hbox{$\; \buildrel > \over \sim \;$}}

\maketitle

%%\label{firstpage}

\begin{abstract}

We address the question of how well the density profile of galaxy clusters can
be determined by combining strong lensing and velocity dispersion data. We use
cosmological dark matter simulations of clusters to test the reliability of the
method, producing mock catalogues of tangential and radial gravitational arcs
and simulating the radial velocity dispersion profile of the cluster brightest
central galaxy. The density profiles of the simulated clusters closely follow
the NFW form, but we find that the recovered values of the inner slope are
systematically underestimated, by about 0.4 in the mean, if the lens is assumed
to be axially symmetric. However, if the ellipticity and orientation of the
iso-contours of the cluster lensing potential are taken into account, then the
inner slopes can be recovered quite accurately for a significant subset of the
clusters whose central surface density profiles appear the most regular. These
have lensing potentials with ellipticities in the range $0.15-0.4$. Further
simulations projecting one cluster along many random
lines-of-sight show that, even for lower ellipticities, the central slopes are
underestimated by $\sim 10-35\%$. These simulations closely mimic past observations (see e.g. Sand et al., 2004), 
suggesting that existing estimates of the central slopes may be biased towards low values.
For the remaining clusters, where the lensing
potential is strongly perturbed by active merging or by substructure, the
correct determination of the inner slope requires a more accurate model for the
lens.  When the halo profile is modelled by a generalised NFW profile, we find
that the inferred scale radius and characteristic density, unlike the inner
slope, are generally poorly constrained, since there is a strong degeneracy
between these two parameters.

\end{abstract}


\section{Introduction}
\label{intro}

Over the past two decades a paradigm for the formation of cosmic structure, the
$\Lambda$CDM model, has gradually emerged. In this model, the contents of the
universe are dominated by a mixture of cold dark matter and ``vacuum'' or
``dark'' energy, and structure grows by the gravitational amplification of
initial fluctuations imprinted during an early period of inflation by quantum
processes. This paradigm, first explored theoretically in the 1980s and early
1990s (e.g. Peebles 1982, Davis et al. 1985, Efstathiou, Sutherland and Maddox
1990, Martel, 1991), has been stringently tested in the past few years by a
combination of measurements of anisotropies in the cosmic microwave background
radiation and the clustering of galaxies in the local universe (Spergel et al. 
2003, 2006, Sanchez et al. 2005, Tegmark et al. 2006). These and related
measurements have directly tested the model over an impressive range of scales,
$\sim$$(10- 3000)$Mpc, and cosmic expansion factor, $z\sim(0-1000$). These
tests, however, probe only the linear, or at most, the mildly non-linear
properties of the model.

While the $\Lambda$CDM model makes specific predictions for the
distribution of dark matter in the strongly non-linear regime,
devising and implementing appropriate empirical tests is
complicated. Much attention has focused on the nature of the mass
distribution near the centres of dark halos which can, in principle,
be probed through observations of galaxies and clusters. The
theoretical expectations are now reasonably well established through
extensive N-body studies carried out over the past decade. The
simulations show that halos develop a ``cuspy'' profile near the
centre, that is, the spherically averaged density profile continues to
rise to the resolution limit of the simulations (Navarro, Frenk \&
White, 1996, 1997, hereafter NFW). According to these simulations, 
halos of all masses follow approximately the same density law
independently of the values of the cosmological parameters, with
density falling off as $r^{-3}$ in the outer parts and diverging as
$r^{-1}$ towards the centre:  

\begin{equation}
  \rho(r)=\frac{\rho_{\rm s}}{(r/r_{\rm s})(1+r/r_{\rm s})^2} \; ,
  \label{eq-nfw}
\end{equation}
where $r_{\rm s}$ and $\rho_{\rm s}$ denote the scale radius and
characteristic density. 

A singular behaviour of the halo density profile is not entirely
unexpected given the scale-free nature of gravity and the featureless
form of the cold dark matter power spectrum. The NFW simulations were
able to resolve the mass distribution on scales larger that about 5\%
of the virial radius. Subsequent simulations of higher resolution have
confirmed the general conclusions of NFW although the exact nature of
the central cusp is still uncertain \citep{Moore98}. The largest
existing simulations which resolve the mass distribution down to
$\sim(0.5-1)\%$ of the viral radius (which encloses less than $\sim
1\%$ of the total mass of the halo) show that the profiles have an
inner logarithmic slope shallower that $-1.3$ but the value of the
central slope is not yet established (Power et al. 2003, Navarro et al.
2004, Diemand et al. 2005).

Observational tests of the inner structure of dark matter halos have
so far proved inconclusive. Several factors complicate the comparison
with observations. Foremost amongst them is the fact that the
evolution of the baryonic component of halos is likely to have
affected the distribution of the dark matter in ways that are still
poorly understood. The growth of a normal, bright galaxy at the centre
of a halo is likely to change its concentration but the size and even
the sign of any effect is controversial (Navarro, Eke \& Frenk 1996,
Gnedin et al. 2004). For this reason, much observational effort has
focused on studying the rotation curves of faint dwarfs and
low-surface brightness galaxies in the hope that their apparently low
baryon content may minimise the dissembling effects of the luminous
matter. Unfortunately, the physical mechanisms responsible for the low
baryon content are unknown and so the possibility that they may have
disturbed the dark matter cannot be discounted. In addition to these
fundamental problems, attempts to recover the density profile of halos
from rotation curve data suffer from numerous practical difficulties
(van den Bosch and Swaters 2001, de Blok, Bosma \& McGaugh 2003, Simon
et al. 2003, 2005, Swaters et al. 2003, Hayashi \& Navarro 2005).

Constraints on dark matter halo density profiles are more
straightforward to derive in galaxy clusters than in individual
galaxies. Although in clusters the mass distribution is also likely to
have been affected by the growth of the central galaxy, clusters
present a number of measurable properties that are simpler to
interpret than rotation curves. In the inner parts, at radii $\sim
10\%$ of the virial radius, the X-ray emission from the hot
intracluster plasma is easy to observe and the temperature profile of
the gas can be measured reliably using Chandra or XMM. There are
now several examples of clusters for which the dark matter density
profile inferred from such data and the assumption of hydrostatic
equilibrium is well fit by an NFW profile in the range $\sim (0.1-1)
r_{vir}$, where $r_{vir}$ denotes the virial radius (e.g. Allen,
Ettori \& Fabian 2001, Schmidt, Allen \& Fabian,
2001; Pratt \& Arnaud 2002, 2003; Ettori \& Lombardi 2003).  It is
difficult with X-ray data alone to probe the dark matter profile
further in mainly because clusters that appear relaxed tend to have
``cooling flows''. In these clusters the X-ray emission is often
disturbed rendering suspect the assumption of hydrostatic
equilibrium. Nevertheless, Lewis, Stocke \& Buote (2002), Lewis, Buote
and Stocke (2003) and Buote \& Lewis (2004) have found 2 examples of
clusters (A2029 and A2589) in which the X-ray emission from the core
appears regular and for which they conclude that the halo structure is
consistent with an NFW profile well inside 0.1$r_{vir}$. Similar
results have been obtained by Allen, Schmidt \& Fabian (2002) and
Arabadjis, Bautz \& Garmire (2002).

In addition to X-ray data, galaxy clusters offer another powerful
diagnostic of their dark matter distribution: gravitational
lensing. Weak lensing of background galaxies is now routinely
exploited as a means to reconstruct the mass distribution in the outer
parts of clusters (Mellier 1999). Using this technique, Dahle,
Hannestad \& Sommer-Larsen (2003) found that the average density
profile of 6 massive clusters is in good agreement with the NFW
formula at radii $r \gtrsim 0.1 r_{vir}$. In the inner parts, the
effects of lensing are no longer linear but the mass distribution can
still be constrained through strong lensing effects. The most common
of these is the production of tangential arcs which have now been
observed in a large number of rich clusters (Mellier 1999). The
location of a tangential arc is determined by the projected mass
density interior to the arc. If a background galaxy appears at special
locations in the source plane, its image can be distorted into a
radial arc whose position depends on the local derivative of the
cluster mass density profile. On rare occasions, clusters produce both
tangential and radial arcs. From an analysis of tangential and radial
arcs in A383, Smith et al. (2001) found an inner profile steeper than
NFW at a radius of $\sim 1\% r_{vir}$. Combining strong and weak
lensing features, Kneib et al. (2003) found that Cl 0024\_1654 is well
fit by an NFW profile from about $0.1 r_{vir}$ to well beyond
$r_{vir}$.  Using a related approach, Gavazzi et al. (2003) and Gavazzi (2005)
also found 
that the inner regions of MS2137.3-2353 can be fit with an NFW profile
although their data seemed to favour an isothermal profile in this
region. 

If, in addition to tangential and radial arcs, information about the
potential of the central galaxy is available, then much stronger
constraints on the inner density profiles of both the dark and
luminous components can be placed (Miralda-Escud\'e, 1995). Systems of
this kind offer the best opportunity to determine the nature of the
dark matter distribution in the centres of clusters and thus, uniquely, 
to provide a stringent test of the cold dark cosmogony in the strongly
non-linear regime. Miralda-Escud\'e's proposal has recently been
implemented in practice by Sand et al. (2002, 2004). In their first
paper, Sand et al. considered a single system, but in their second
paper, they identified a sample of 6 galaxy clusters with tangential
arcs, three of which also have radial arcs. They then measured the
velocity dispersion profile of the central galaxy in each
case. Combining these data, they inferred much flatter inner slopes
than predicted by the NFW form, concluding that their sample is
inconsistent with the cold dark matter predictions at the 99\%
confidence level. This conclusion calls into question the validity of
the standard cosmological model on small scales unless some complex
interaction between dark and visible material can account for the
difference between the inferred mass profile and that seen in the
N-body simulations.

The results of Sand et al. support the idea that important cosmological
information can be inferred from cluster structural properties. However, their
conclusions are based on a fairly small sample of clusters. Moreover, Sand et
al. made an important simplifying assumption in their analysis which, as we
shall see, is applicable only to a restricted class of objects, and, in general,
can bias the inferred inner slope towards flatter values.  This is the
assumption that the lensing cluster can be treated as an axially symmetric
system. As Bartelmann \& Meneghetti (2004) have shown, the position of the
radial and tangential critical lines and thus the inferred dark matter profile
depend very sensitively on any ellipticity of the mass distribution or,
equivalently, on the presence of external shear. Using an analytical mass model,
Bartelmann \& Meneghetti showed that even small deviations from axial symmetry
could relax the constraints on the cluster mass distributions, allowing cuspy
profiles to be consistent with the data.

While exposing the strong dependence of the inferred mass profile on the assumed
shape of the mass distribution, Bartelmann \& Meneghetti's analytical study was
not able to address the degree of asymmetry or the strength of the shear fields
expected for realistic cold dark matter halos and to establish for which types
of clusters the method proposed by Miralda-Escud\'e and applied Sand et al. is
applicable under simplified assumptions. Neither did it investigate the
possibility of extending the method to a broader class of clusters by increasing
the number of model parameters in the fit to the observational data. These 
questions can only be addressed using N-body simulations. This is the subject of
this paper. Here, we use high-resolution N-body simulations of cluster halos
grown from cold dark matter initial conditions in a full cosmological setting to
test directly the constraints that can be set on the mass profiles from
{combined strong lensing and velocity dispersion data}. These clusters naturally
possess NFW profiles. A model galaxy with {realistic} properties is placed at
the centre of each simulated cluster and imaginary background galaxies are
lensed, giving rise to tangential and radial arcs. We then carry out an
analogous exercise to that of Sand et al. We find that if we assume axial
symmetry, as Sand et al. did, we generally infer central slopes that are
substantially flatter than the NFW value, except for extremely round
clusters. However, when we abandon the assumption of axial symmetry and,
instead, fit the data with elliptical lensing potentials, we are able to recover
the correct, cuspy profiles for a much larger number of numerically simulated
clusters.

The remainder of this paper is organised as follows. In Section~2, we briefly
review the basic lensing concepts used later in the paper. In Section~3, we
describe our lens model and in Section~4 the N-body simulations. The analysis of
the simulations is carried out in Section~5 which presents our main
results. Section~6 is dedicated to a direct comparison to the analysis of Sand
et al. (2004). We conclude with the summary and discussion of Section~7.

\section{Basic lensing equations}
\label{basic_equations}
In this section we define the lensing variables and lensing equations
that we will need later on. We use the thin lens approximation throughout.

We start by defining the optical axis as a line running from the
observer through an arbitrary point on the lens plane towards the
background sources. We take the points where the axis intercepts
the lens and source planes as the origins for local cartesian
coordinate systems. 

 We use the symbols $\bvec{\xi}$ and $\bvec{\eta}$ to denote the 2-component
positions of points on the lens and source planes respectively.  By
choosing a length scale on the lens plane $\xi_0$, we can then define
dimensionless coordinates $\bvec{x}=\bvec{\xi}/\xi_0$ on the lens
plane. Similarly for the source plane we define dimensionless
coordinates $\bvec{y}=\bvec{\eta}/\eta_0$. For convenience we set
$\eta_0=\xi_0D_{\rm s}/D_{\rm l}$, where $D_{\rm s}$, $D_{\rm l}$ are
the angular diameter distances between observer and source and
observer and lens respectively. We define $D_{\rm ls}$ as the angular
diameter distance between the lens and source.

The lens equation, relating the position of an image on the lens plane
to that of the source on the source plane is then
\begin{equation}
  \bvec{y}=\bvec{x}-\bvec{\alpha}(\bvec{x}) \, ,
  \label{eq:leq}
\end{equation}
where $\bvec{\alpha}(\bvec{x})$ is the reduced deflection angle at position
$\bvec{x}$ relative to the optical axis.
The reduced deflection angle is given by the gradient of the 2D lensing potential,
$\psi(\bvec{x})$,
\begin{equation}
  \bvec{\alpha}(\bvec{x}) = \bvec{\nabla} \psi(\bvec{x}) \, ,
\label{eq:angpot}
\end{equation}
\citep[see e.g.][]{SC92.1}.

The lens convergence $\kappa(\bvec{x})$ and shear $\bvec{\gamma}(\bvec{x})$ can
be derived from the deflection angle:
\begin{eqnarray}
  \nonumber
  \kappa(\bvec{x}) & = & \frac{1}{2} \left(  \frac{\partial \alpha_1}{\partial x_1} +
  \frac{\partial \alpha_2}{\partial x_2} \right) \, , \\
  \label{eq:lensprop}
  \gamma_1(\bvec{x}) & = & \frac{1}{2} \left(  \frac{\partial \alpha_1}{\partial x_1} -
  \frac{\partial \alpha_2}{\partial x_2} \right) \, , \\
   \nonumber
  \gamma_2(\bvec{x}) & = & \frac{\partial \alpha_1}{\partial x_2} = \frac{\partial \alpha_2}{\partial
  x_1} \, , 
\end{eqnarray}
where the subscripts $(1,2)$ denote the Cartesian vector components. It can be
easily shown that the convergence corresponds to the lens surface density in units
of the critical surface density,
\begin{equation}
  \kappa(\bvec{x})=\frac{\Sigma}{\Sigma_{\rm cr}} \;,
\end{equation}
where
\begin{equation}
  \Sigma_{\rm cr}=\frac{c^2}{4 \pi G}\frac{D_{\rm s}}{D_{\rm l}D_{\rm ls}} \;.
\end{equation}  

The local imaging properties of the lens are described by the Jacobian
matrix of the lens mapping. The inverse of the determinant is the
lensing magnification. The Jacobian matrix has two eigenvalues, which
give the local distortion of the image in the radial and in the
tangential directions respectively for an axially symmetric
lens. These are written in terms of $\kappa$ and $\gamma$ as
\begin{eqnarray}
  \nonumber
  \lambda_r(\bvec x) & = & 1-\kappa(\bvec{x})+\gamma(\bvec{x}) \;,\\
  \lambda_t(\bvec x) & = & 1-\kappa(\bvec{x})-\gamma(\bvec{x}) \; .
  \label{eq:eigen}
\end{eqnarray}  

The radial and tangential arcs are seen near the centres of some galaxy
clusters. These arcs are strongly magnified images of background
galaxies. They form around the radial and the tangential critical
lines, the loci of which are determined by the zeros of the radial or
tangential eigenvalues of the Jacobian matrix.

\section{The lens model}

\subsection{Modelling the cluster density profile}
\label{sect:model}

 For the lens model we take a two component system consisting of a dark matter halo and
a central massive galaxy. The dark matter halo is
modelled with a density profile of the form:
\begin{equation}
  \rho(r)=\frac{\rho_{\rm s}}{(r/r_{\rm s})^\beta(1+r/r_{\rm s})^{3-\beta}} \;.
  \label{eq:nfwgen}
\end{equation}
which is a generalisation of the NFW profile given in equation \ref{eq-nfw}.
 This has three free parameters, namely the inner logarithmic slope,
$\beta$, the scale radius, $r_{\rm s}$, and the characteristic density
$\rho_{\rm s}$. The NFW formula corresponds to $\beta=1$.

We assume the central brightest cluster galaxy (BCG hereafter) has a
\cite{JA83.1} profile,
\begin{equation}
  \rho_{\star}(r)=\frac{\rho_{\star J}}{(r/r_{\rm J})^2(1+r/r_{\rm J})^2} \;,
  \label{eq:bcg}
\end{equation}
where $r_{\rm J}$ is the Jaffe radius and $\rho_{\star J}$ is a galaxy
characteristic density. \cite{SA03.1} derive a value of $r_{\rm J}\sim
60$ kpc by fitting the surface brightness profile of the BCG in Abell
383 with a Jaffe profile.  We assume this value for our modelling of
the central galaxy.

The lensing properties are straightforwardly derived from these density profiles
\citep[see e.g.][]{BA04.1}.  The total density profile of the cluster is
obtained by summing the profiles of the dark matter halo and of the BCG. In
doing that, we assume that the dark matter density is unchanged by the growth
of the central dominant galaxy.  

\subsection{Axially symmetric model}
\label{axial_model}  
We start by discussing axially symmetric models. 
For axially symmetric lenses, the lensing potential is independent of
the position angle with respect to the lens centre. If we choose the
optical axis to pass through the lens centre, this implies that
$\psi(\bvec x)=\psi(|\bvec{x}|)$.

The deflection angle is given by 
\begin{equation}
  \alpha(x)=\frac{\dd \psi}{\dd x}=\frac{m(x)}{x} \, ,
  \label{eq:anglesy}
\end{equation}
where $m(x)$ is the dimensionless lens mass within a circle of radius $x$,
\begin{equation}
  m(x)=\frac{M(x)}{\pi \xi_0^2 \Sigma_{\rm cr} } \, ,
\end{equation}
and $M(x)$ is the projected mass in physical units enclosed by radius $x$. 

Using Eqs.~(\ref{eq:lensprop}) and (\ref{eq:eigen}), the eigenvalues of the
Jacobian matrix of the lens mapping can be written as
\begin{eqnarray}
  \lambda_{\rm t}(x) & = & 1-\frac{m(x)}{x^2} \;,\\
  \lambda_{\rm r}(x) & = & 1-\frac{\dd}{\dd x}\left[\frac{m(x)}{x} \right]\;.
\end{eqnarray} 
 These latter two equations imply that:  (i) that the position of a
tangential gravitational arc constrains the projected mass enclosed by
the tangential critical line  and (ii) the location of the radial arc
provides a measurement of the derivative of the projected mass at the
radial critical curve.
 
\subsection{Pseudo-elliptical model}
\label{pseudo_model}

As described in \cite{ME03.1}, a pseudo-elliptical generalisation of any
axially symmetric lens model can be easily obtained by deforming the lensing
potential $\psi$ so that the iso-contour lines become ellipses.  If $\psi(x)$
is the lensing potential of an axially symmetric lens model, an ellipticity
$e=1-b/a$ can be introduced by substituting the radial coordinate $x$ with

\begin{equation}  
  x \rightarrow \overline{x} \equiv \sqrt{\frac{x_1^2}{(1-e)}+x^2_2(1-e)}\, ,
\end{equation}
where $a$ and $b$ are the ellipse major and minor axes, respectively. Such
transformation deforms circular iso-potential contours into ellipses whose major
axis coincides with the $x_2$-axis.

The cartesian components of the deflection angles are obtained by taking
the gradient of the lensing potential,
\begin{eqnarray}
  \alpha_1 & = & \frac{\partial \psi}{\partial x_1} =
  \frac{x_1}{(1-e)\overline{x}}\hat\alpha(\overline{x}) \\
  \alpha_2 & = & \frac{\partial \psi}{\partial x_2} =
  \frac{x_2(1-e)}{\overline{x}}\hat\alpha(\overline{x}) \, ,
  \label{eq:ellgen}
\end{eqnarray}
where $\hat\alpha$ denotes the deflection angle for the axially symmetric
lens model, given in Eq.~(\ref{eq:anglesy}).   

The components of the deflection angles in a reference frame rotated by
an angle $\theta$ are straightforwardly obtained by applying the rotation
matrix
\begin{eqnarray}
\left(
\begin{array}{c}
  \alpha'_1 \\
  \alpha'_2 \\
\end{array} \right)
 &=& 
\left(
\begin{array}{cc}
  \cos \theta & \sin \theta \\
  -\sin \theta & \cos \theta \\
\end{array} \right)
\left(
\begin{array}{c}
  \alpha_1 \\
  \alpha_2 \\
\end{array} \right) \;.
\end{eqnarray}

Using Eqs.~(\ref{eq:lensprop}) and (\ref{eq:eigen}), the radial and the
tangential eigenvalues can be computed at any position on the lens plane.

We recognize that, as shown by \cite{ME03.1}, dumbbell-shaped mass distributions
originate from elliptically distorted lensing potentials for
$e>0.2-0.3$. However, as we shall show later, our goal is to model properly the
shape of the critical lines rather than the shape of the lens isodensity
contours. As many numerical tests confirm, pseudo-elliptical models make this
possible even for relatively large values of $e$ (see e.g
Fig.~\ref{fig:fitexample}).

\subsection{Velocity dispersion profile}
\label{sect:vdisp}
We combine now the lensing constraints on the cluster mass profile with
those derived from the dynamics of stars of the BCG. Assuming both
the cluster and the BCG are spherical or nearly spherical, the
dynamics can be described using the spherical Jeans equation:
\begin{equation}
  \frac{1}{\rho_{\star}(r)}\frac{\dd [\rho_{\star}(r) \sigma_{r}^2(r)]}{\dd r} +
2 \frac{\delta(r)  
  \sigma_{r}^2(r)}{r} = - \frac{GM_{3D}(r)}{r^2} \;,
\end{equation}
where $\sigma_r$ is the stellar radial velocity dispersion, $M_{3D}$ is the
three-dimensional mass enclosed at radius $r$, and
\begin{equation}
  \delta(r) = 1 - \frac{\sigma_\theta^2(r)}{\sigma_r^2(r)} 
\end{equation}
is the anisotropy parameter of the velocity distribution at each point, with
$\sigma_\theta$ denoting the tangential component of the stellar velocity
dispersion. Following \cite{SA03.1}, we assume isotropic orbits and set
$\delta=0$.

The radial velocity dispersion is determined from the Jeans equation:
\begin{equation}
  \sigma_r^2(r)=\frac{G}{\rho_{\star}(r)}\int_r^\infty \frac{M_{3D}(r') \rho_{\star}(r')}{r'^2}\dd
  r' \;.
\end{equation}
By projecting along the line of sight, we obtain the projected velocity
dispersion profile,  
\begin{equation}
  \sigma_p^2(\xi)=\frac{\int_\xi^\infty \frac{\rho_{\star}(r)\sigma_r^2(r)r \dd
  r}{\sqrt{r^2-\xi^2}}}{\int_\xi^\infty \frac{\rho_{\star}(r)r\dd r}{\sqrt{r^2-\xi^2}}}
  \;.
\end{equation}

In order to apply the previous equations, we had to make the
assumption that the BCG and the galaxy cluster can be approximated as
a spherical systems.  We are aware that, when applying this method to
real clusters, this could be inappropriate. However, as discussed in
the following sections, this approximation is consistent with our
modelling of the BCG for the numerically simulated clusters.

\section{Numerical simulations}
\label{sect:numerical}

\subsection{N-body simulations}
\label{subsec:nbody}

   The dark matter halos used in this paper were drawn from a sample
of 10 cluster mass halos  simulated by the Virgo Consortium
as part of a project to study the central density profiles of dark
matter halos over a range of halo masses. A description of how the initial
conditions were set up is given in \cite{Navarro04}.  All ten clusters
were used in \cite{Gao04a} and eight of them feature in \cite{Gao04b}.

   The cluster halos were selected from the $\Lambda$CDM-512 dark
matter simulation described in \cite{Yoshida2001} which has a volume
of 479(Mpc/$h$)$^3$, where $h=0.7$.  A halo catalogue for the entire
simulation volume was made by running the friends-of-friends group
finding algorithm \citep{Davis85} with a linking length of 0.164 times the inverse
cube root of the particle number density.  The halos in the catalogue
were ranked by mass and the 10 most massive halos with masses less
than $10^{15}h^{-1}M_\odot$ were selected to be resimulated with high
mass resolution.

High resolution initial conditions were set up for each of the ten
halos with a particle mass of $5.12\times10^{8}h^{-1}M_\odot$ for all
particles which end up inside about three virial radii from the
cluster centre.  The more distant material which interacts with the
forming cluster through gravitational tidal forces was represented by
more massive particles with a mass increasing approximately linearly
with distance from the cluster region.  The gadget-1.1 code
\citep{Springel01} was used to evolve the initial conditions to
redshift zero.  None of the more massive `tidal' particles fell into
any of the clusters.

 The numerical parameters used in these simulations were chosen
according to the criteria of \cite{PO03.1} to ensure that the circular
velocity profile is accurate to 10\% to within 1\% of $r_{200}$, the
radius of a sphere centred on the density maximum of the halo with a mean
interior density of 200 times the critical density.  The gravitational
softening length used by the gadget code was $5{\rm kpc}/h$.

 Eight of the ten halos were used in this paper. Where we need to refer to a
 particular simulated cluster we use names cl1, cl2 etc. For our analysis, we
 use the simulation snapshots at $z_l=0.24$.

\subsection{Ray-tracing simulations}
\label{sect:raytr}

The ray-tracing simulations were carried out as follows. First, we
select those particles which are contained in a cube of $1.5$ Mpc$/h$
side-length centred on the halo. The particle positions are
projected along the coordinate axes, giving three separate
two-dimensional distributions of particles. For the purposes of this
paper we treat these separate projections as though they were
independent clusters. This effectively increases the sample size
from 8 to 24 clusters. We use the notation cl1.1, cl1.2, cl1.3 to
refer to the three projections of cluster cl1 etc. 

 In order to avoid strong discontinuities between neighbouring cells,
which might introduce noise in the calculation of the deflection
angles, we interpolate the projected particle positions on to a
regular grid of $512\times512$ cells using the {\em Triangular Shaped
Cloud} method \citep{HO88.1}. The resulting surface density maps
$\Sigma_{i,j}$ are used as lens planes in the following lensing
simulations.

A bundle of $1024\times1024$ rays is traced through a regular grid covering
the central quarter of the lens plane. This choice is driven by the requirement
to study the central part of the cluster in detail, where the lens critical
lines are located and where the tangential and radial arcs form. The deflection
angles are calculated as described in \cite{ME00.1} and
\cite{ME01.1}. First, a grid of $256\times256$ ``test'' rays is defined. For
each of these rays the reduced deflection angle is calculated by summing the
contributions from each element of the surface density map,
\begin{equation}
  \bvec{\alpha}_{h,k}=\frac{D_{\rm ls}}{D_{\rm s}}\frac{D_{\rm
  l}}{\xi_0}\frac{4G}{c^2}\sum_{i,j} 
  \Sigma_{i,j}A \frac{\bvec 
  {x_{h,k}}-\bvec {x_{i,j}}}{|\bvec {x_{h,k}}-\bvec {x_{i,j}}|} \;,
\end{equation}
where $A$ is the area of one pixel on the surface density map and $x_{h,k}$
and $x_{i,j}$ are the positions on the lens plane of the ``test'' ray $(h,k)$
and of the surface density element $(i,j)$. We adopt $\xi_0=0.75$Mpc$/h$ as
the scale length on the lens plane, which corresponds to the side-length of the region
through which the rays are traced. The reduced deflection angle of
each of the $1024\times1024$ ``regular'' rays is then determined by bi-cubic
interpolation between the four nearest ``test'' rays. 
 
In order to mimic the presence of the BCG at the cluster centre, we use the
method described by \cite{ME03.2}. We model the galaxy using the Jaffe
profile, as described in Sect.~\ref{sect:model}. The BCG contribution to the
deflection angle of the ray crossing the lens plane at the distance $r$ from
the galaxy centre is given by
\begin{equation}
  \alpha_\mathrm{J}(r)=\frac{r_{\rm J}}{\xi_0}\kappa_\mathrm{J}\,\left[
    \pi-\frac{2(r/r_{\rm J})\,\mathrm{acosh}[(r/r_{\rm
    J})^{-1}]}{\sqrt{1-(r/r_{\rm J})^2}}
  \right] \;,
\label{eq:jaffe_angle}
\end{equation} 
where $\kappa_{\rm J}=\rho_{\star J} r_{\rm J} \Sigma_{\rm cr}^{-1}$
\citep{BA04.1}.
The total deflection angles are obtained by summing the contributions from the
smooth mass distribution of the cluster and from the BCG.

The position of each ray on the source plane, which we place at redshift
$z_s=1$, is determined using the lens equation (\ref{eq:leq}). Here we
distribute a large number of source galaxies. These are modeled as ellipses
with axis ratios randomly drawn with equal probability from $[0.5,1]$. They
have random orientation and an equivalent diameter of $r_{\rm e}=1''$. The
sources are distributed over a region corresponding to one quarter of the field
of view where rays are traced. We first start with a regular grid of $32\times
32$ galaxies. Since in our analysis we intend to use highly magnified arcs, we
increase the number density of sources towards the high-magnification regions
of the source plane by adding sources on sub-grids whose resolution
is increased towards the lens caustics \citep[see e.g.][]{BA98.2}. 

By collecting rays whose positions on the source plane fall within a single
source, we reconstruct the images of the background galaxies. Arc properties
are determined as follows. First, three characteristic points are identified
in the image, namely 1) the image of the source centre, 2) the image point at
the largest distance from the point 1) and 3) the image point at the largest
distance from the point 2). We define the length $L$ of the arc through the
circle segment within points 2) and 3). To determine the image width $W$, we
search for a simple geometrical figure with equal area and length, whose perimeter
matches that of the image. For this fitting procedure, we consider ellipses,
circles, rectangles and rings. For the various cases, the image width is
approximated by the minor axis of the ellipse, the radius of the circle, the
smaller side of the rectangle or the width of the ring, respectively.

Tangential and radial arcs are distinguished in the resulting arc catalogue by
measuring the tangential and radial eigenvalues of the Jacobian matrix at the
arc centre point 1).  Images are classified as tangential (radial) arcs if the
local tangential (radial) magnification exceeds the radial (tangential) one by a
fixed factor $f$. For our analysis we use $f=5$. This is a conservative choice
that allows elongated images that are effectively close to the respective
critical lines to be reliably identified as tangential or radial arcs. For
lower values of $f$, some tangential features may be wrongly identified as
radial arcs. This occurs mainly for images forming along those parts of the
tangential critical line that are relatively close to the radial critical line.

\section{Results}
\label{results}
\subsection{Observables}

The method adopted by \cite{SA03.1} which we too will adopt uses two
different types of data:
\begin{enumerate}
\item the position of radial and tangential gravitational arcs to constrain
  the location of the lens critical lines;
\item the stellar velocity dispersion profile of the brightest cluster galaxy, 
to constrain the total mass within the inner region of
  the cluster.
\end{enumerate}
We start our analysis by assuming axial symmetry for the lensing
potential, ignoring the actual ellipticity of the simulated
clusters. We then allow for elliptical lensing potentials. 
The most stringent constraints on the cluster density profiles come
from lenses in which both radial and tangential arcs are observed
\citep[see e.g][]{SA03.1}. Thus, we focus our analysis on this
particular class of lens.

\begin{figure}
\plotone{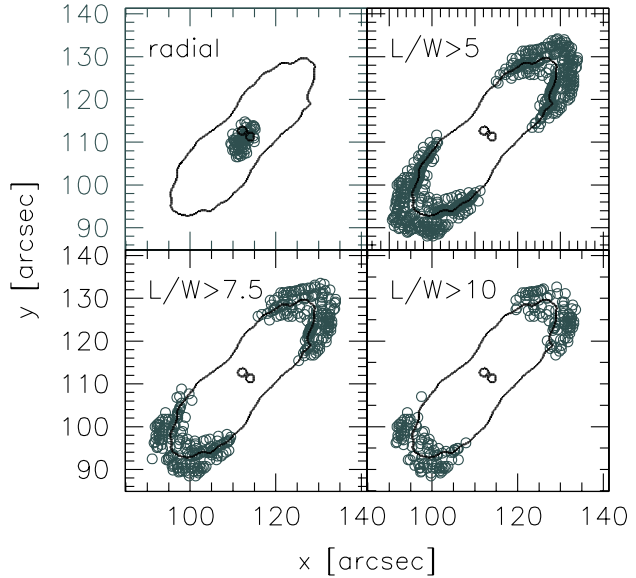}
\caption{The open circles show the location of the arcs for one of our
simulated clusters. The lines common to all panels are the
radial and tangential critical lines. The top left panel shows the
location of radial arcs while the remaining panels show the tangential
arcs for samples selected by their length to width ratios.}
\label{fig:allarcs}
\end{figure}

\subsubsection{Radial and tangential arcs}

First, we explore how well the positions of radial and tangential arcs
can constrain the location of the respective critical lines. This is
clearly shown in the four panels of Fig.~\ref{fig:allarcs}, where the
critical lines for one of our numerically simulated galaxy clusters
are shown.  The arcs, which we determine from the ray-tracing simulation
described above, are shown as open circles in each of the four
panels.  First, we note that in the top left panel, the radial arcs
identified by our criterion are located very close to the radial
critical line. The remaining panels show tangential arcs only and these
are spread over a thick region surrounding the tangential critical line. We
plot only those tangential arcs in each panel which exceed a minimal
length-to-width ratio in order to show that the larger the minimal $L/W$ of
the arcs chosen, the smaller the spread of arc positions around the
critical lines. Second, we note also that tangential arcs tend
to form along those parts of the critical curves which are furthest
from the cluster centre. The reason for this has been discussed
already in some earlier papers \citep[see e.g][]{BA95.1,ME01.1}. Close
to the tangential critical line $\lambda_{\rm t}=1-\kappa-\gamma \sim
0$, hence $\gamma \sim 1-\kappa$. The radial magnification is then
$\mu_{\rm r}=\lambda_{\rm r}^{-1} \sim [2(1-\kappa)]^{-1}$. Therefore,
the smaller the convergence $\kappa$ is, the more the tangential arcs
are radially demagnified, resulting in an arc with a larger
length-to-width ratio. Since the convergence is approximately a
decreasing function of the cluster-centric distance, tangential arcs
with large length-to-width ratio form preferentially close to the
critical points at largest distance from the cluster centre.

It is necessary to use tangential arcs with large length-to-width
ratios in order to constrain the position of tangential critical curve
accurately.  However, by choosing those arcs, at least for clusters
having elongated or elliptical critical lines, we will constrain only
those parts of the critical curves where the convergence is small. As
will be discussed in the following sections, this has to be properly
taken into account when trying to fit the arc positions to the lens
critical curves.

For our analysis we focus on tangential arcs with $L/W>10$. We
simulate simultaneous observations of radial and tangential arcs by
randomly selecting pairs of radial and tangential images from the arc
catalogue generated by our ray-tracing simulations.

\subsubsection{Velocity dispersion profiles}

As explained in Sect.~\ref{sect:raytr}, the presence of the BCG is
mimicked by simply adding the surface density profile of the galaxy onto
that of the cluster. In fact, the growth of the BCG might influence the
distribution of the dark matter in the cluster centre and this approximation
may be wrong in reality but this is beyond the scope of this paper.

\begin{figure}
\plotone{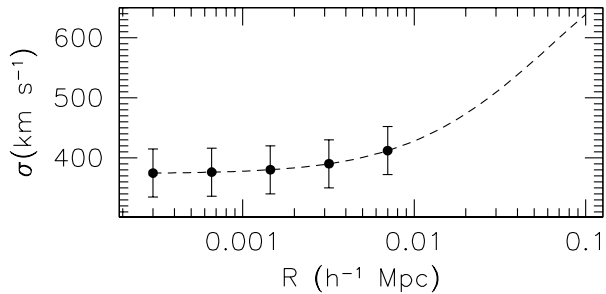}
\caption{The velocity dispersion profile of one of the galaxy clusters
in our sample obtained by combining the best fit density profile of
the cluster and the density profile of the BCG. In order for the
velocity dispersion profile on scales below $10$ kpc to flatten, the
mass of the BCG has been fixed to $10^{12}\,h^{-1}\,M_\odot$ within
the Jaffe radius.  The points indicate the `measured' values that will
be used in the fitting procedure. The errorbars mimic observational
uncertainties of the measurements. }
\label{fig:vdisp_prof}
\end{figure}

Since the simulated clusters do not contain stars, we cannot derive
the projected velocity dispersion from the simulations. Moreover,
velocity dispersion measurements are possible only in the very central
part of the cluster up to radii of the order of $\sim 10$ kpc. Despite
the high mass resolution, this represents the limit of reliability of
the mass profiles which can be inferred from our simulations
\citep{PO03.1}.  Therefore we model the velocity dispersion profiles
as follows. First, we fit the density profile of the numerically
simulated cluster with a generalised NFW model given by
Eq.~(\ref{eq:nfwgen}). Then, taking the best fit parameters $(r_{\rm
s},\rho_{\rm s},\beta)_{fit}$ we extrapolate the cluster density
profile, $\rho_{\rm cl}(r)$, to the inner region, $r<10$ kpc. The BCG
density profile is given by Eq.~(\ref{eq:bcg}). We calculate the
enclosed mass $M_{3D}(r)$ by integrating the total density obtained by
summing the cluster and the BCG density profiles,
\begin{equation}
  M_{3D}=4 \pi \int_0^r (\rho_{\rm cl}+\rho_{\rm BCG}) r^2 \dd r \;.
\end{equation} 
Finally we derive the projected velocity dispersion profile as described in
Sect.~\ref{sect:vdisp}. 

As an example, we show in Fig.~(\ref{fig:vdisp_prof}) the simulated velocity
dispersion profile for one of the clusters in our sample. The points indicate
the constraints used in the following fitting procedure. We assign to each
``measurement'' an error $\Delta\sigma \sim 40$ km s$^{-1}$, mimicking
typical observational uncertainties \citep[see e.g.][and references
therein]{SA03.1}.

The density profile of the BCG is fully determined by the mass
enclosed by the Jaffe radius. We choose the mass of the BCG so as to
produce a velocity dispersion profile which is nearly flat at radii
$r\lesssim 10$ kpc, implying that the mass of the BCG is large enough
to dominate the cluster centre, as is the case for the galaxy clusters
investigated by \cite{SA03.1}. At radii much less that the Jaffe
radius the Jaffe profile becomes isothermal, $\rho_{\star} \propto
r^{-2}$.  So if the galaxy dominates the cluster mass at its core,
the velocity dispersion profile is expected to be flat close to the
cluster centre, and the velocity dispersion is then approximately related
to the circular velocity, by
\begin{equation}
v^2_{\rm rot}=\frac{GM_{3D}(r)}{r} \;,
\end{equation}
by $\sigma_r=v_{\rm rot}/\sqrt{2}$

We are able to reproduce flat velocity dispersion profiles in all our clusters by choosing the mass of the BCG in the range $10^{12}\ldots5\times10^{12}\,h^{-1} M_\odot$. Such masses are compatible with velocity dispersions measured in several BCGs \citep[see e.g.][]{BE07.1}.

\begin{figure}
\plottwo{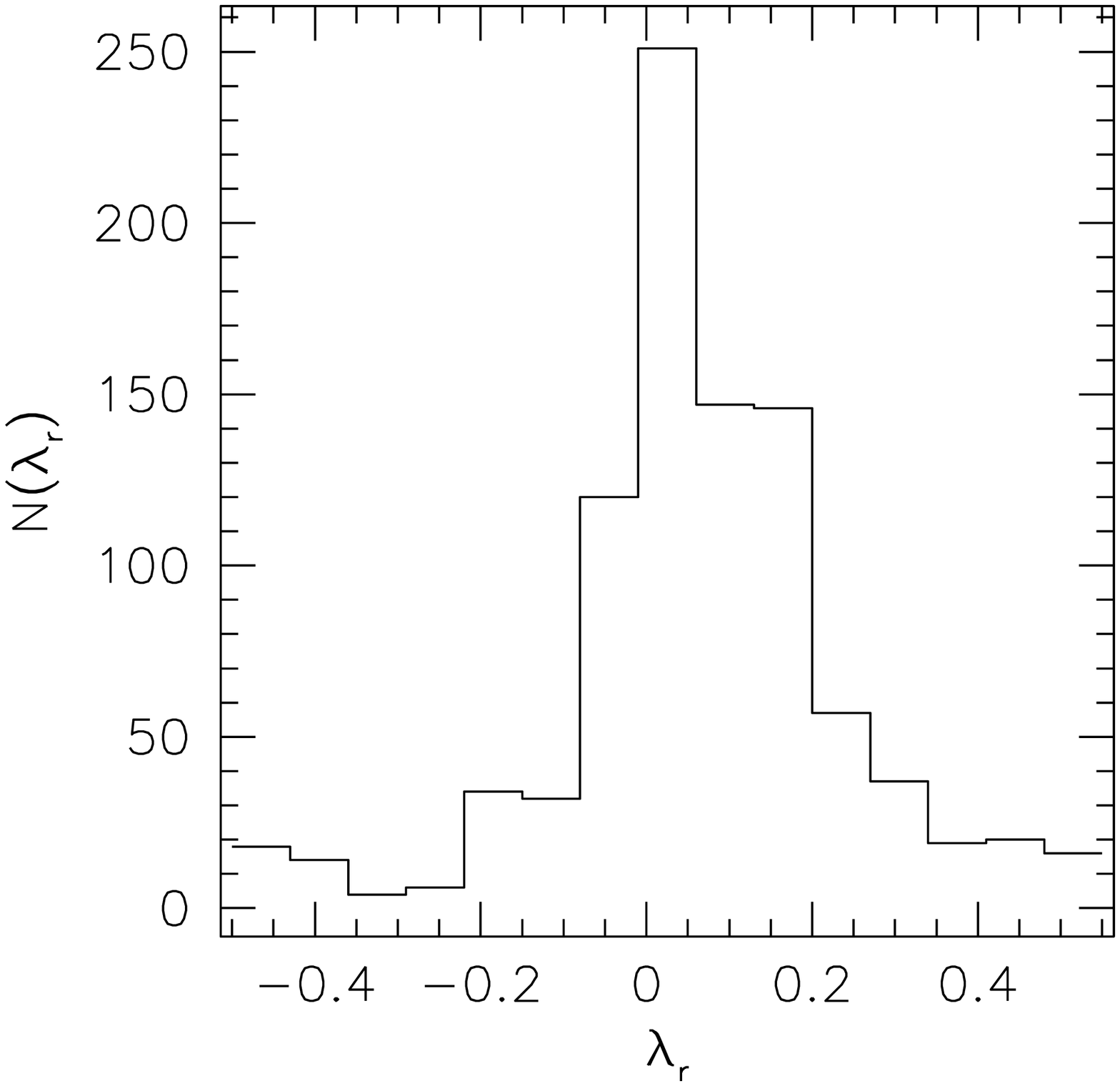}{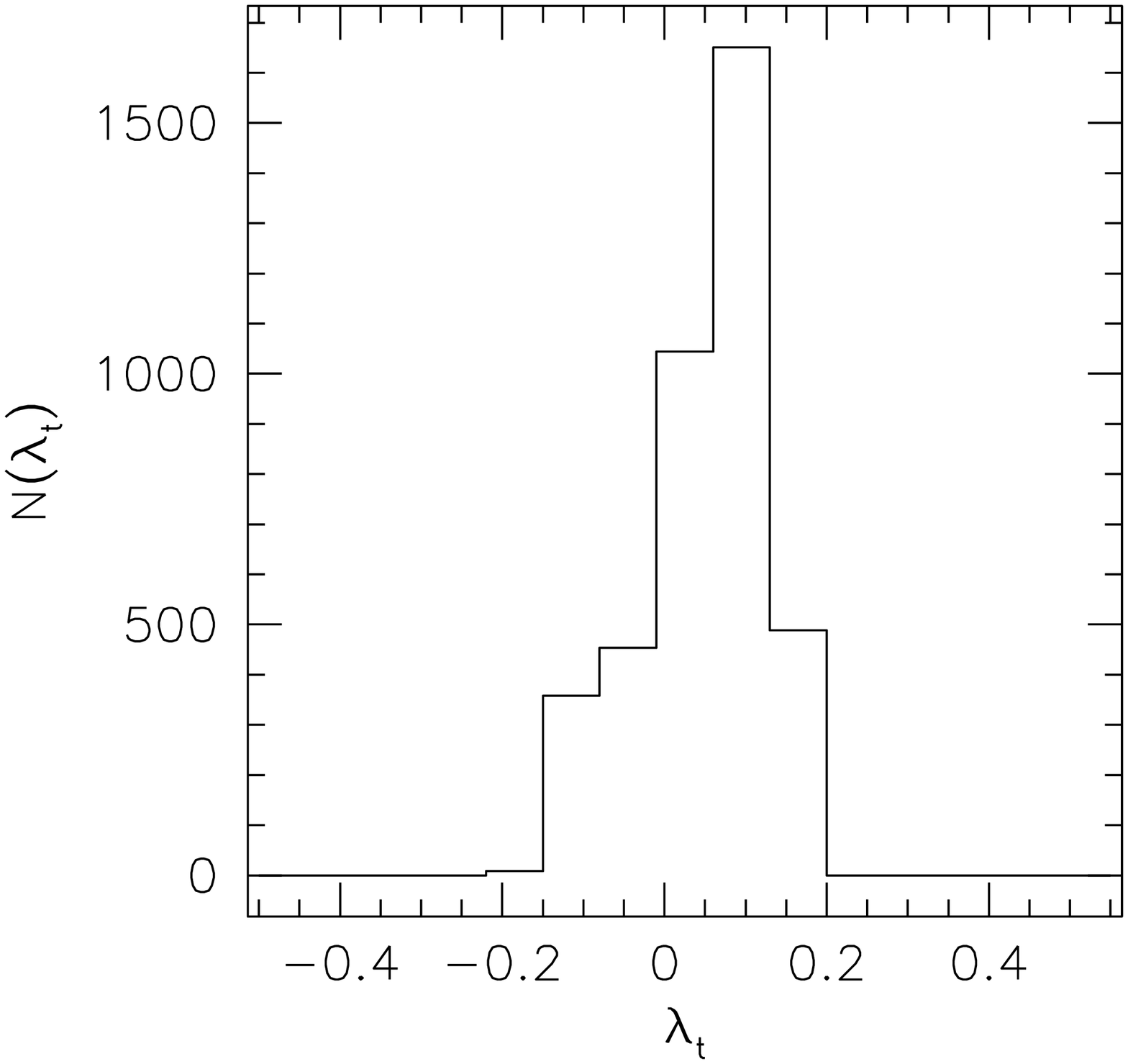}
\caption{Distributions of the values of the radial (left panel) and of the
  tangential (right panel) eigenvalues measured at the positions of the radial
  and of the tangential arcs found in all the simulations. }
\label{fig:eigdistr}
\end{figure}

\subsection{Application to the simulated clusters}

From the list of images obtained from the ray-tracing simulations we randomly
selected pairs of radial and tangential arcs. As discussed earlier, these are
used to constrain the location of the lens critical lines. The complete set of
data which we try to fit with our model is then given by the pair of
gravitational arcs and by the velocity dispersion profile measured at five
different radii in the range $[0-10]\,h^{-1}\,$kpc.

We perform three kinds of fit to the data. Firstly, we repeat the fit
made by \cite{SA03.1}, assuming axially symmetric lensing potentials
while keeping the scale radius of the fitting model fixed at $r_{\rm
s}=400\,h^{-1}\,$kpc, and varying the two other parameters: the halo
characteristic density $\rho_{\rm s}$ and the inner slope
$\beta$. Secondly, we continue to assume axial symmetry, but we let all three
parameters be free. Thirdly, we allow for ellipticity in the lensing
potential of the fitting model. 

For simplicity, we decided not to consider the mass of the BCG as a
free parameter in our fits.  Moreover, when using elliptical models,
we assume we know the ellipticity and the orientation of the
iso-contours of the lensing potential by some other means (from other
independent observations). Therefore, this fitting model also has only
three free parameters, i.e. the parameters characterising the halo
density profile. The mass of the BCG and the ellipticity and the
position angle of the lens are set to their true values.

The fit is done by minimising a $\chi^2$ variable. This is constructed
by
combining the lensing and the velocity dispersion constraints,
\begin{equation}
  \chi^2(r_{\rm s},\rho_{\rm s},\beta)=\chi^2_{lens}(r_{\rm s},\rho_{\rm
  s},\beta)+\chi^2_{\sigma_p}(r_{\rm s},\rho_{\rm s},\beta) \;. 
\label{eq:chi2}
\end{equation}

\begin{figure}
\plottwo{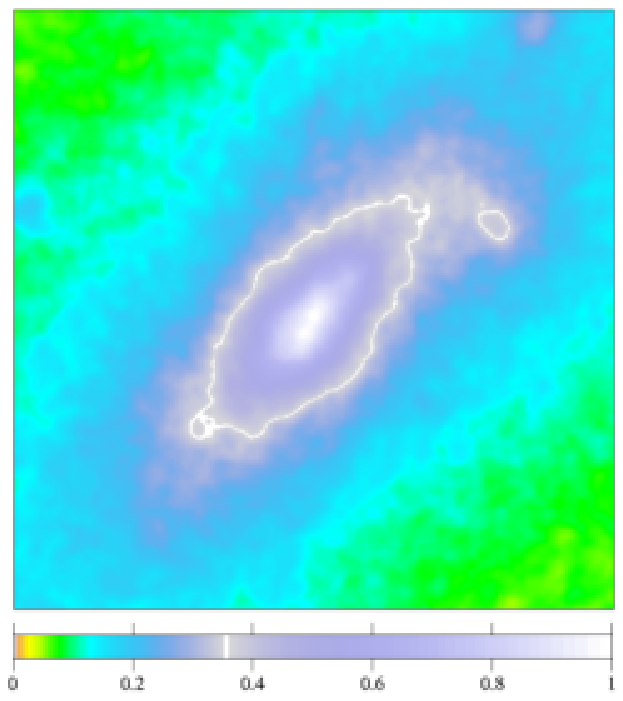}{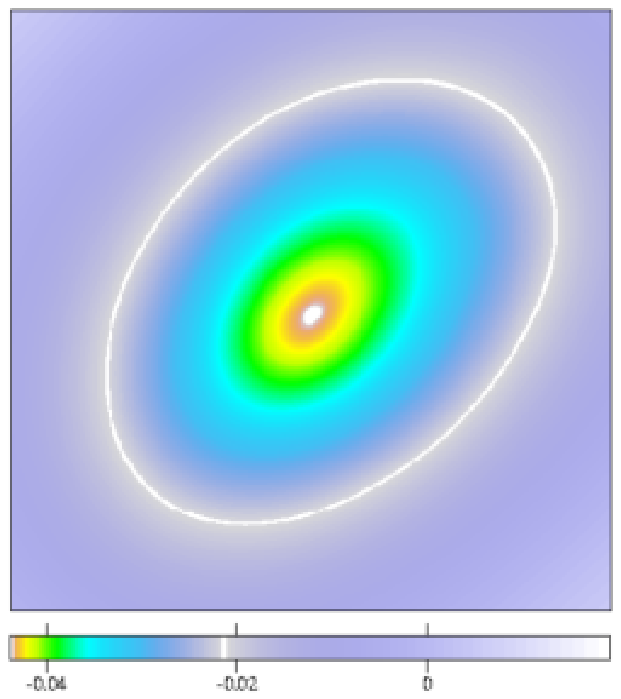}
\caption{Convergence (left panel) and lensing potential (right panel) of the same
  cluster whose critical lines are shown in Fig.~\ref{fig:allarcs}. The side
  length is $\sim 112''$. }
\label{fig:potexample}
\end{figure}

\begin{figure}
%\plottwovert{figures/cl3.vdisp_fit.eps}{figures/cl3.pair48.arcs.allmeth_new.eps}
\plottwovert{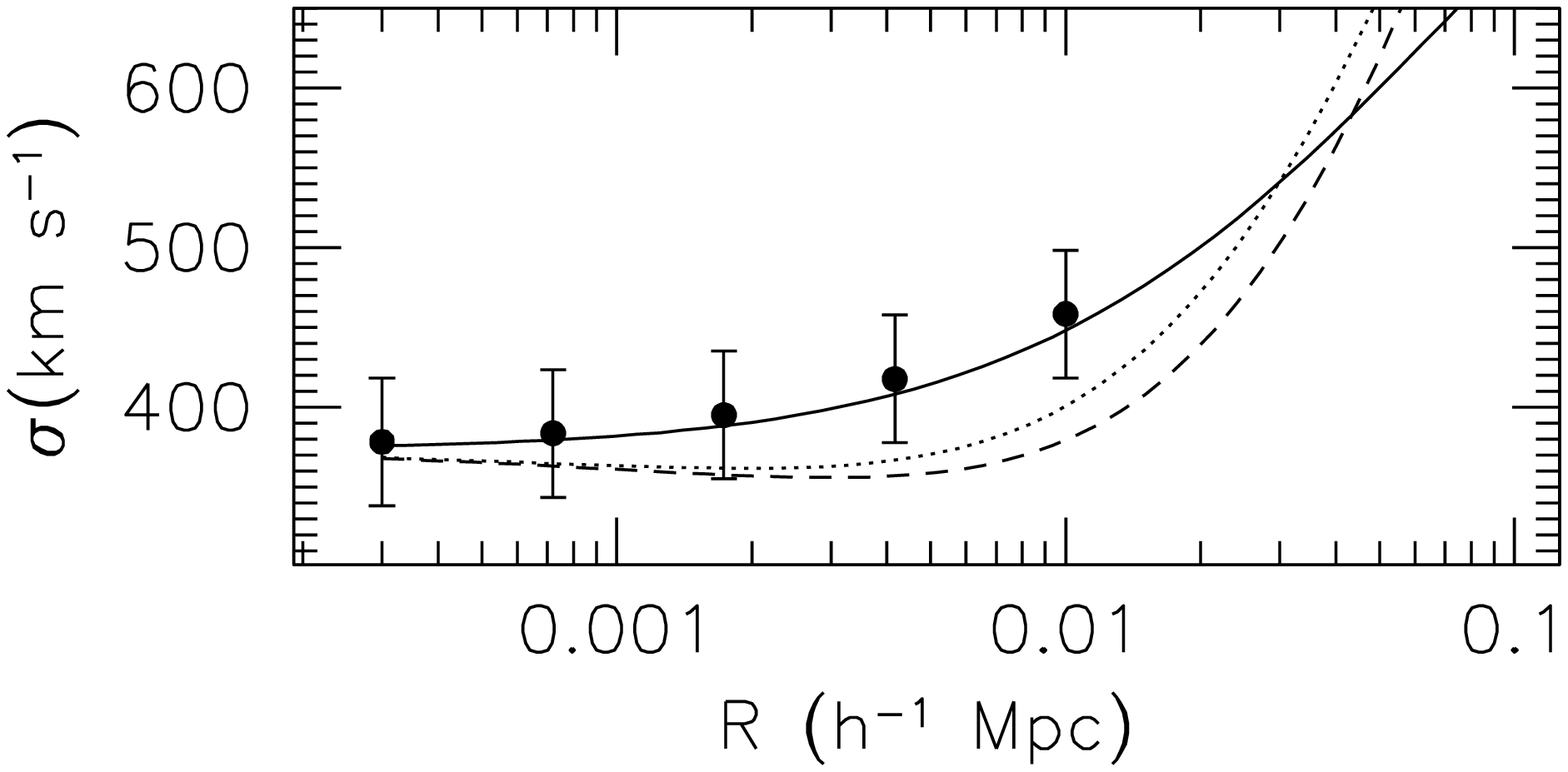}{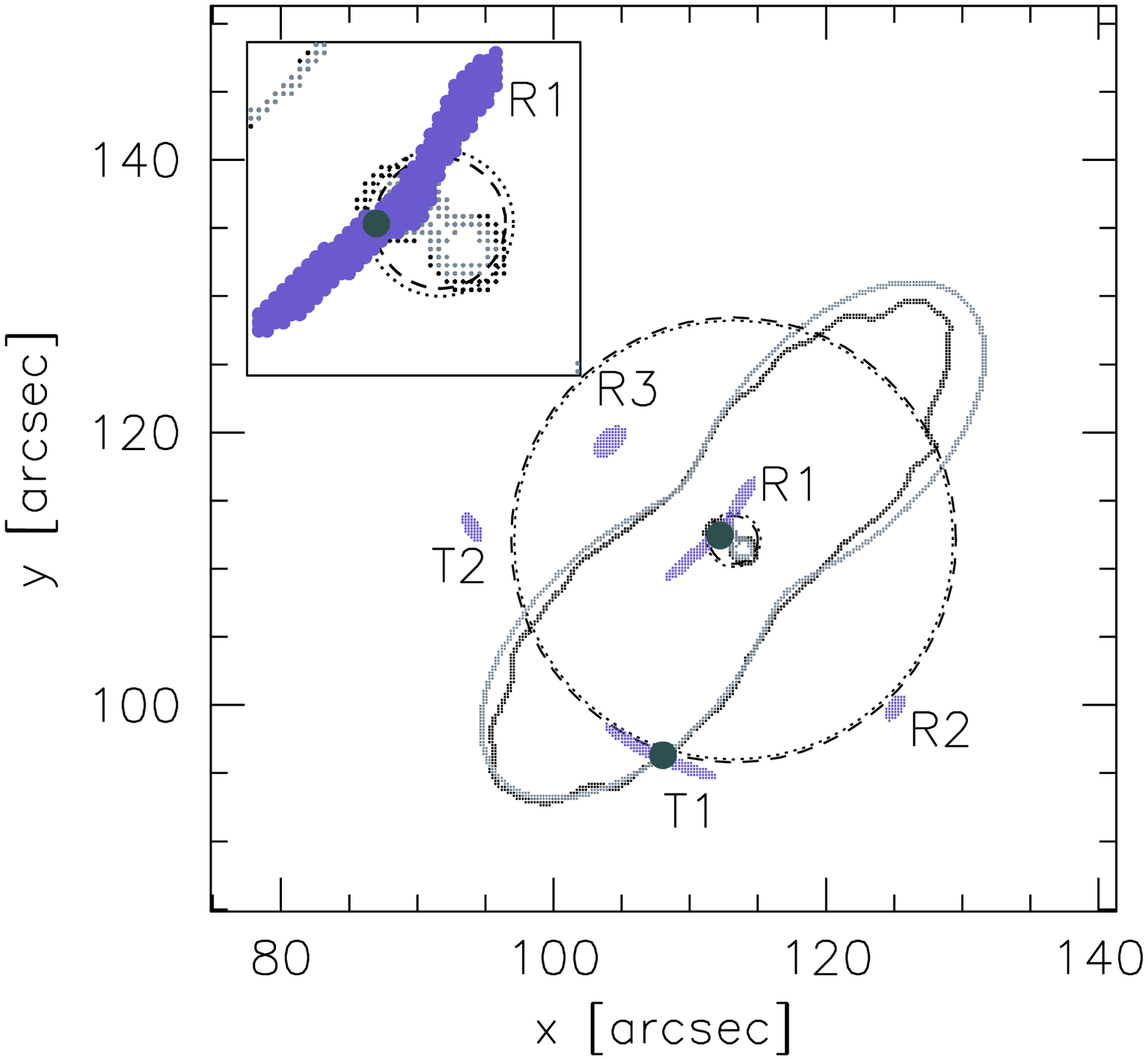}
\caption{Best fit models for one of the simulated clusters in our
sample. The velocity dispersion of the model BCG and our assumed 
observational error bars
are shown in the top panel. The solid line shows the elliptical model
fit, the dotted and dashed lines show axially symmetric fits with
the scale radius either fixed or free respectively.  The bottom panel
shows the lens critical lines (black solid lines) and the arcs used to
constrain their position (images T1 and R1, marked by the dark-gray
dots, indicating the arc centre). A zoom, corresponding to a region of
$9"\times9"$ around the radial arc R1, is displayed in the inset at 
the upper left corner. The images T2, R2 and R3 are 
multiple images of the same sources which are lensed into the
tangential arc T1 and into the radial arc R1. The velocity dispersion
profile and the critical lines of the best fit models, obtained by
assuming an elliptical lensing potential, an axially symmetric lensing
potential and an axially symmetric lensing potential with $r_{\rm
s}=400\,h^{-1} \mathrm{kpc}$, are given by the solid, the dotted and
the dashed lines, respectively. }
\label{fig:fitexample}
\end{figure}

The first term, which concerns the lensing constraints, is defined as
follows. The lens critical lines mark the location where the
eigenvalues of the Jacobian matrix are zero (see
Eq.~\ref{eq:eigen}). If we assume that the critical point locations
coincide with the arc positions, the best fit model is that which
minimises the value of the tangential and radial eigenvalues at
the positions of the tangential and radial arcs,
respectively. The first term in Eq.~(\ref{eq:chi2}) is then
\begin{equation}
  \chi^2_{lens}(r_{\rm s},\rho_{\rm s},\beta)=\left[\frac{\lambda_t(r_{\rm
  s},\rho_{\rm s},\beta)}{\Delta \lambda_t}\right]^2 +
  \left[\frac{\lambda_r(r_{\rm s},\rho_{\rm s},\beta)}{\Delta
  \lambda_r}\right]^2 \;,   
\end{equation}
where $\Delta \lambda_t$ and $\Delta \lambda_r$ reflect the
uncertainties in the determination of the critical lines through the
position of the radial and tangential arcs. We set $\Delta
\lambda_r$ and $\Delta \lambda_t$ equal to the respective eigenvalues
of the Jacobian matrix measured in the numerical simulation at the arc
location. This is motivated by the fact that the larger the
eigenvalue, the further the position of the arc from the lens critical
line.  A histogram showing the distributions of the radial and
tangential eigenvalues measured at the position of the radial and
tangential arcs is shown in Fig.~\ref{fig:eigdistr}. The uncertainty
in position is typically larger for radial arcs. Observationally, the
determination of the location of the radial critical curve is further
complicated by the presence of the BCG, which prevents the detection
of radial images close to the cluster centre. In this paper, we
neglect this problem, and use all the radial images, even those which
might be impossible to observe due to the light of the BCG. Note that
our definition of $\chi^2_{lens}$ differs from that adopted by other
authors. \cite{SA03.1} directly fit the position of the lens critical
line determined visually from the arc properties. In practice, the two
methods should lead to the same result.  Observationally, the
estimation of the errors $\Delta \lambda_t$ and $\Delta \lambda_r$ is
difficult and requires calibration with numerical simulations. On the
other hand, by fitting the tangential and the radial eigenvalues,
instead of the location of the critical line, the required computing
time is substantially reduced, which is mandatory since we repeat the
analysis on thousands of virtual lensing systems. We defer a more
detailed comparison with the results of
\cite{SA03.1} to a subsequent section.

The second term in Eq.~(\ref{eq:chi2}) is given by
\begin{equation}
  \chi^2_{\sigma_p} = \sum_i\left[\frac{\sigma_{p,i}-\hat\sigma_{p,i}(r_{\rm
  s},\rho_{\rm s},\beta)}{\Delta
  \sigma_{p}}\right]^2 \;,
  \label{eq:chi2vd}
\end{equation}
where $\sigma_{p,i}$ and $\hat\sigma_{p,i}$ denote the measured and the
expected values of the velocity dispersion at the radius $\xi_i$,
respectively, and the sum is over all the available measurements of
$\sigma_p$. As previously stated, the uncertainty in the measurement of the
velocity dispersion is fixed at $\Delta \sigma_p=40$ km$s^{-1}$.

We find the set of profile parameters, $(r_{\rm s},\rho_{\rm s},\beta)$, which
minimises $\chi^2$. We repeat the same calculations for $100$ pairs of
radial and tangential arcs, obtaining $100$ different ``best fit''
determinations of the cluster density profiles.

As stated earlier, when fitting with the elliptical model, we assume
we know the ellipticity and the position angle of the iso-contours of the
lensing potential. These are measured in the simulations by finding the
principal axes of the cluster's lensing potential as follows. We first
evaluate the lensing potential on a grid of $N_g=1024\times1024$ points by
inverting Eq.~(\ref{eq:angpot}). Then, we calculate the tensor
\begin{equation}
  \Psi_{ij} = \sum_k^{N_g} \psi_k (\bvec{x_{\mit k}}^2 \delta_{ij}-x_{k,i} x_{k,j}) \;,
\end{equation}
where $k$ labels the grid cells, $\bvec{x_k}$ is the position vector of the
$k$-th cell, with components $x_{k,i}$, $\psi_k$ the corresponding value of
the lensing potential and $\delta_{ij}$ is the Kronecker
delta. The eigenvectors and the eigenvalues of the tensor $\Psi_{ij}$ give
the orientation and the ellipticity of the iso-contours of the lensing
potential. Since both ellipticity and orientation change with radius, we
only consider the region enclosing the lens critical lines.    

\begin{figure}
\plotone{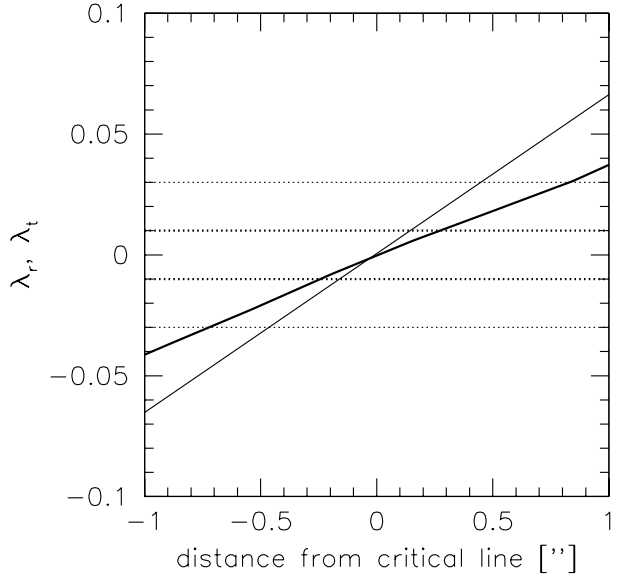}
\caption{The thick and the thin solid lines show the tangential and the radial eigenvalues along the lines connecting the cluster center to the arcs T1 and R1 of Fig.~\ref{fig:fitexample}. The plot zooms over a region of $\pm1"$ around the tangential and the radial critical lines. The thick and thin dotted horizontal lines correspond to $\Delta \lambda_t=\pm 0.01$ and $\Delta \lambda_r=\pm 0.03$, respectively.}
\label{fig:ldist}
\end{figure}

When applying the method to real clusters, the ellipticity and
position angle of the lens should be treated as free
parameters. However, to reduce the computational requirement for our
simulations, we will assume that these are known quantities. In some
cases, stringent constraints on the cluster shape in the central
regions, i.e. around the critical lines, can be derived from
complementary observations. For example, in the case of relaxed
objects, constraints can be imposed from the $X$-ray emission of the
intracluster gas which, assuming that the gas is in hydrostatic
equilibrium, gives the most direct probe of the projected
gravitational potential. However, the equilibrium assumption may be a
poor approximation in many cases.  An alternative approach would be to
combine weak and strong lensing data in order to reconstruct the
cluster potential \citep{BR04.1,CA06.1}. Indeed, recent numerical
tests demonstrate that non-parametric methods can be used to measure
the lensing potential with enough accuracy \cite[see e.g.][]{CA06.1}.

\begin{table}
\centering
\caption{The best fit parameters for the cluster shown in
  Fig.~\ref{fig:potexample}, found by using the constraints showed in
  Fig.~\ref{fig:fitexample}.}
\begin{tabular}{|c|c|c|c|c|}
\hline
\hline
 & $r_{\rm s}$ & $\rho_{\rm s}$ & $\beta$ & $\chi^2$/dof \\ 
 & [$h^{-1}$ kpc] & [$h^2\,M_\odot\,\mathrm{Mpc}^{-3}$] & & \\ 
\hline \hline
input model  & $224$ & $3.54\times10^{15}$ & $0.88$ & \\
\hline
elliptical fit            & $262$ & $3.30\times10^{15}$ & $0.83$ & 0.925 \\ 
axially sym. fit          & $260$ & $1.29\times10^{16}$ & $0.11$ & 1.976 \\ 
$r_{\rm s}$ fixed fit     & $400$ & $7.84\times10^{15}$ & $0.13$ & 2.094 \\ 
\hline
\hline

\end{tabular}
\label{table:bestfit}
\end{table}
  
\subsection{Regular clusters}

\begin{figure*}
\plotthree{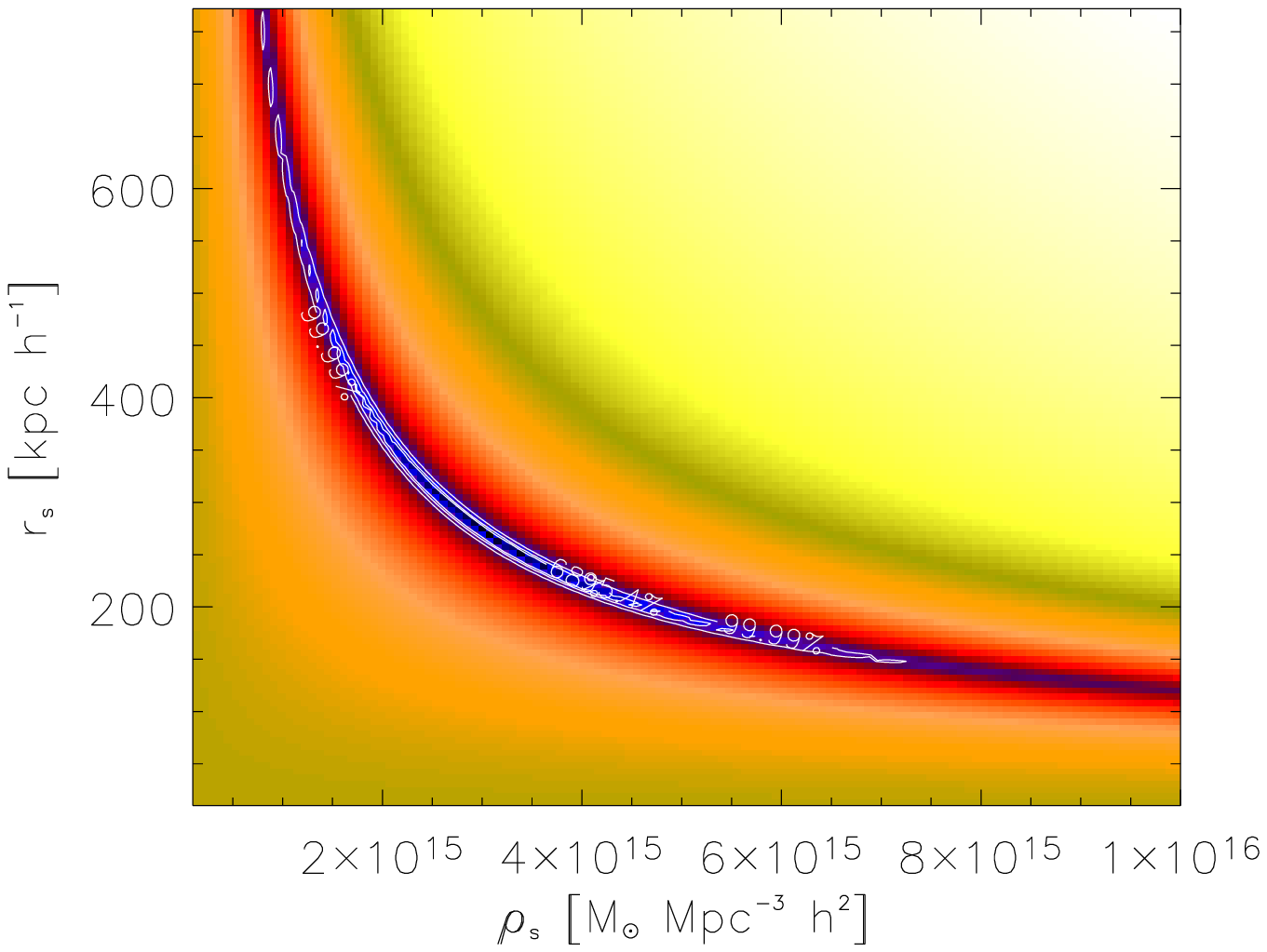}{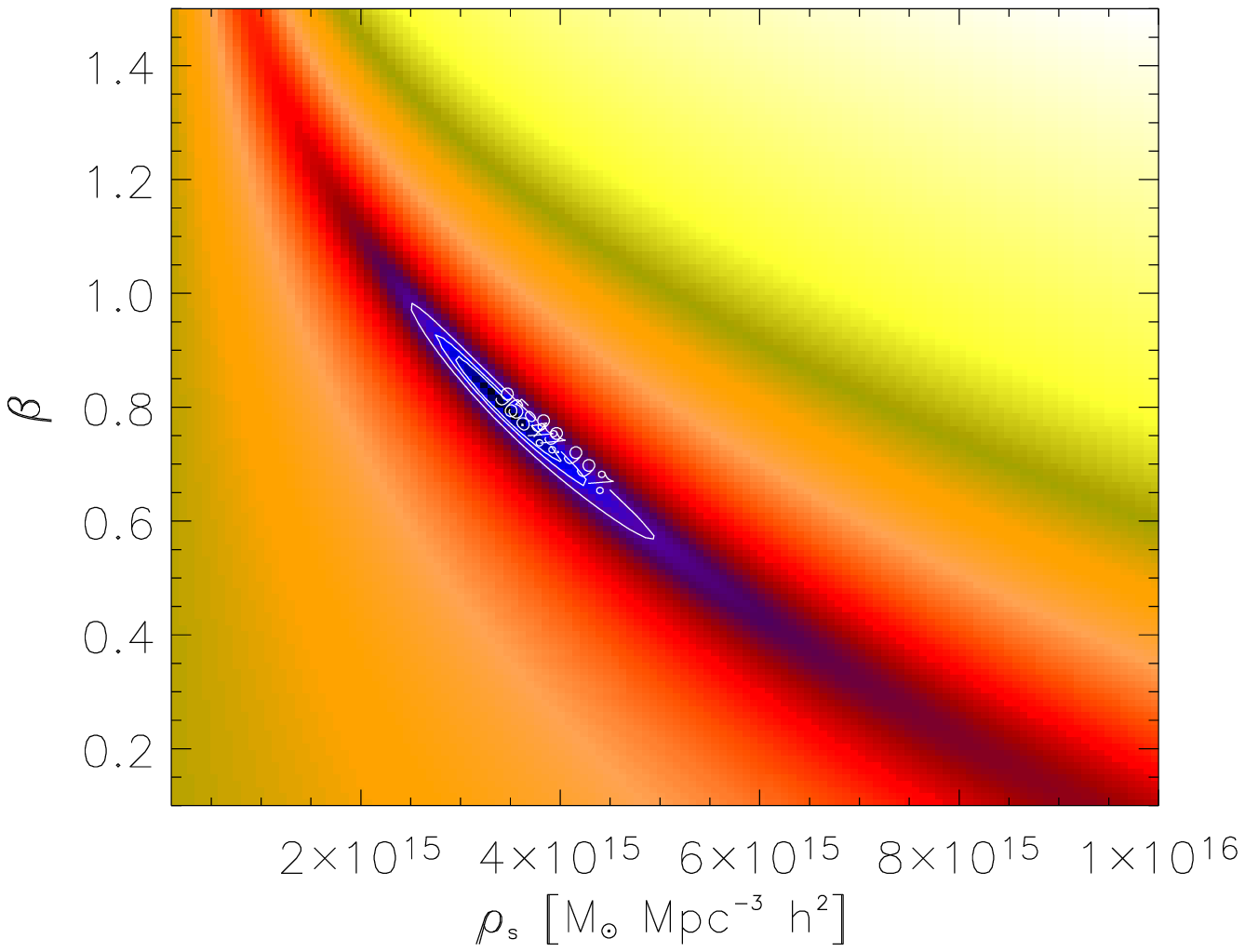}{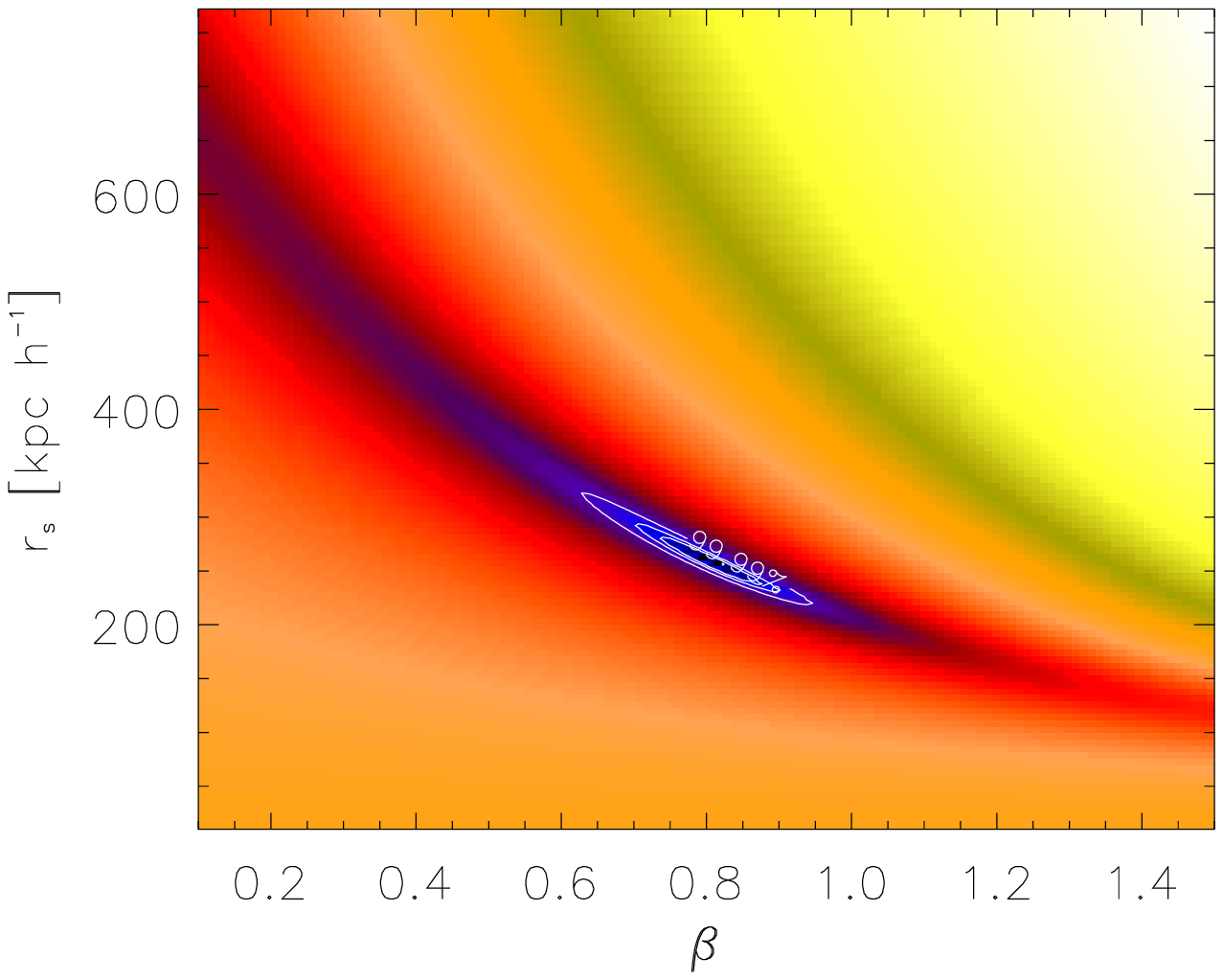}
\caption{Confidence levels in the $\rho_{\rm s}-r_{\rm s}$, $\rho_{\rm
    s}-\beta$ and $\beta-r_{\rm s}$ planes (left, central and right panels,
  respectively). The input model is the cluster illustrated in
  Fig.~\ref{fig:potexample}. The `observational' constraints used are displayed in
  Fig.~\ref{fig:fitexample}. The
  contours corresponding to probability levels of $68.3\%$, $95.4\%$ and
  $99.7\%$ are shown by the light curves. }
\label{fig:chi2cont}
\end{figure*}

We now discuss the results obtained by analysing our simulated clusters.  
We have divided the cluster sample into two sub-samples. In
this subsection we consider only clusters whose mass distribution in
the inner region, critical for strong lensing, shows no evidence of
substructure. We refer to these as ``regular'' clusters. In
the next subsection we consider the remainder which have perturbed core
regions. We call this subsample the ``peculiar'' clusters.

Regular clusters typically have a lensing potential whose
iso-contours can be well fit with ellipses. These clusters are ideal
for applying our method.  Before showing the overall properties of
regular clusters, we first discuss in detail a particular example.

Fig.~\ref{fig:potexample} shows the convergence map and lensing
potential for the same object featured in Fig.~\ref{fig:allarcs} where
the positions of the critical lines are
displayed. Fig.~\ref{fig:fitexample} shows the best fit velocity
dispersion profile (top panel) and critical lines (bottom panel) of
this cluster, obtained by fitting the pair of arcs marked by the
dark-gray dots (T1 and R1) and the velocity dispersion data given by
the black dots with errorbars. A summary of the best fit
parameters, compared to their true values, is given in
Table~\ref{table:bestfit}.

The results show the importance of taking into account the ellipticity
when fitting the velocity dispersion and the lensing data. This
cluster is well described by a lensing potential with an ellipticity
$e=0.33$ and position angle $\theta=46 \deg$. The pair of arcs was
chosen to provide a good constraint on the position of the critical
lines ($\Delta \lambda_r \sim 0.03$, $\Delta \lambda_t \sim
0.01$). When a model with the appropriate ellipticity for the lensing
potential is used, the best fit parameters turn out to be very close
to the true values for the simulated cluster (see
Table~\ref{table:bestfit}). The predicted position of the lens
critical lines and the velocity dispersion profile are also in excellent
agreement with the true ones.  If axial symmetry is assumed however, the fit
is totally wrong. In particular, the inner slope of the density
profile is grossly underestimated ($0.11 - 0.13$ instead of $0.88$).

\begin{figure*}
\plotside{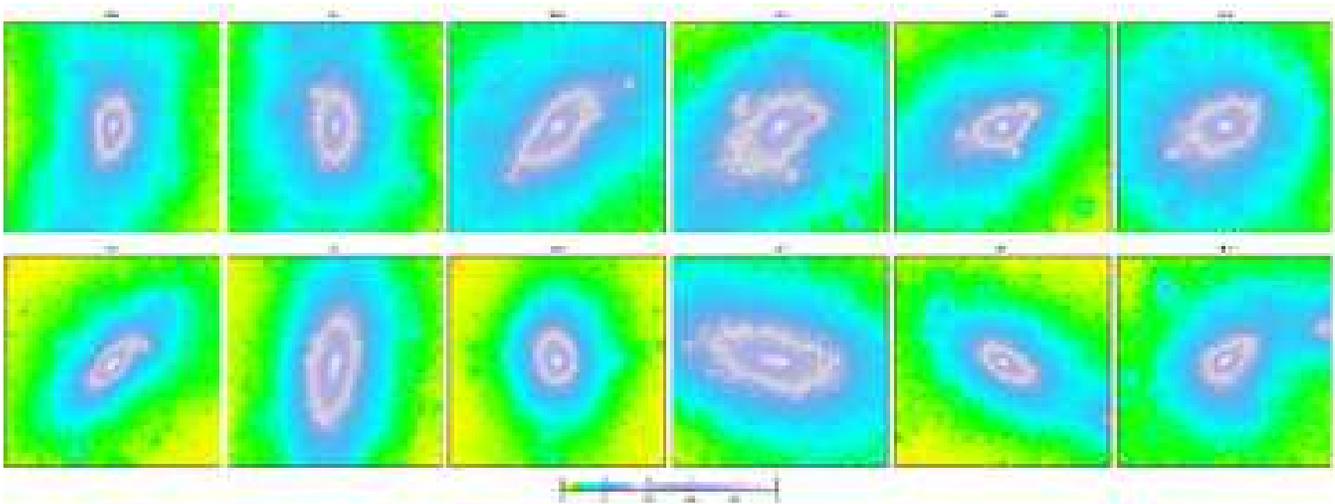}
\caption{Convergence maps of the clusters classified as ``regular''. The
  side-length of each panel is $226''$.}
\label{fig:kregular}
\end{figure*}

We can interpret this result as follows. The mass within the inner
region of the cluster is constrained by the velocity dispersion
data. This implies that the central surface density is approximately
fixed. At the position of the tangential arc $\kappa+\gamma \sim
1$. Imposing axial symmetry while keeping the same density profile
strongly reduces both $\kappa$ and $\gamma$ at the position of the
arc, and the tangential critical line moves inward. In order to
generate a critical line which still passes through the arc, the
axially symmetric model is forced to have a larger projected mass
within the circle passing through the arc. This requires a larger mean
surface density inside the tangential critical line. But, since the
surface density at the cluster centre is fixed, the parameters adjust
themselves to flatten the surface density profile. The flattening is
limited only by the constraint given by the position of the radial
arc. Indeed, the radius of the radial critical curve tends to become
larger as $\beta$ becomes smaller. Since for the case shown in
Fig.~\ref{fig:fitexample} the constraints on the position of the
critical lines are strong, the best fit is obtained at the expense of
the weaker velocity dispersion constraints (and results in a larger $\chi^2$, as given in the last column of Table~\ref{table:bestfit}). Thus the recovered best
fit density profile is characterised by the smallest central density
which is still compatible with the data. This keeps the size of the
radial critical line small. The resulting velocity dispersion profile
for the axially symmetric fit generally underestimates the true
profile. With weaker constraints imposed by the tangential and
radial arcs, the best fit axially symmetric model will be generally
characterised by larger values of $\beta$. 

To simplify the comparison between the positional uncertainties of the critical lines and the errors used in our implementation of $\chi^2_{lens}$, we show in Fig.~\ref{fig:ldist} the profiles of the tangential (thick line) and of the radial (thin line) eigenvalues along the lines connecting the cluster center to the arcs T1 and R1 of Fig~\ref{fig:fitexample}, as derived from the ray-tracing simulation. We zoom into a region of width $\pm 1"$ around the critical lines. We see that the interval $\Delta \lambda_t\sim \pm 0.01$ (thick dotted lines) corresponds to $\pm 0.3"$ around the tangential critical line. Similarly, $\Delta \lambda_r \sim \pm 0.03$ corresponds to an uncertainty $\pm 0.45"$ in the position of the radial critical line. Such uncertainties are well in agreement with typical errors on the position of arcs \citep[see e.g. Table 3 of ][]{SA03.1}.

%For example if one assumes
%$\Delta \lambda_t=\Delta \lambda_r=0.1$, the best fit inner slope in
%this example becomes $\beta \sim 0.5$.

\begin{figure*}
  \centering
  \includegraphics[width=3.65cm]{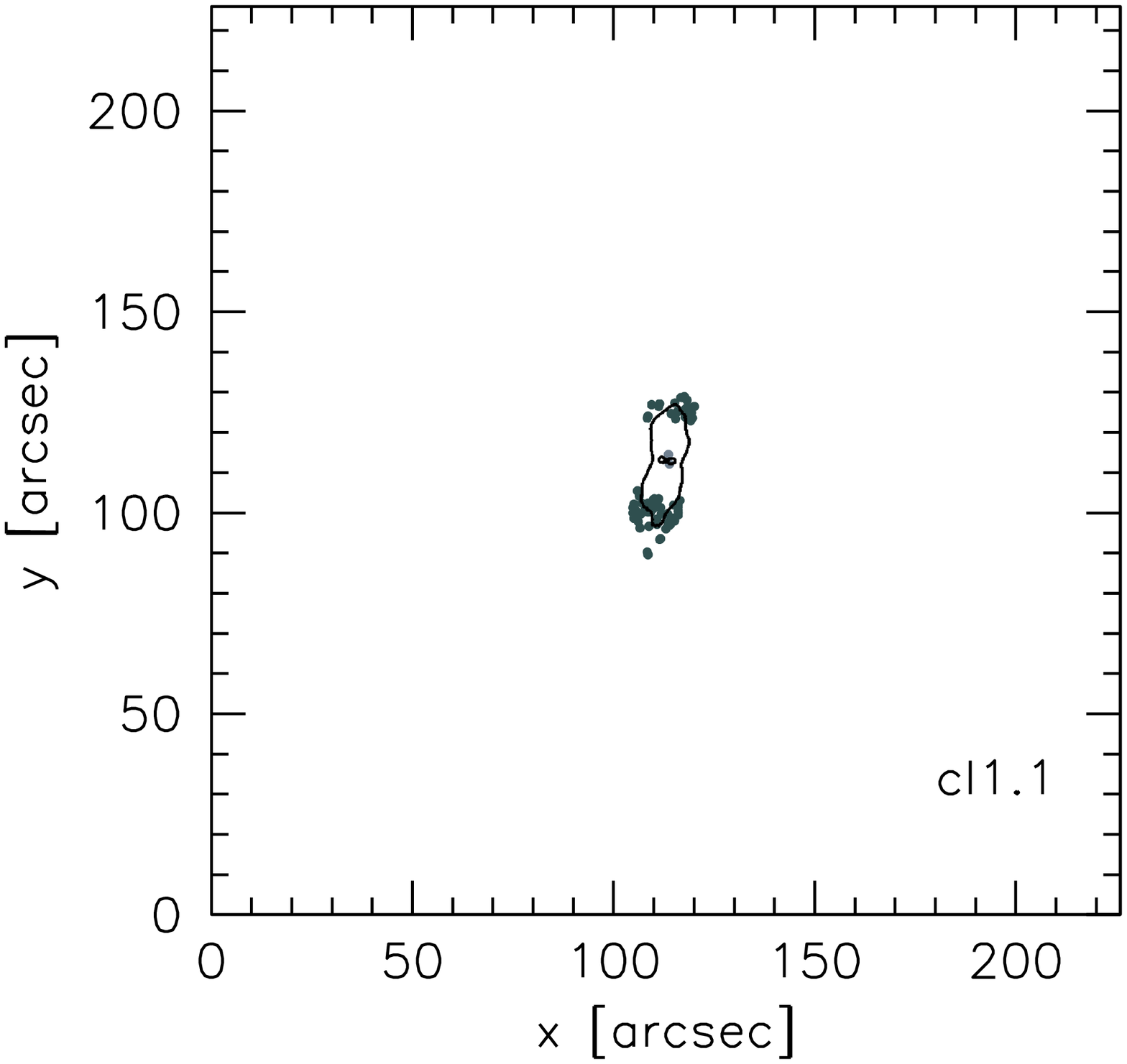}
  \includegraphics[width=3.65cm]{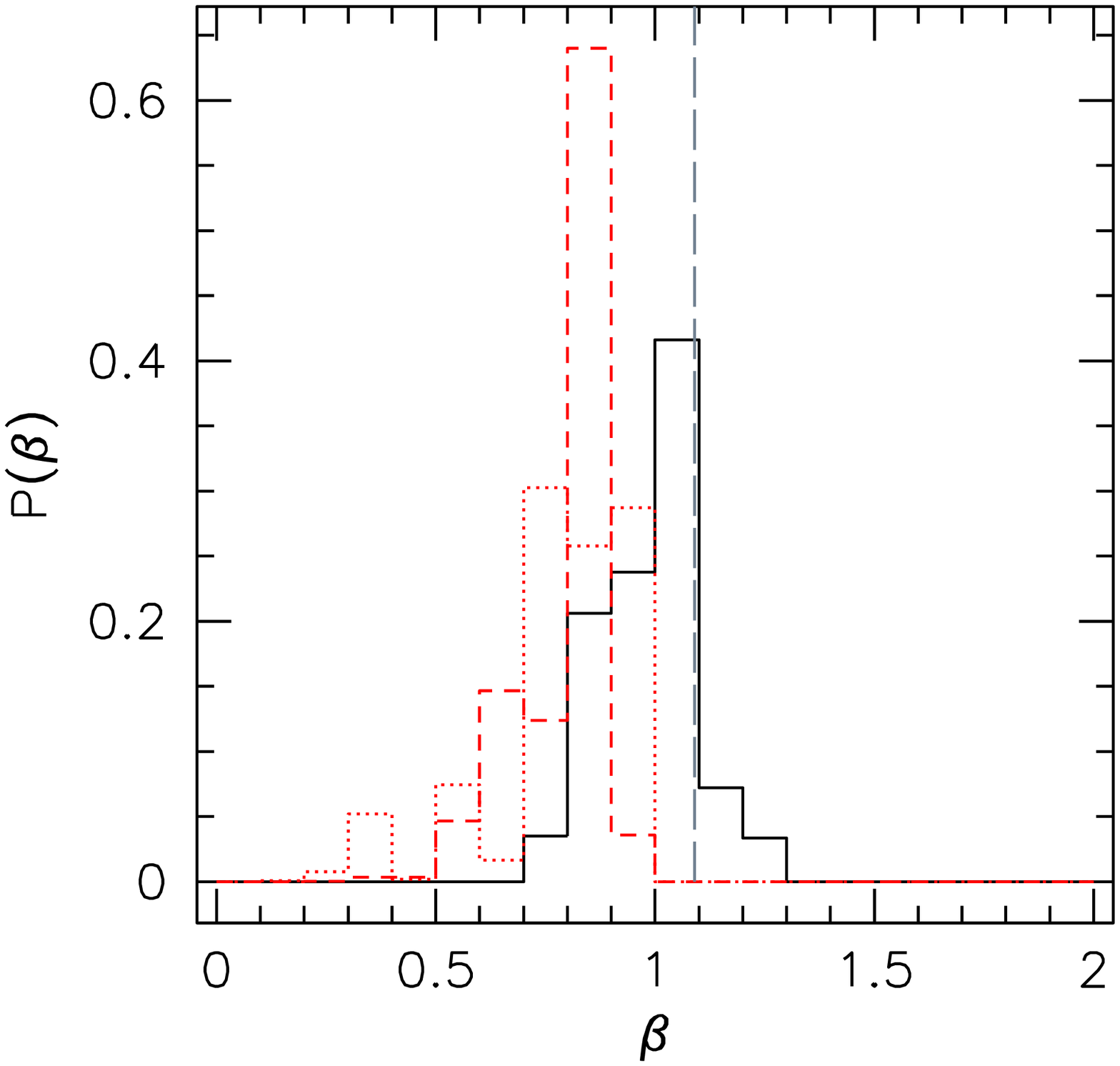}
  \includegraphics[width=3.65cm]{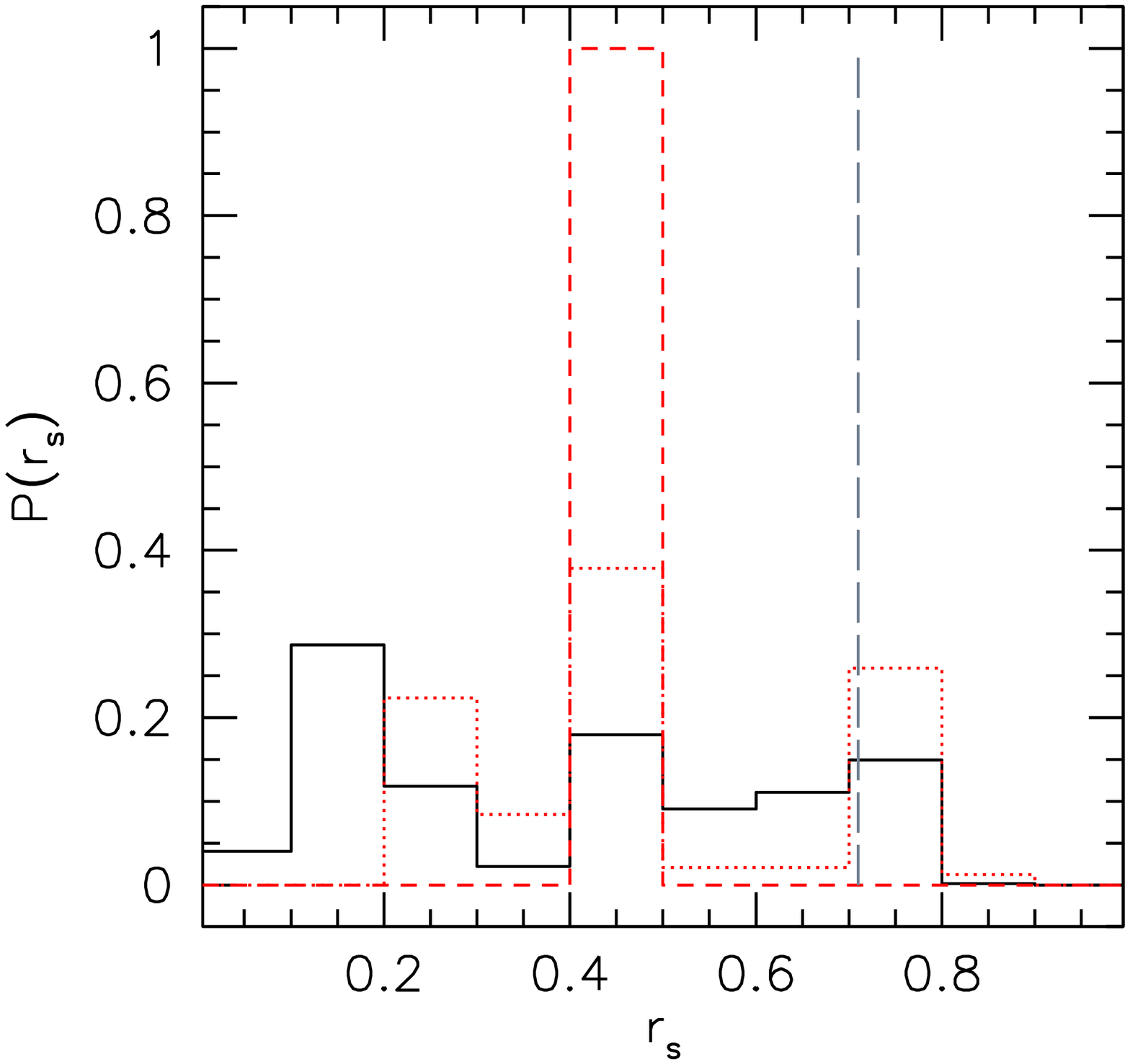}
  \includegraphics[width=3.65cm]{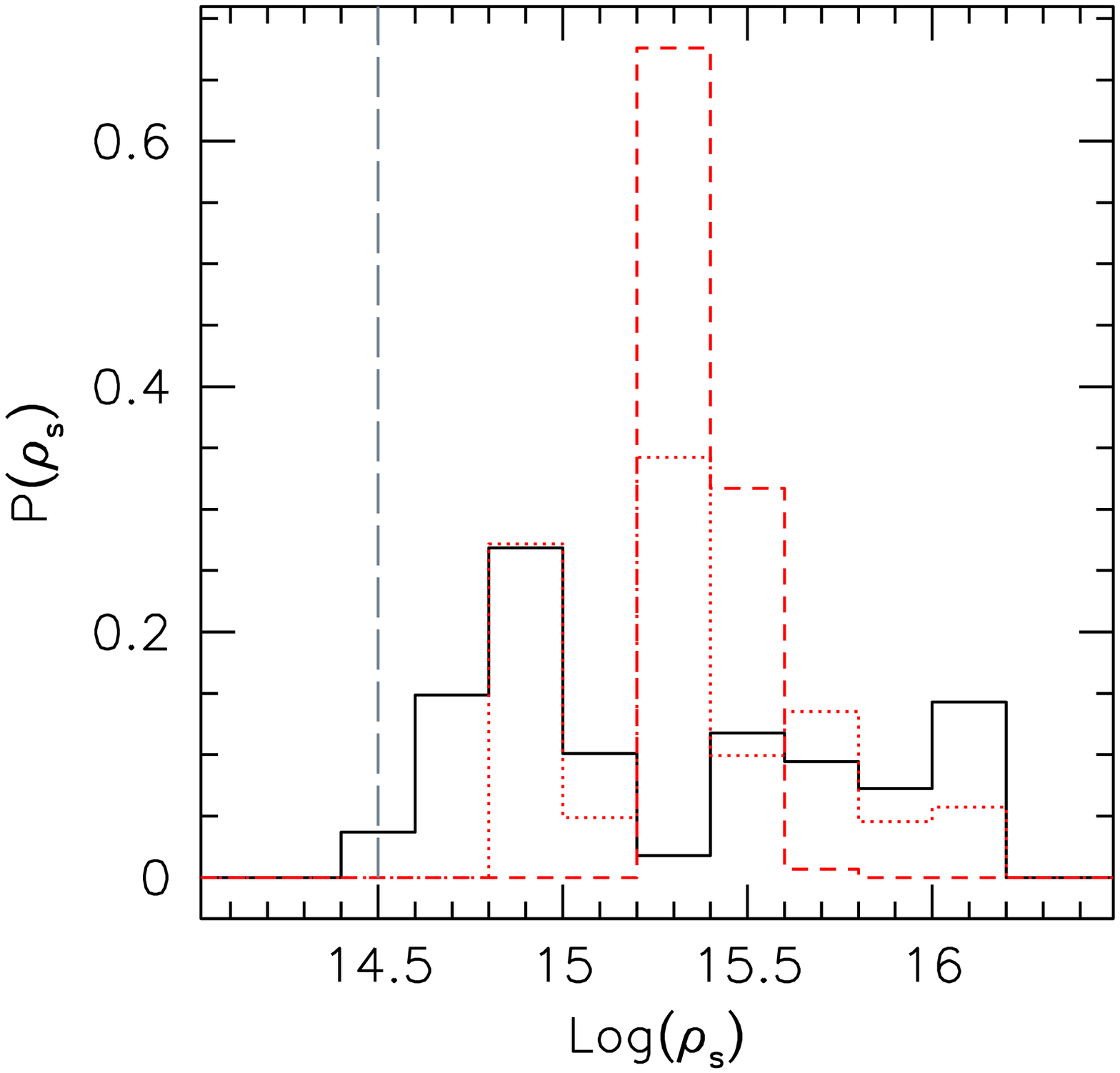} \\
  \includegraphics[width=3.65cm]{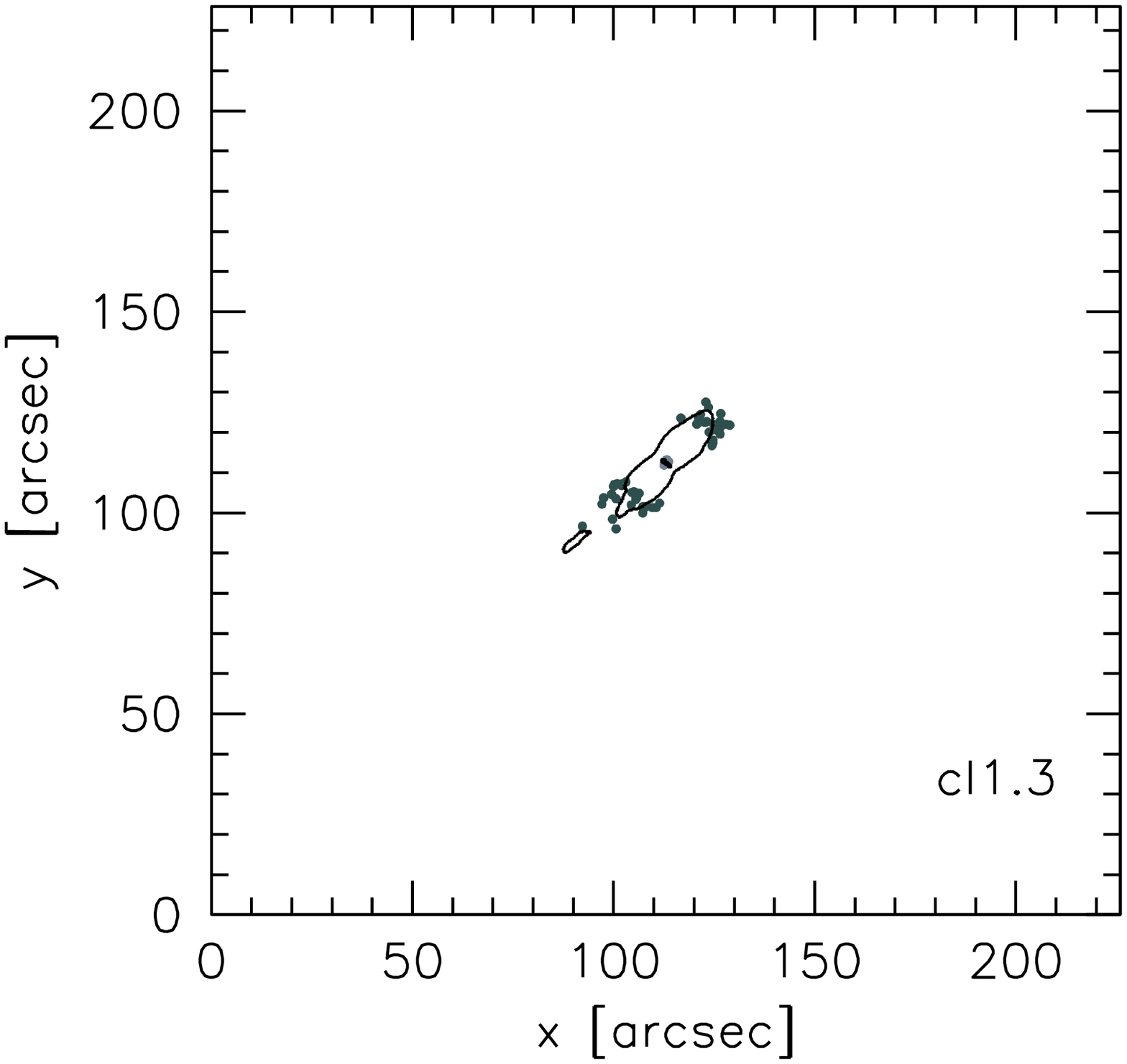}
  \includegraphics[width=3.65cm]{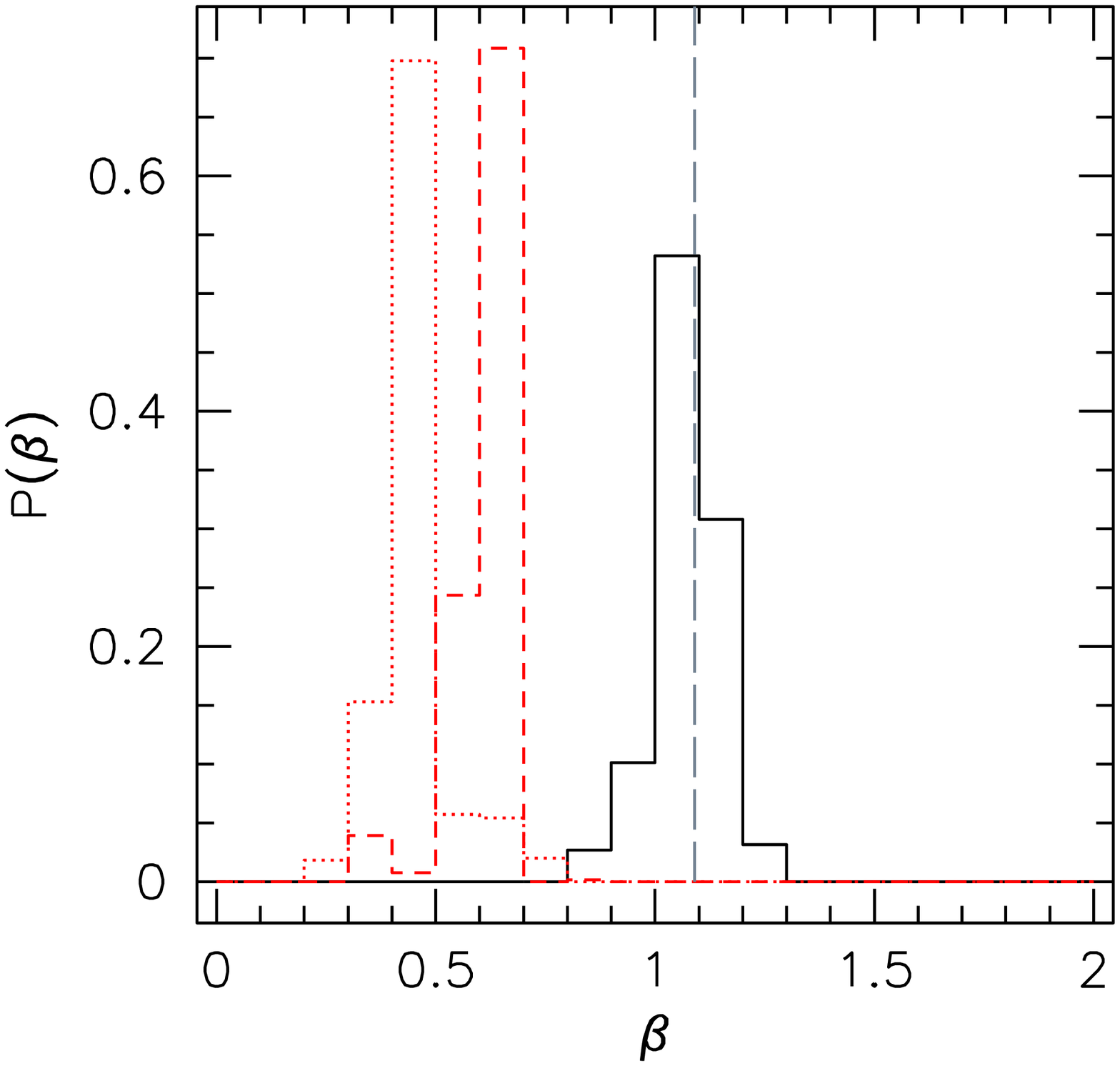}
  \includegraphics[width=3.65cm]{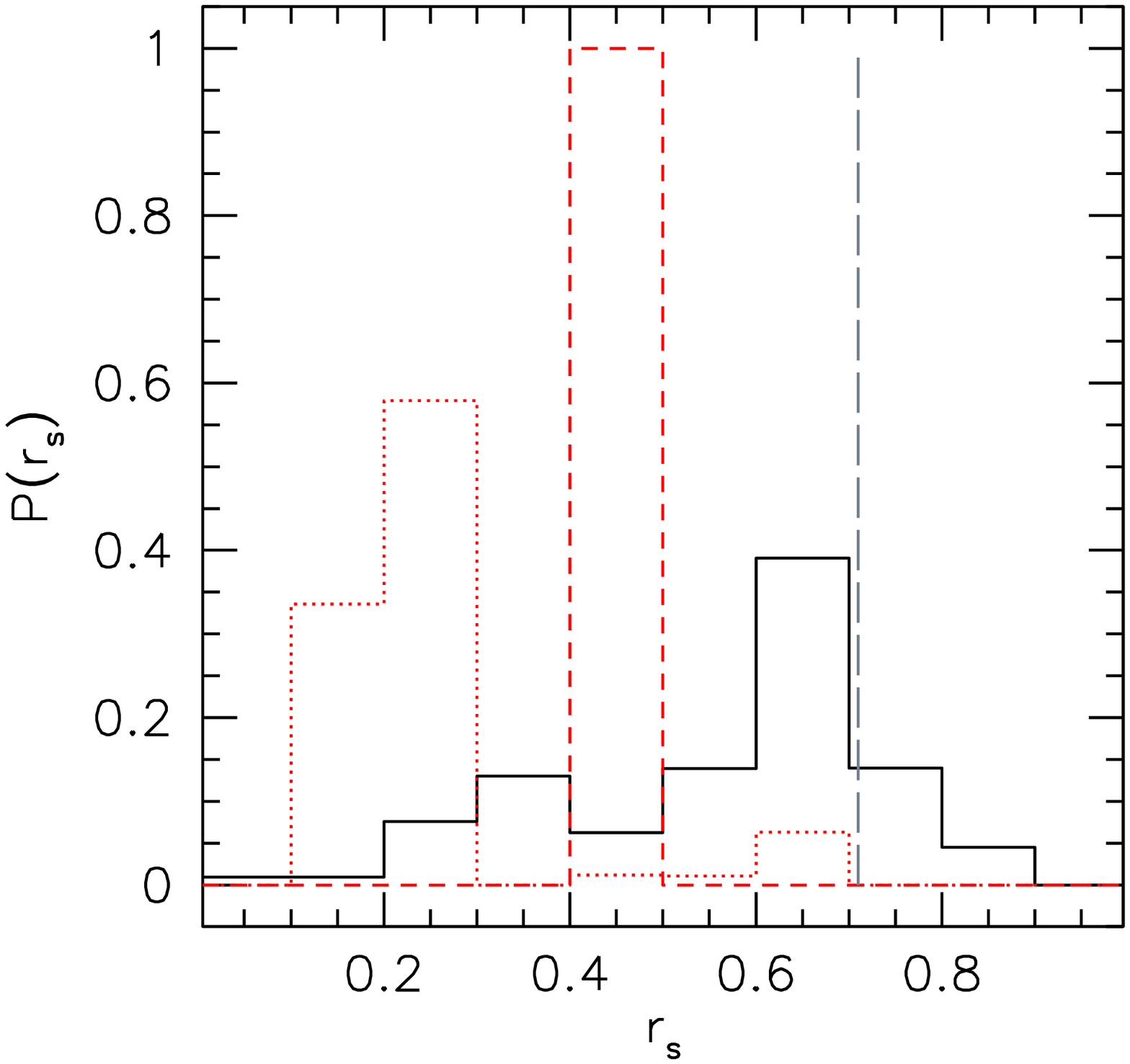}
  \includegraphics[width=3.65cm]{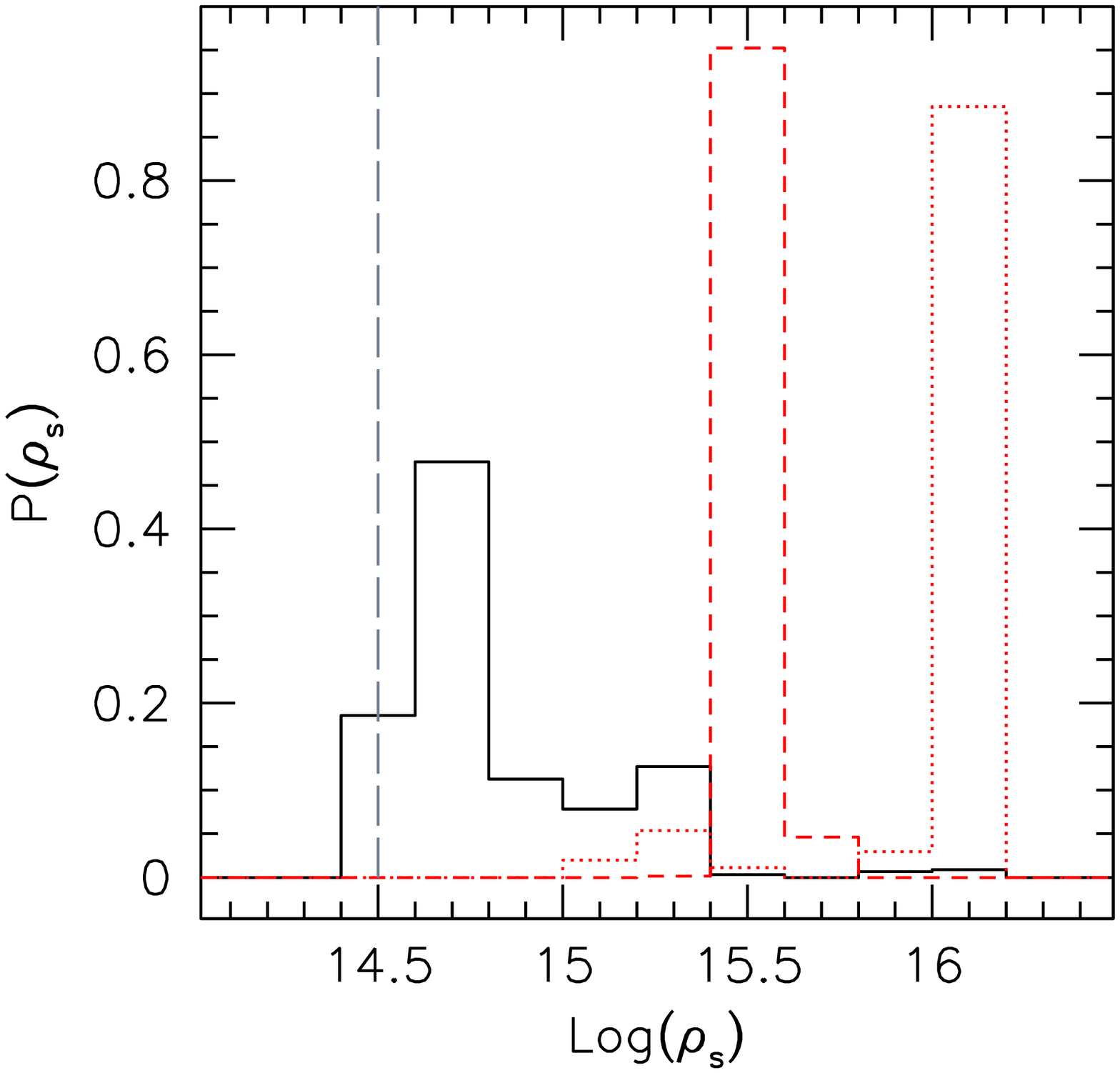} \\
  \includegraphics[width=3.65cm]{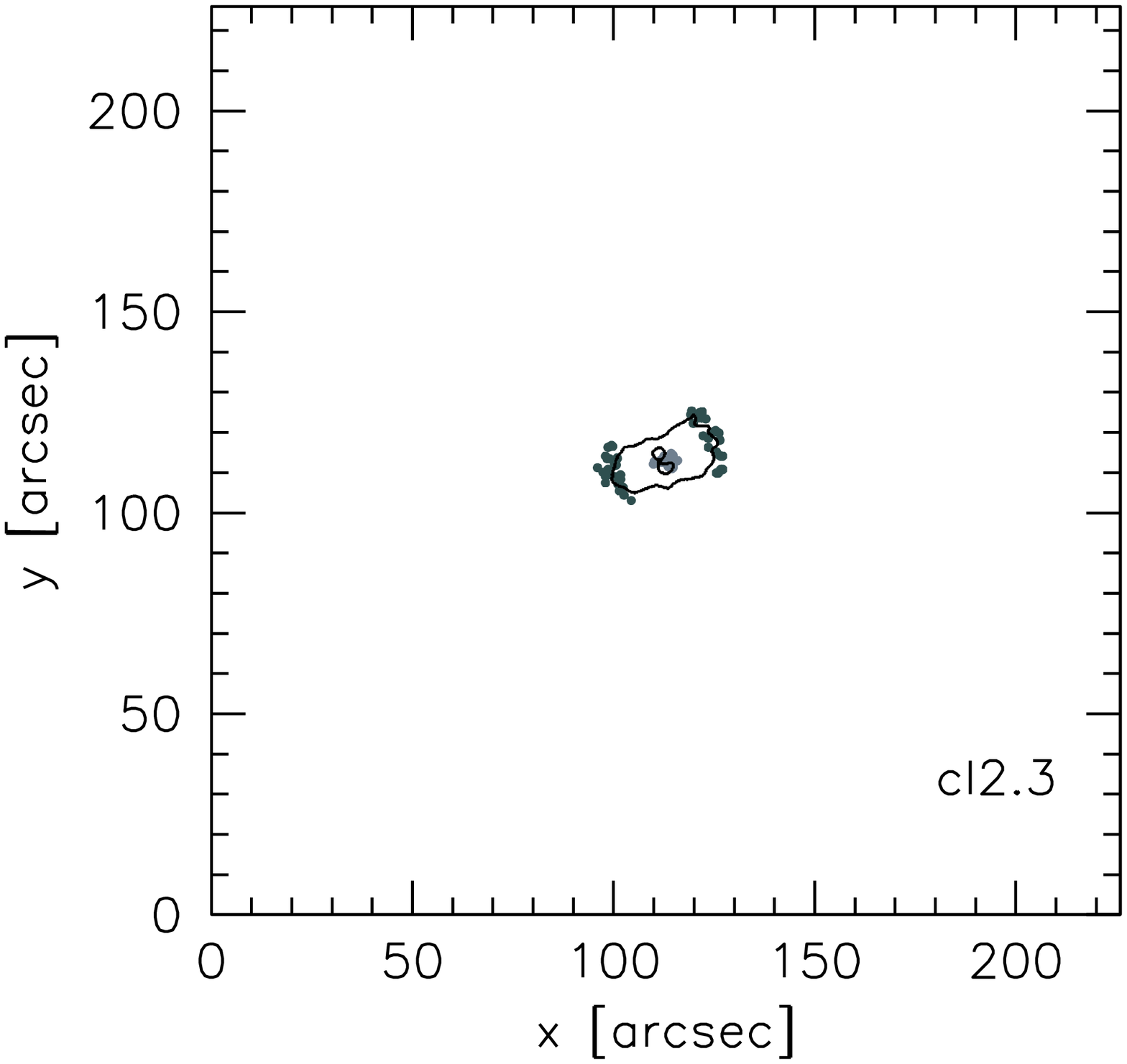}
  \includegraphics[width=3.65cm]{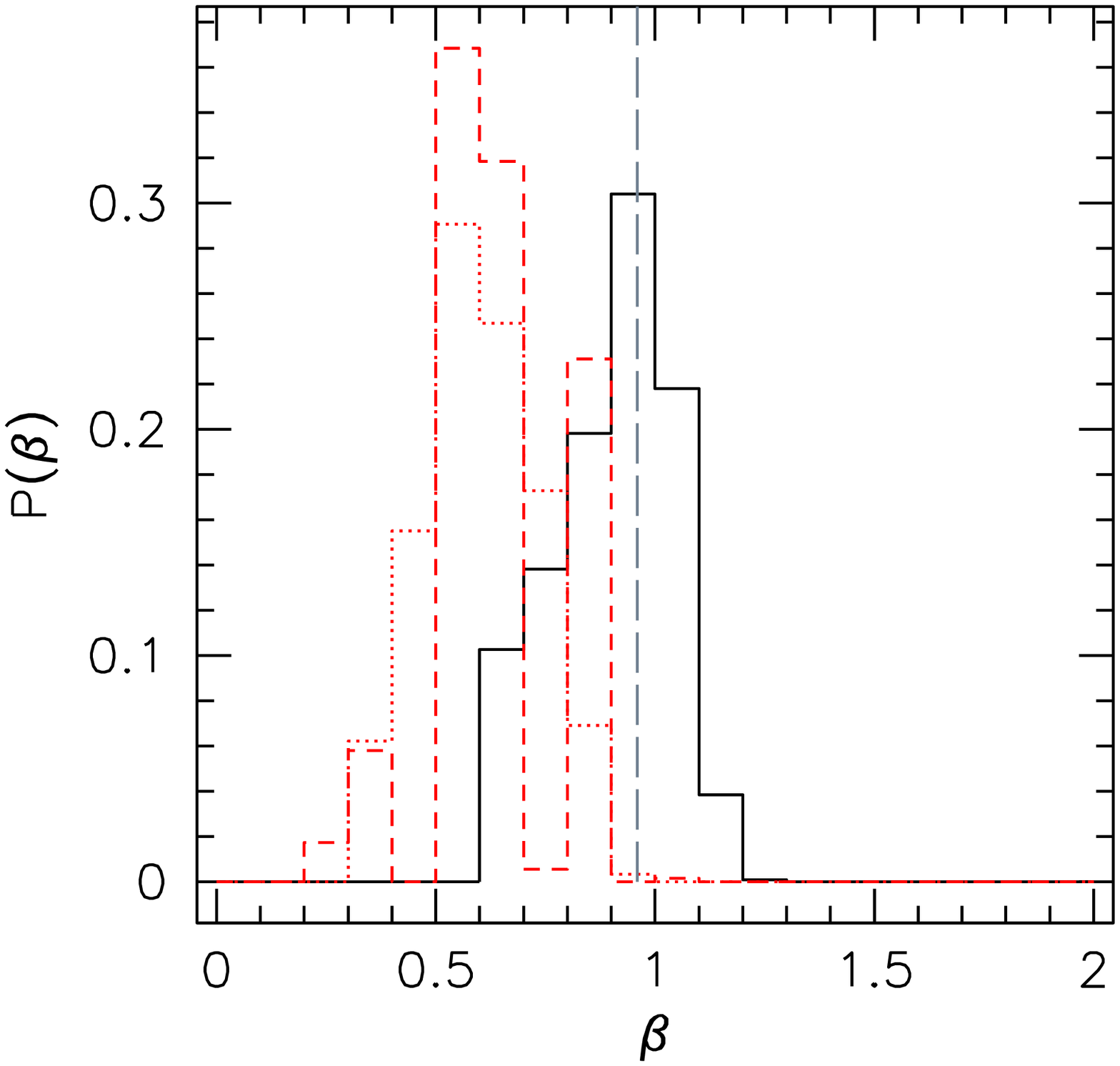}
  \includegraphics[width=3.65cm]{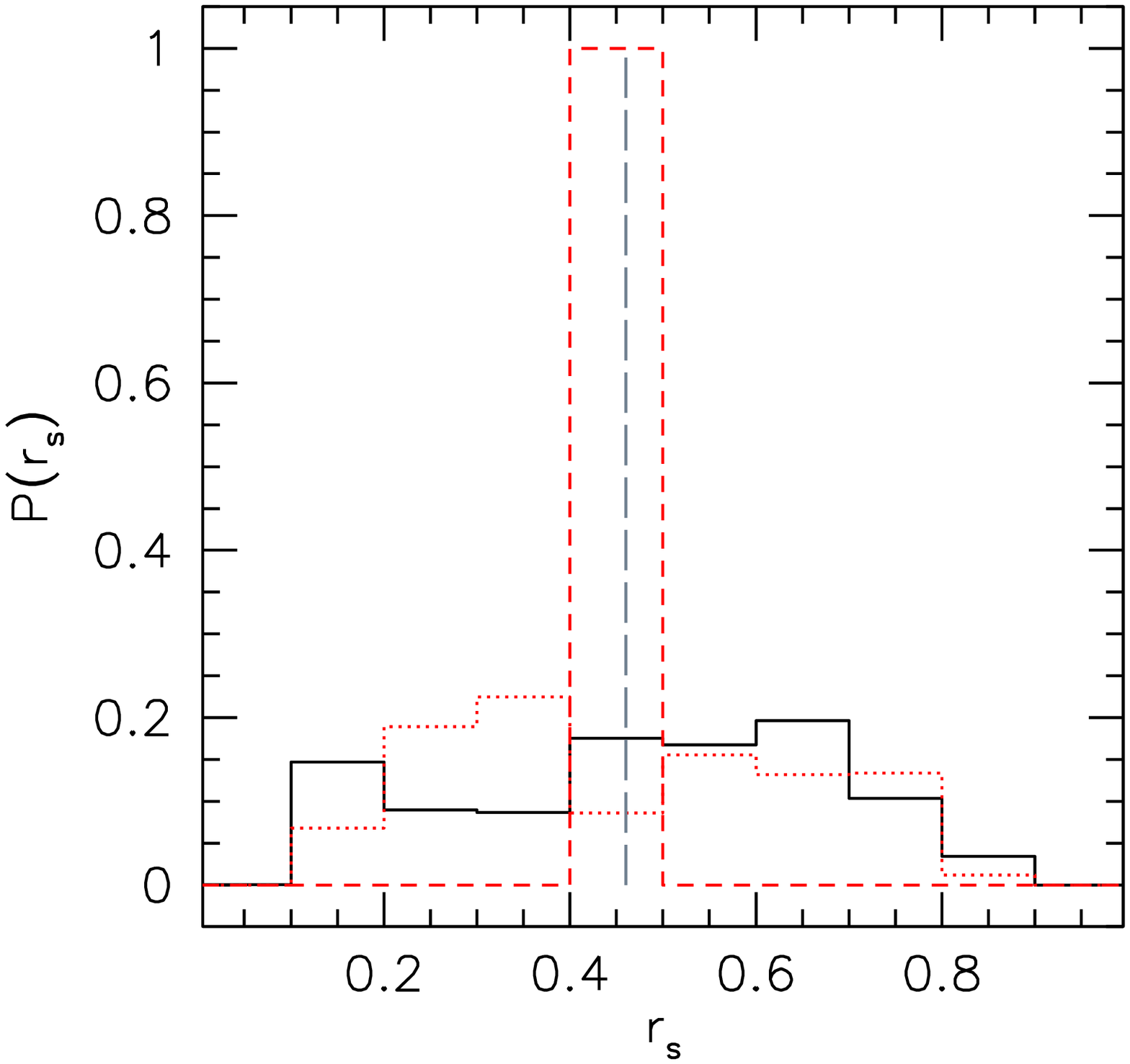}
  \includegraphics[width=3.65cm]{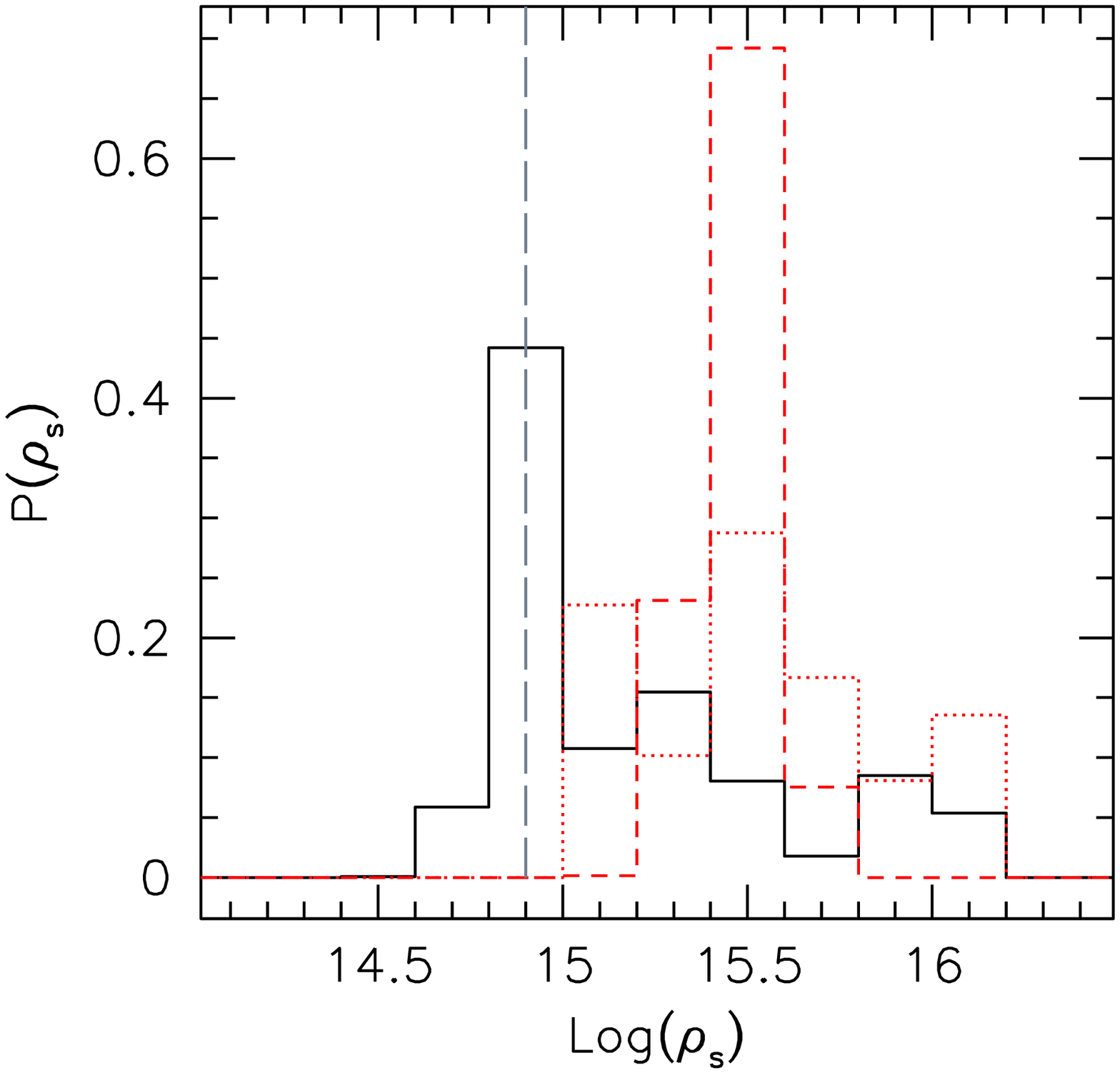} \\
  \includegraphics[width=3.65cm]{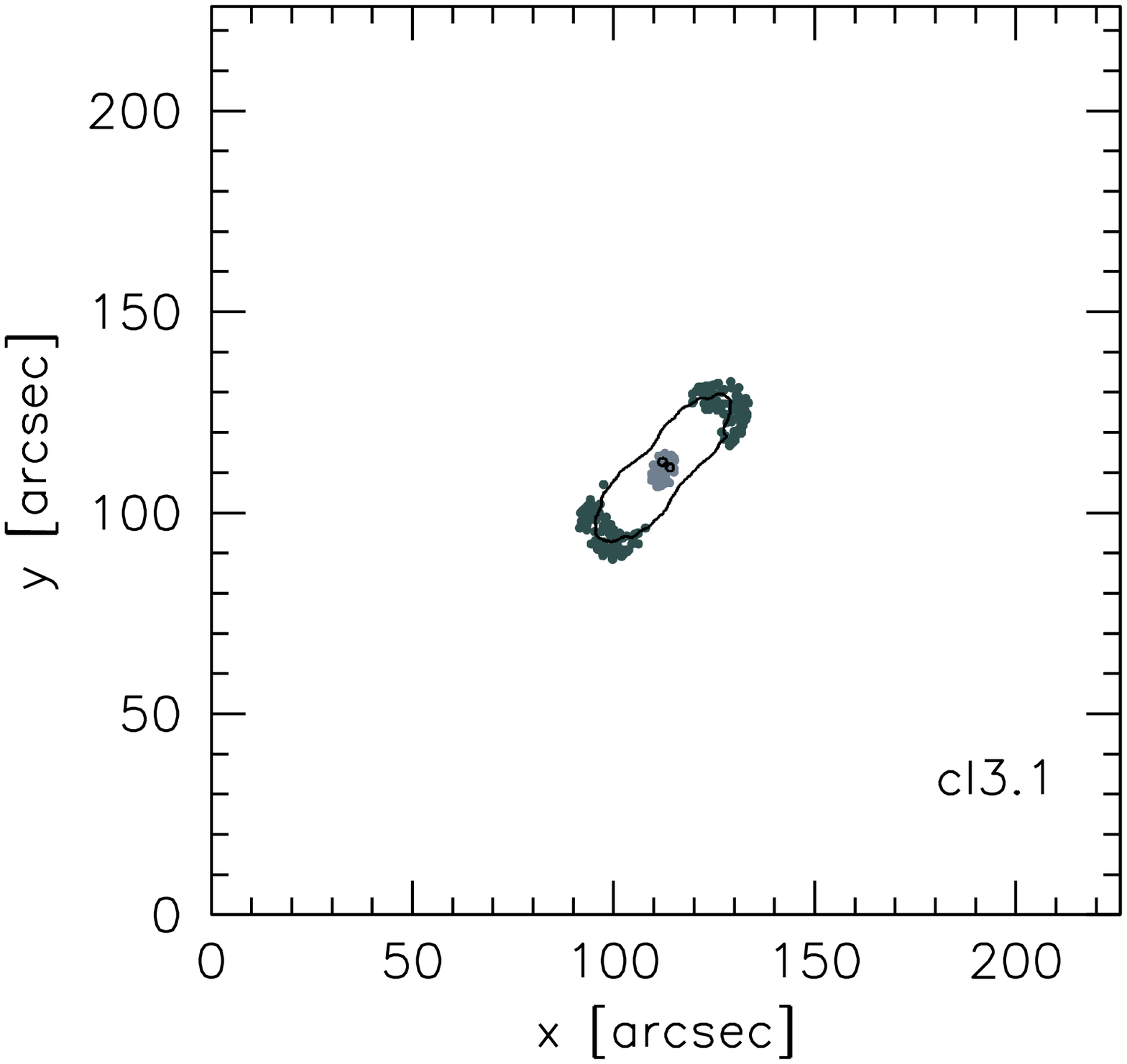}
  \includegraphics[width=3.65cm]{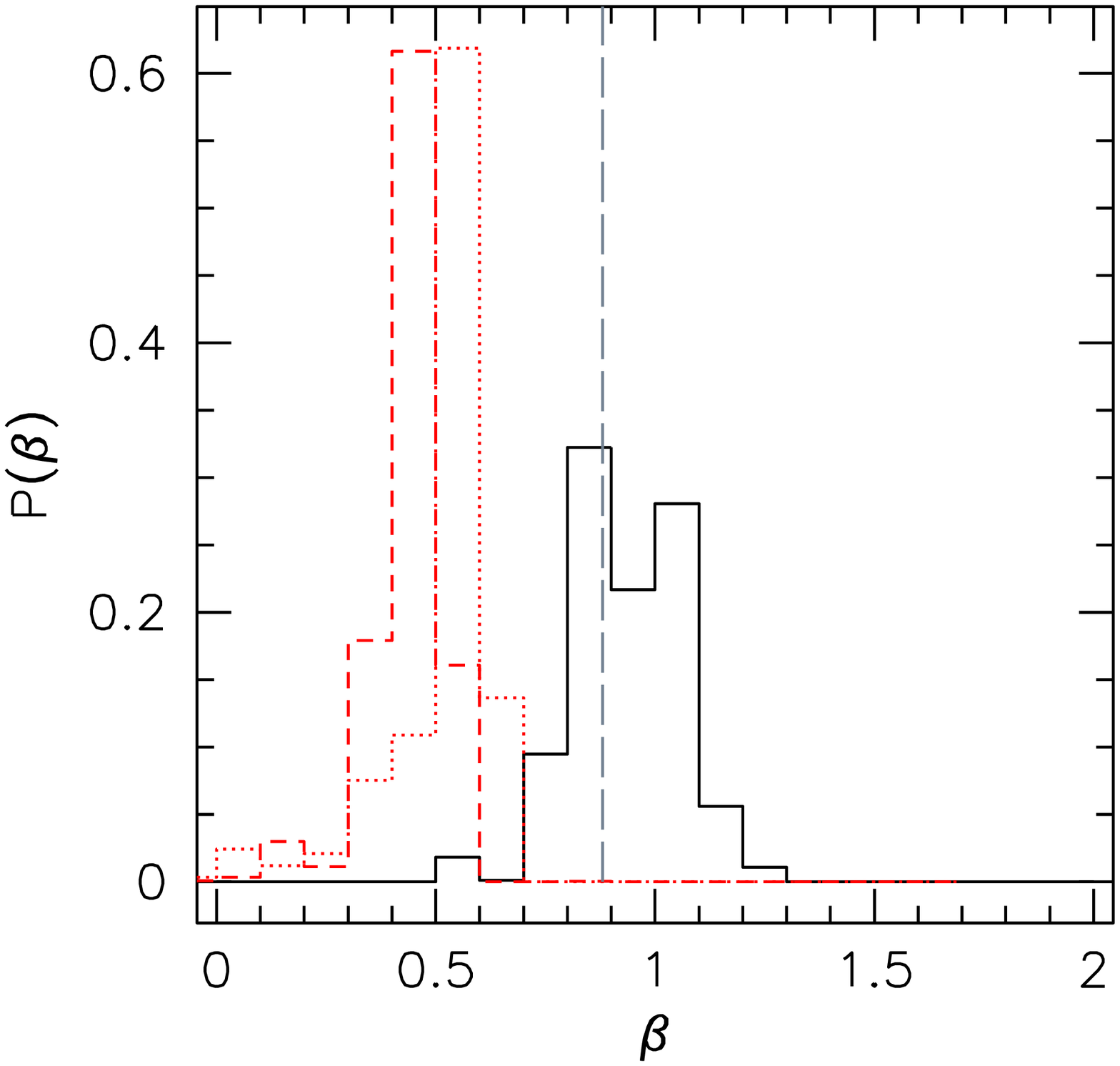}
  \includegraphics[width=3.65cm]{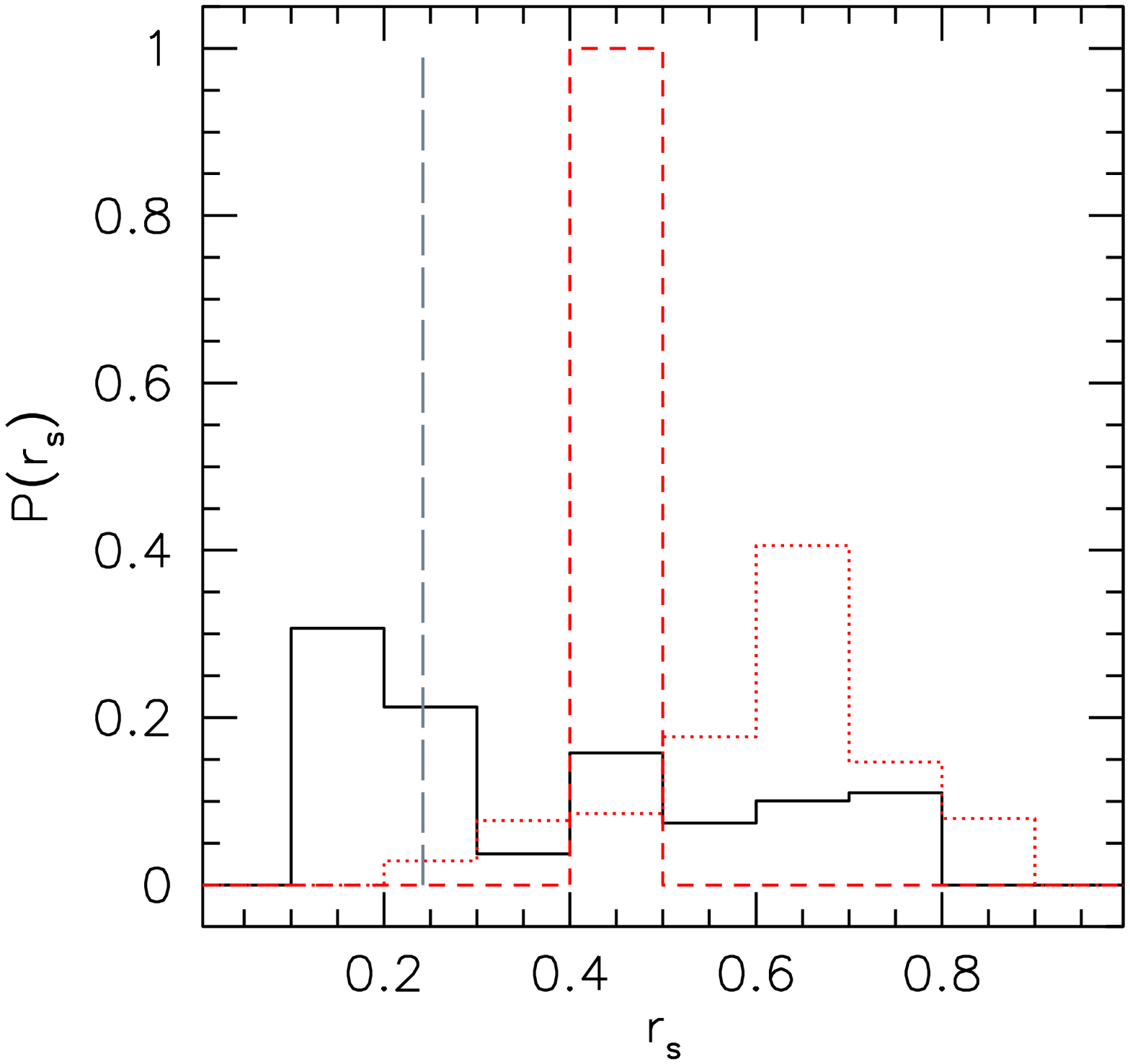}
  \includegraphics[width=3.65cm]{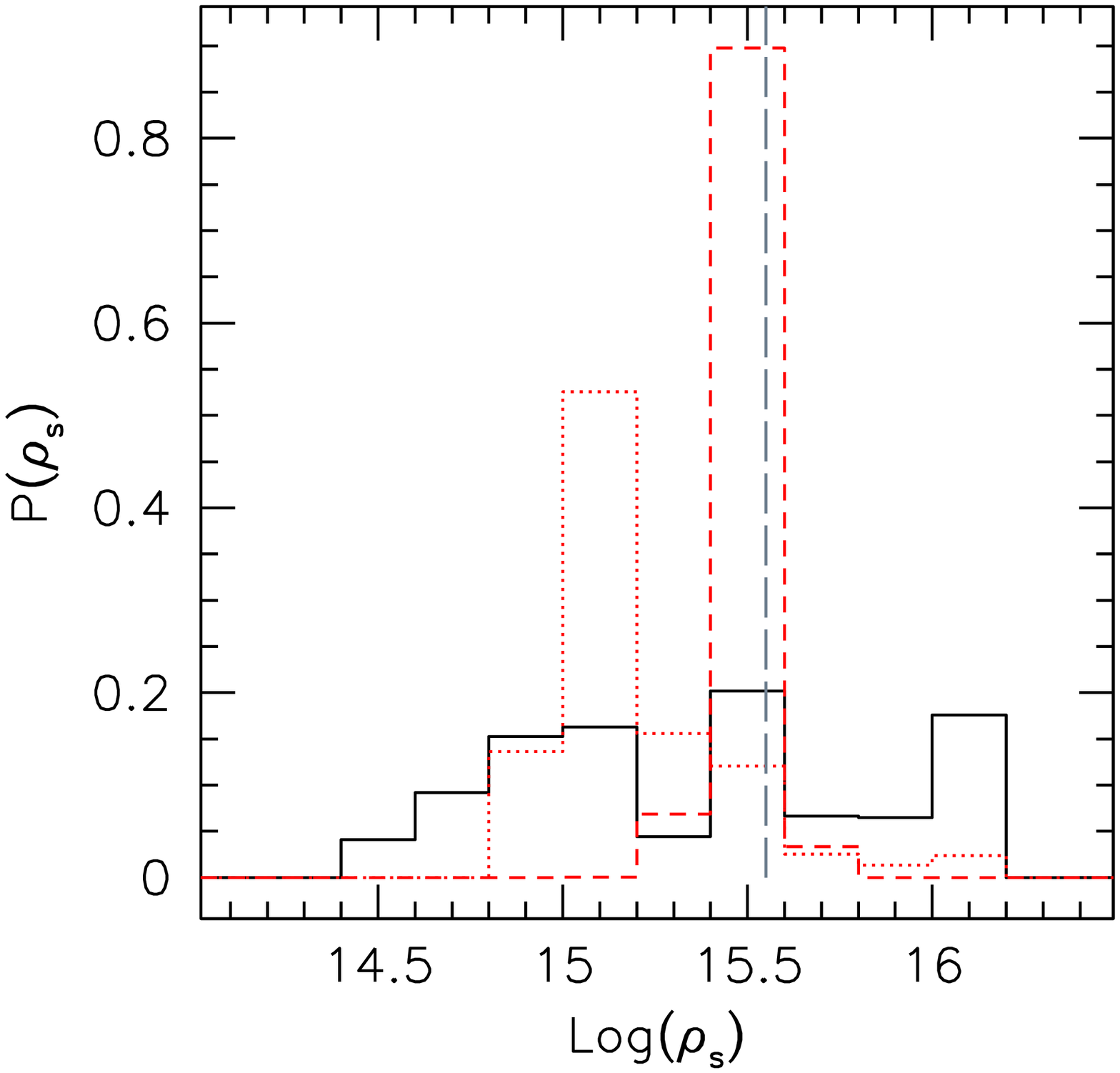} \\
  \includegraphics[width=3.65cm]{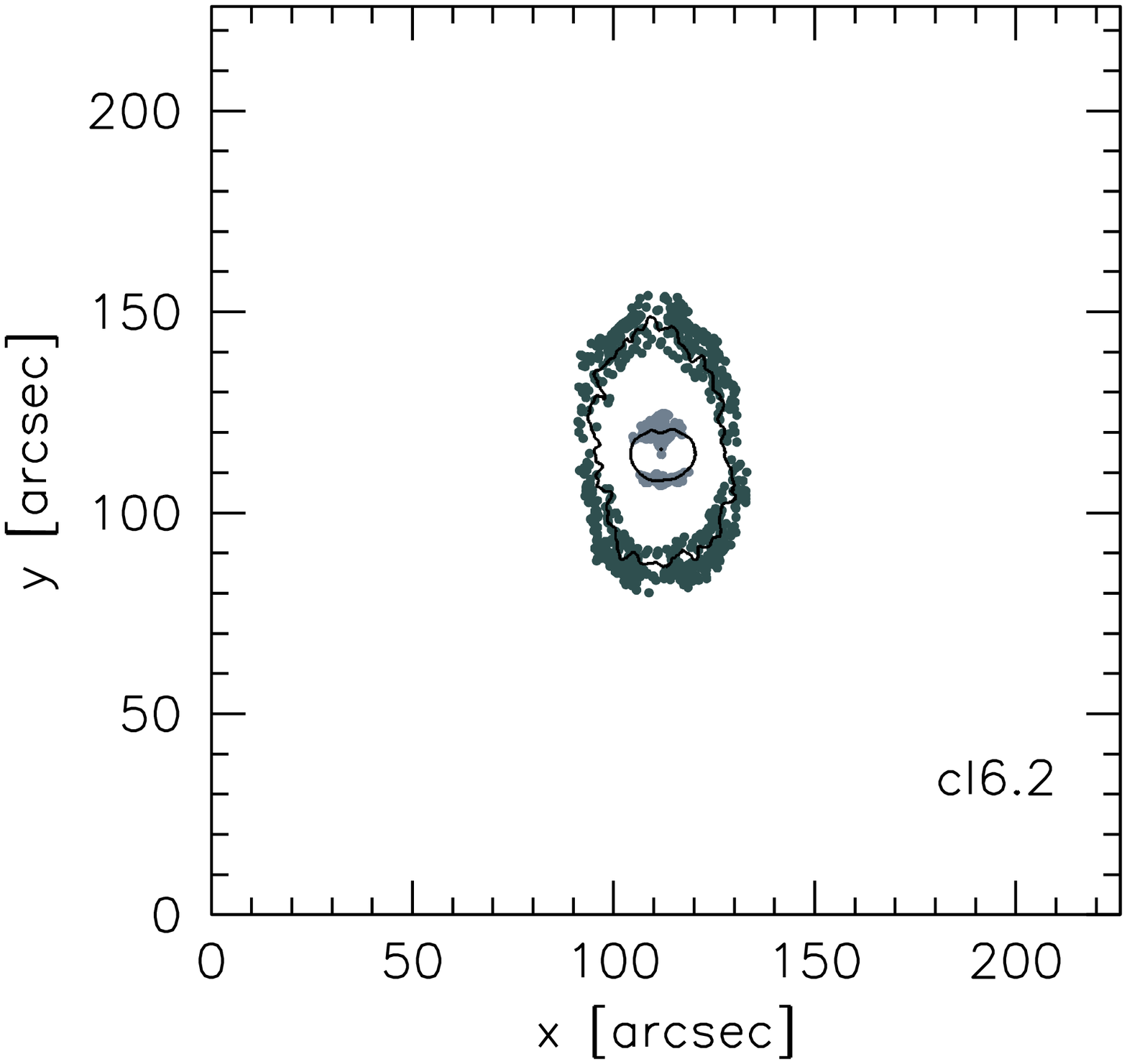}
  \includegraphics[width=3.65cm]{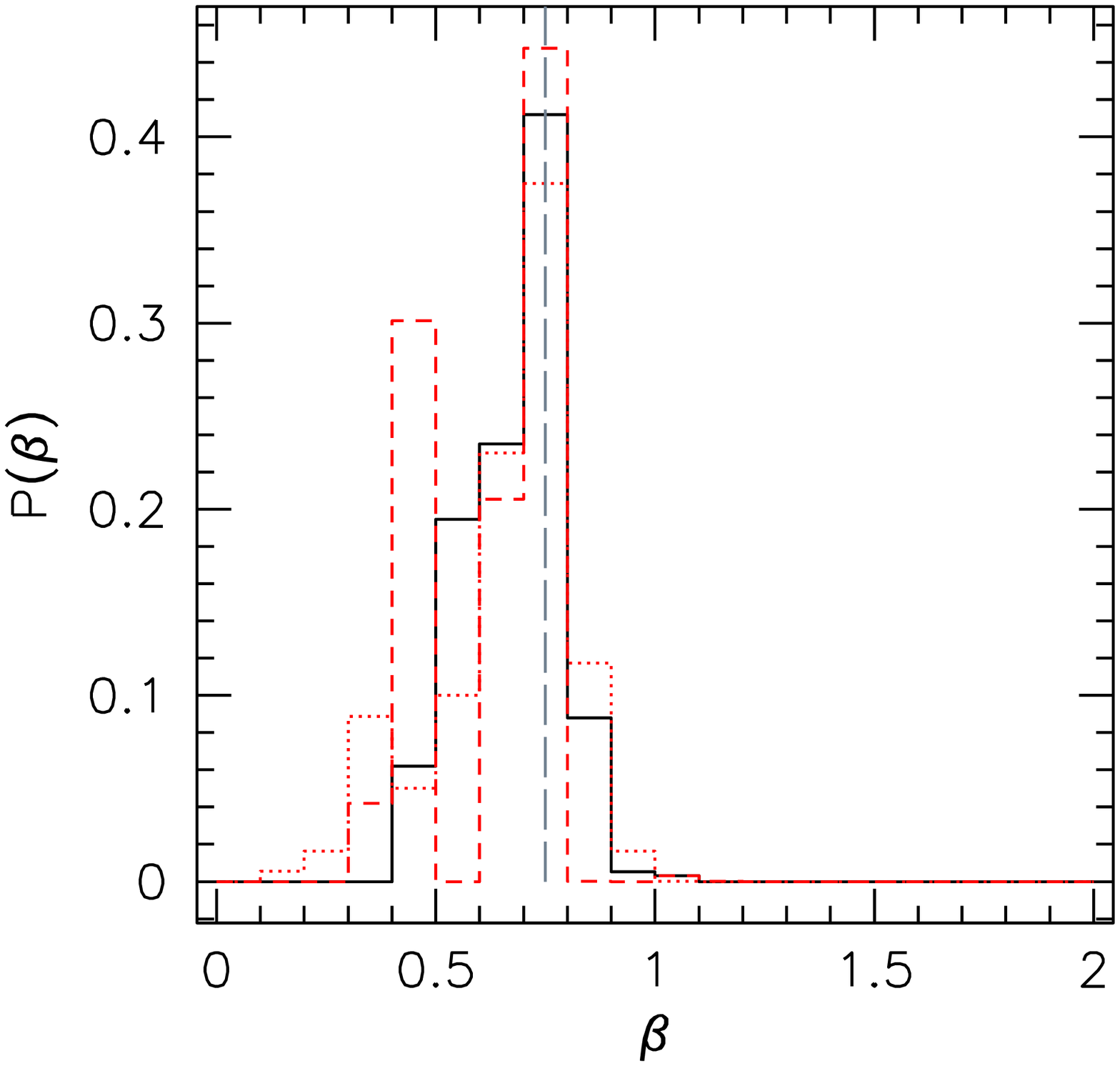}
  \includegraphics[width=3.65cm]{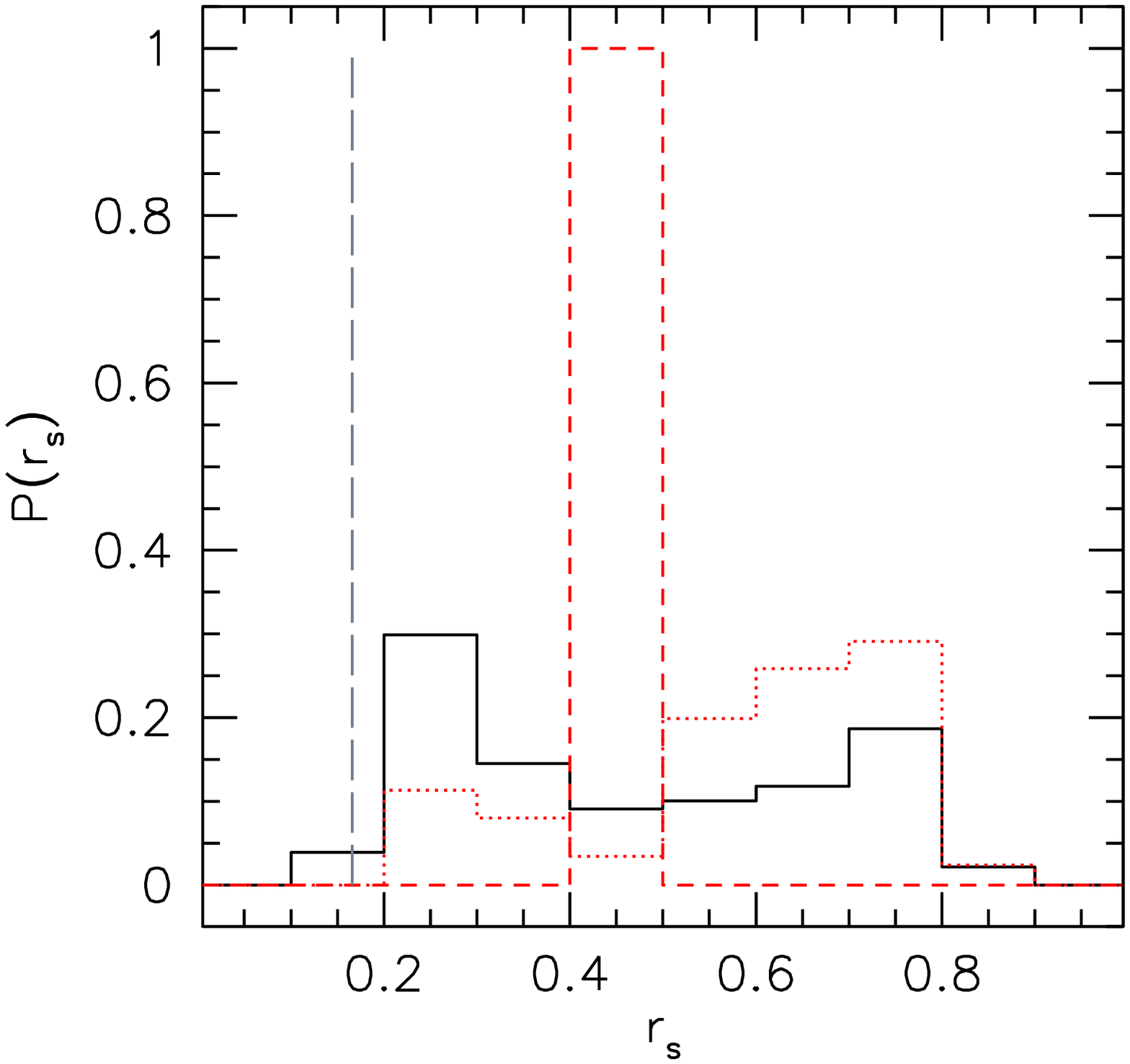}
  \includegraphics[width=3.65cm]{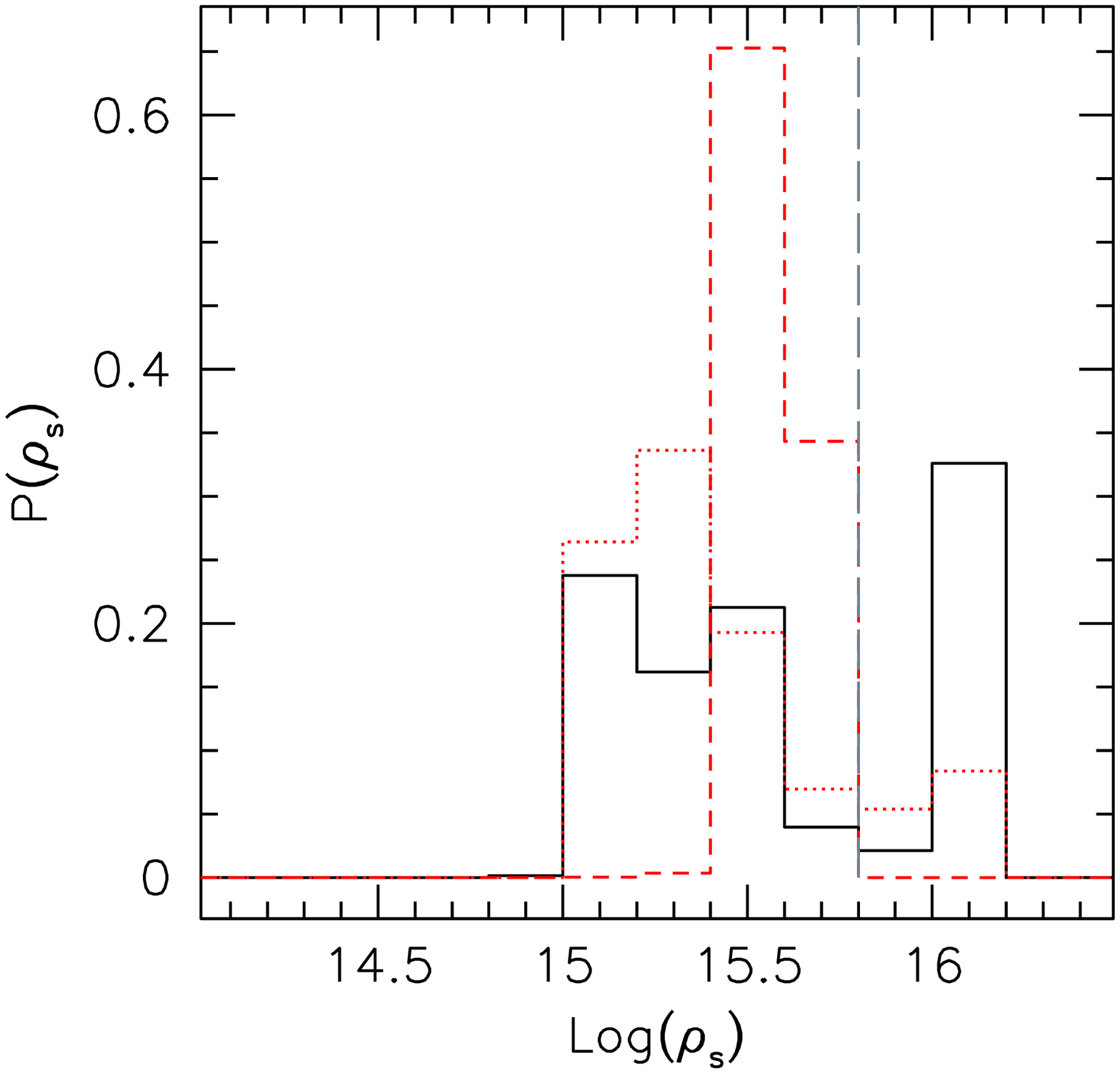} \\
  \includegraphics[width=3.65cm]{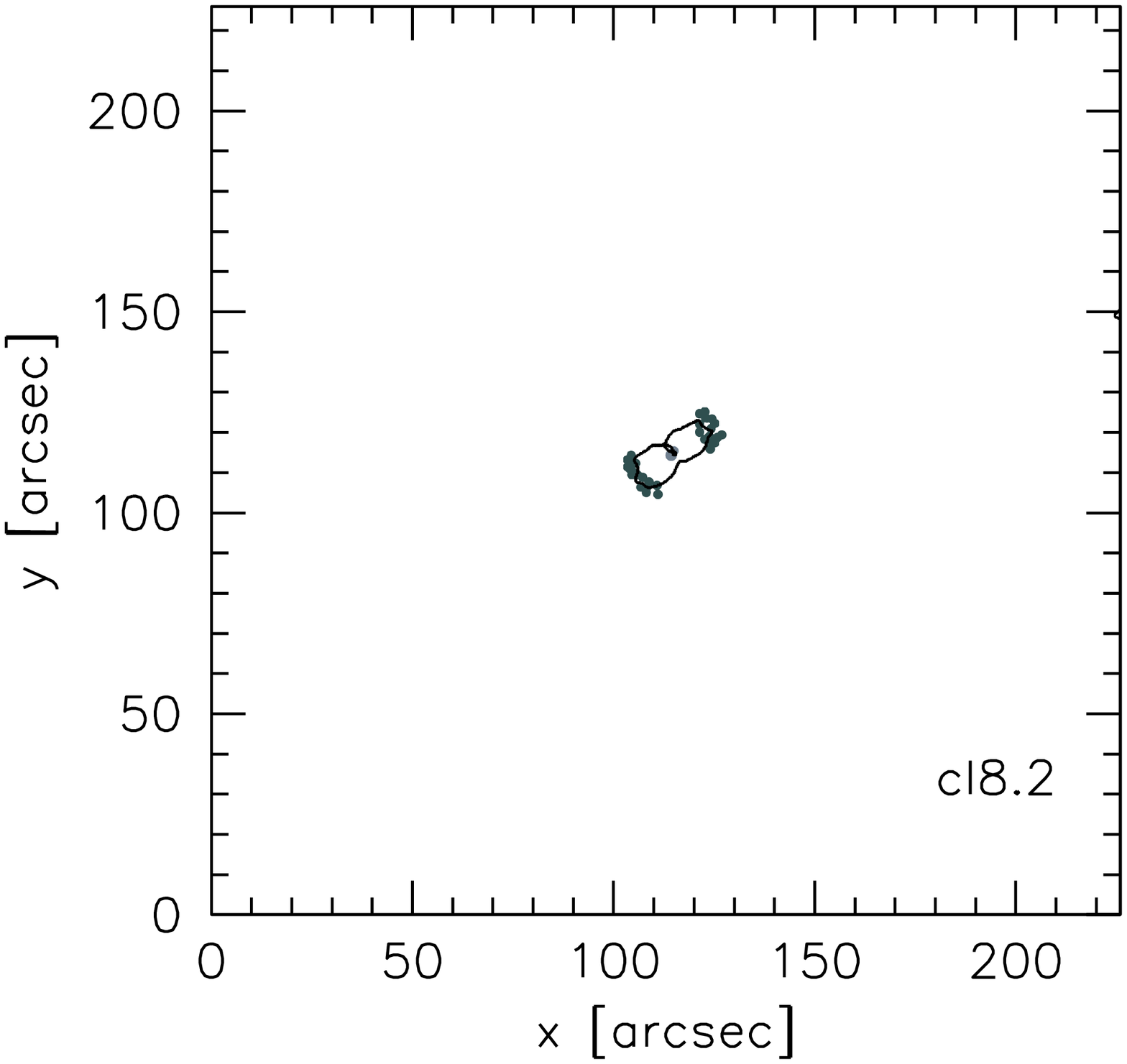}
  \includegraphics[width=3.65cm]{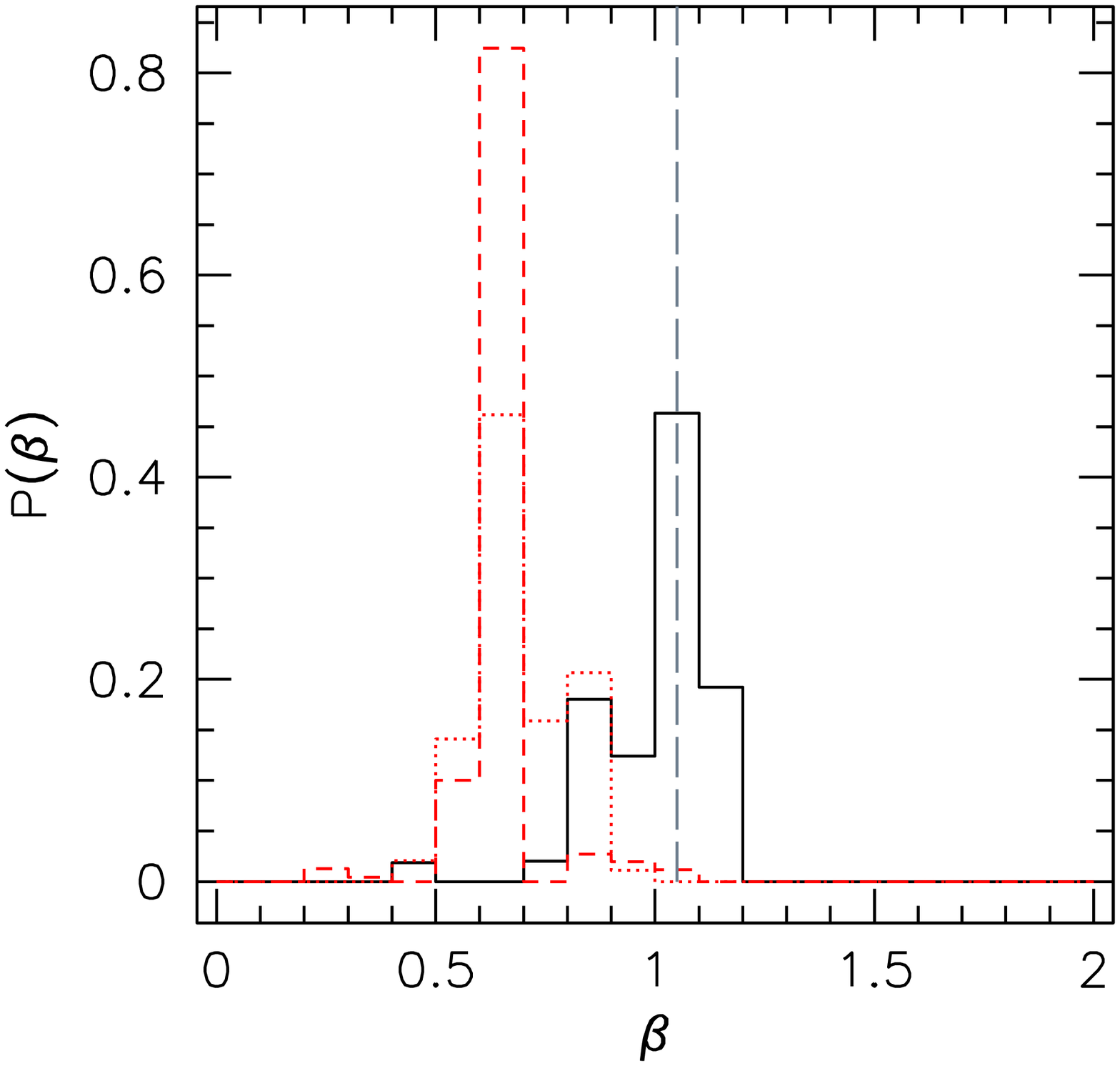}
  \includegraphics[width=3.65cm]{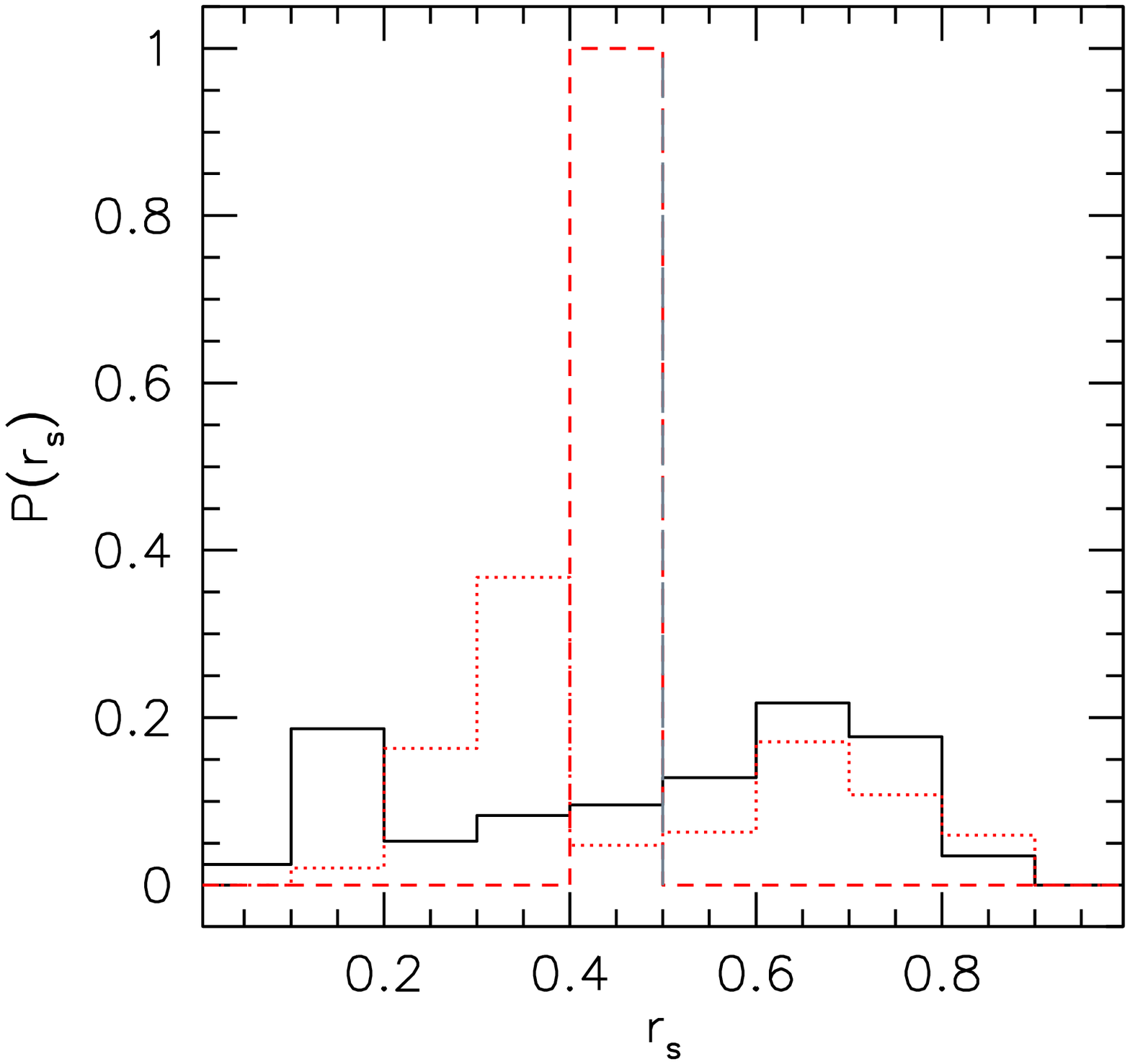}
  \includegraphics[width=3.65cm]{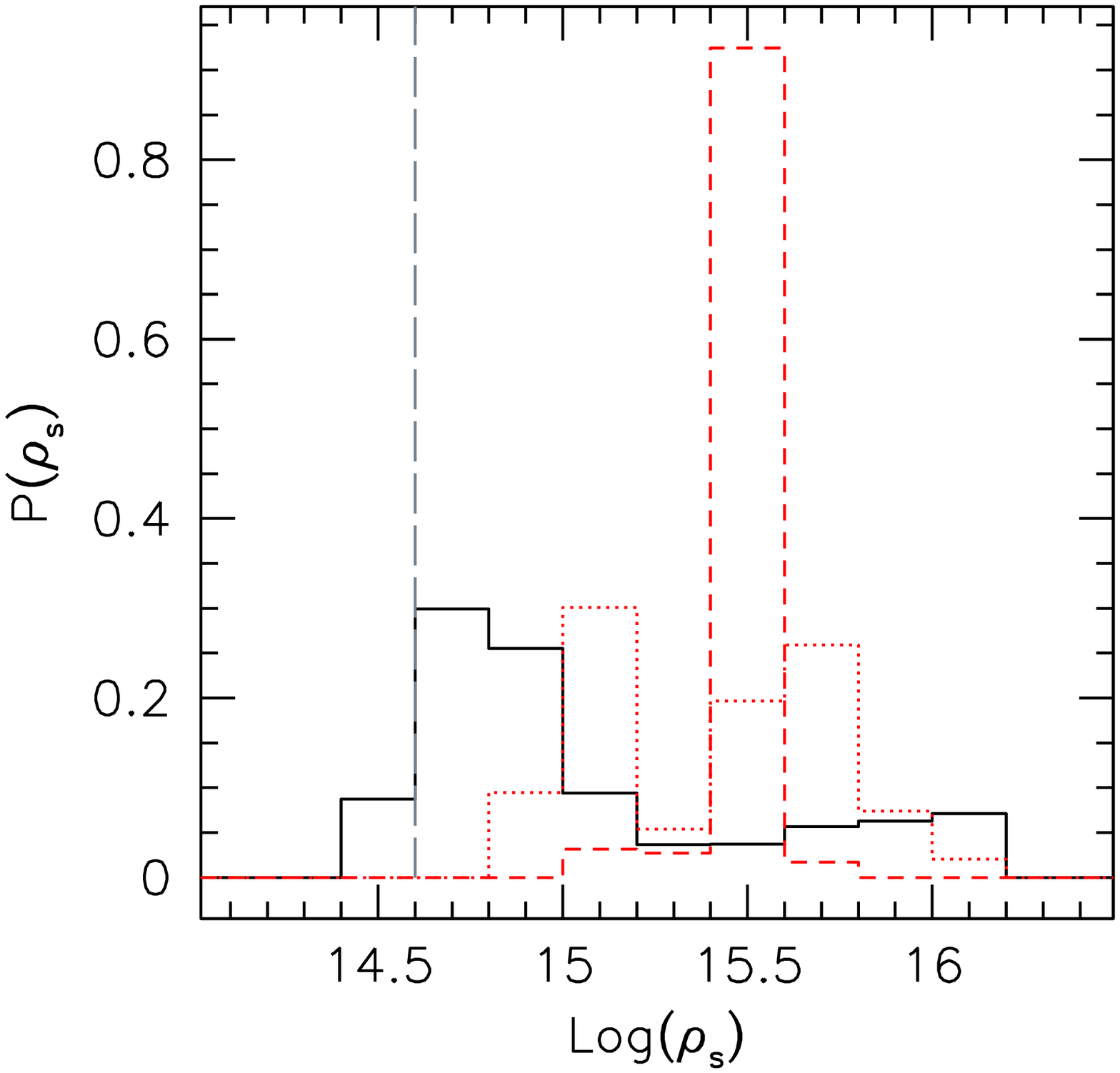}
\caption{First column: critical lines of six of the clusters
  classified as ``regular''. Gray dots indicate the position of tangential and
  radial arcs found in the ray-tracing simulations and used in our analysis.
  Second to fourth column: probability distribution functions of
  $\beta$, $r_{\rm s}$ and $\rho_{\rm s}$ resulting from constraining the
  galaxy cluster density profiles using couples of radial and tangential arcs
  combined with velocity dispersion data. Solid, dotted and short-dashed
  histograms refer to elliptical, axially symmetric and axially symmetric with
  fixed $r_{\rm s}=400 \,h^{-1}\,\mathrm{kpc}$ fitting models, respectively. The dashed
  vertical line in each case indicated the true value of the corresponding
  parameter in the simulation. }
\label{fig:reg_distrib}
\end{figure*}

An important question is how well can we constrain each of
the three parameters characterising the cluster density profile, at
least
using an appropriate elliptical model. We expect considerable degeneracy especially between the parameters $\rho_s$ and $r_s$, since both the lensing and the velocity dispersion data provide constraints in a region typically well inside the scale radius. In Fig.~\ref{fig:chi2cont}, we
show the confidence levels in the $\rho_{\rm s}-r_{\rm s}$, $\rho_{\rm
s}-\beta$ and $\beta-r_{\rm s}$ planes, for the same cluster as
above. In each panel we have fixed the remaining parameter to its best fit value, in order to show the degeneracy that remains even after reducing the number of free parameters in the model. Even with good constraints on the position of the lens critical
lines there is a strong degeneracy between $r_{\rm s}$ and $\rho_{\rm s}$: 
for any value of $\rho_s$ in the range $(10^{15} -  8\times
10^{15})\,M_\odot\,\mathrm{Mpc}^{-3}\,h^2$ there is a corresponding value of
$r_{\rm s}$ in the range $(200 - 800)~\mathrm{kpc}\,h^{-1}$ which
produces a good fit. However, $\beta$ appears to be much better
constrained. Therefore, unless extremely precise measurements of
the position of the lens critical lines are made, the only parameter
of the three-dimensional cluster density profile which is likely to be
well constrained using this method is the inner slope $\beta$. 

As stated earlier, the experiment illustrated above has been repeated
for $100$ pairs of radial and tangential arcs for each cluster. The
probability distribution functions for some of the cluster projections
which have been classified as ``regular'' are shown in
Fig.~\ref{fig:reg_distrib}. Note that, differently from what was done in Fig.~\ref{fig:chi2cont}, the probability distribution functions are now marginalized, i.e. we do not fix any parameter to its best fit value. In all the panels, the vertical
long-dashed lines indicate the ``true'' values of the parameters found
by fitting Eq.~(\ref{eq:nfwgen}) to the three-dimensional density
profile of the cluster. While, as discussed earlier, the scale radius
$r_{\rm s}$ and characteristic density $\rho_{\rm s}$ are poorly
constrained due to the strong degeneracy between these two parameters,
in nearly all the cases which we have studied the probability
distribution functions of the inner slope $\beta$ have peaks which
coincide with the true values, if a lensing potential with the
appropriate ellipticity and position angle are used to model the
cluster. On the other hand, using axially symmetric models generally
leads to underestimate the value of $\beta$. 

\begin{figure*}
\plotthree{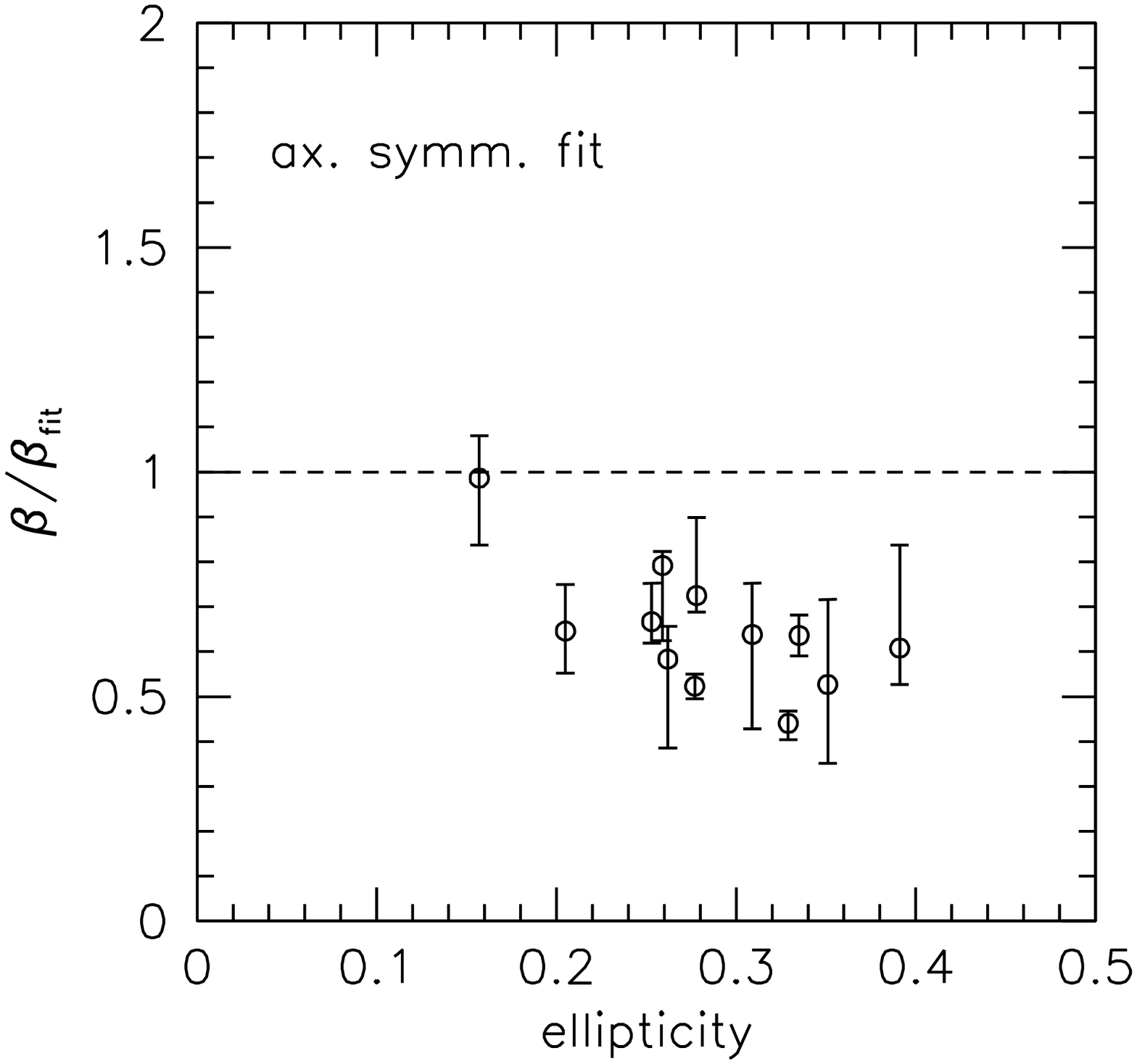}{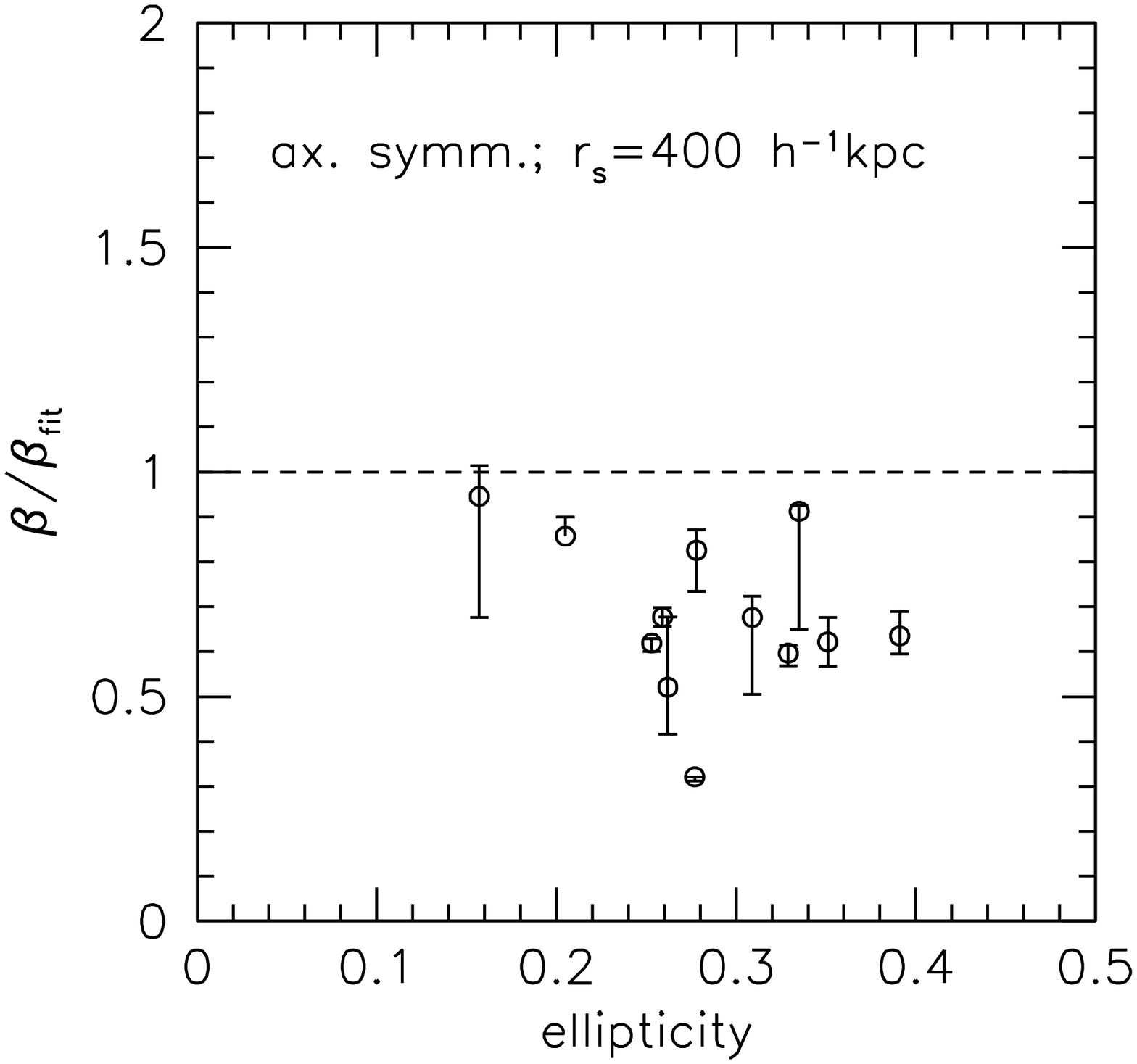}{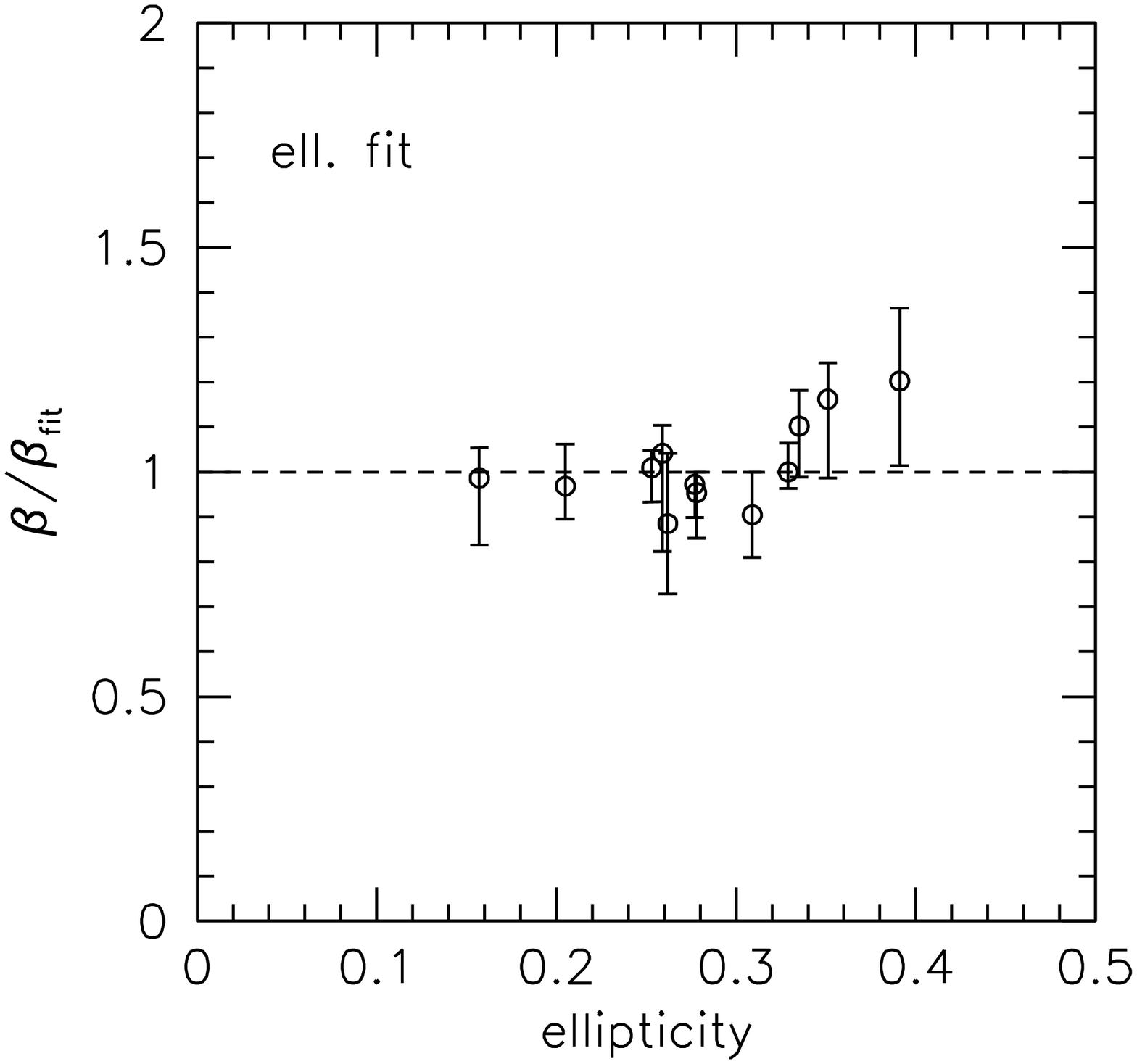}
\caption{Medians of the probability distribution functions of the inner slopes
  for all the clusters classified as ``regular'' as a function of the
  ellipticity of the lensing potential. The medians are normalised to
the
  best-fit inner slope $\beta_{fit}$ of each cluster. The error bars are given
  by the interquartile ranges of the distributions. Each panel corresponds to one
  of the model fits discussed in the text. }
\label{fig:ellipticity}
\end{figure*}

The discrepancy between the true and the most probable $\beta$ when
fitting axially symmetric models is expected to correlate with the
ellipticity of the cluster lensing potential.
Fig.\ref{fig:ellipticity} illustrates this dependence on the
ellipticity. We plot the medians of the probability distribution
functions of $\beta$ for all clusters classified as ``regular'' as a
function of the ellipticity of the cluster lensing potential. The
medians have been normalised to the best-fit slope $\beta_{fit}$ of
each simulated cluster.  The left and the middle panels refer to the
fit with axially symmetric lenses, with the scale radius considered as
a free parameter or assumed to be fixed, at $r_{\rm s}=400\,h^{-1}$kpc
respectively. The right panel shows the results obtained adopting a
model with the correct ellipticity and orientation for the lensing
potential iso-contours. The errorbars show the interquartile range of
each distribution. Under the assumption of axial symmetry, the
measured slope is consistent with the true one only for the cluster
model with the smallest ellipticity; it is underestimated by
$\sim20-70\%$ for cluster models where a larger $e$ was measured.
Unfortunately, most of our clusters have ellipticities in the range
$[0.2 - 0.4]$ and we have only one case with $e\sim 0.15$. On the
basis of these results it is difficult to determine the threshold
below which the ellipticity can be ignored and an axially symmetric
lens model safely used. A larger number of simulated clusters with
small ellipticity would be necessary for this purpose. For example,
for the cluster cl6.2, which has $e=0.15$, we obtain a good estimate
of $\beta$ even when fitting an axially symmetric lens model, but for
the cluster cl2.3, which has an ellipticity only slightly larger,
$e\sim 0.2$, the slope is underestimated by $\sim 20-35\%$ with the
axially symmetric model. On the other hand, when elliptical models are
fit, all the measured values of $\beta$ are consistent with their
true values.

\begin{figure}

\plotone{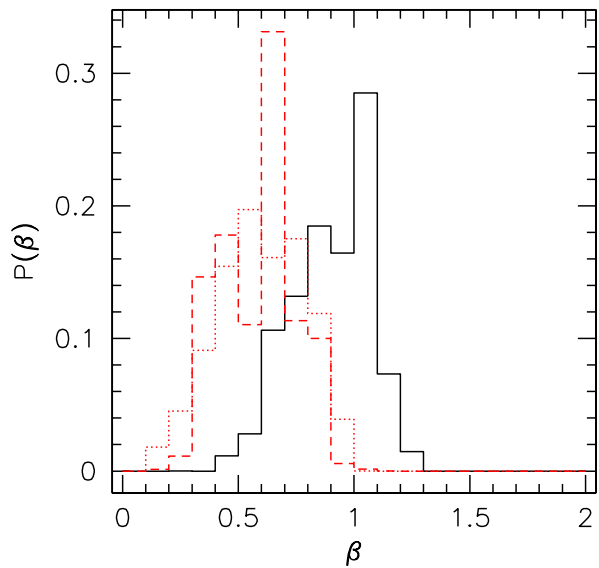}
\caption{Averaged probability distribution function for the sub-sample of
  ``regular'' clusters. Solid, dotted and short-dashed histograms refer to
  elliptical, axially symmetric and axially symmetric with $r_{\rm s}=400
  \,h^{-1}\,\mathrm{kpc}$ fitting models, respectively.}
\label{fig:cumulate}
\end{figure}

Finally, we show in Fig.\ref{fig:cumulate} the probability
distribution function of the inner slope obtained by averaging over
all the clusters in this subsample. The resulting $P(\beta)$ obtained
using elliptical lens models peaks around unity, in agreement with the
mean value of $\beta_{fit}$ for our cluster sample,
$\overline\beta_{fit}\sim0.95$. The distributions obtained by fitting
with axially symmetric lens models have a maximum at $\beta \sim 0.6$.

\subsection{Peculiar clusters}

\begin{figure*}
\plotside{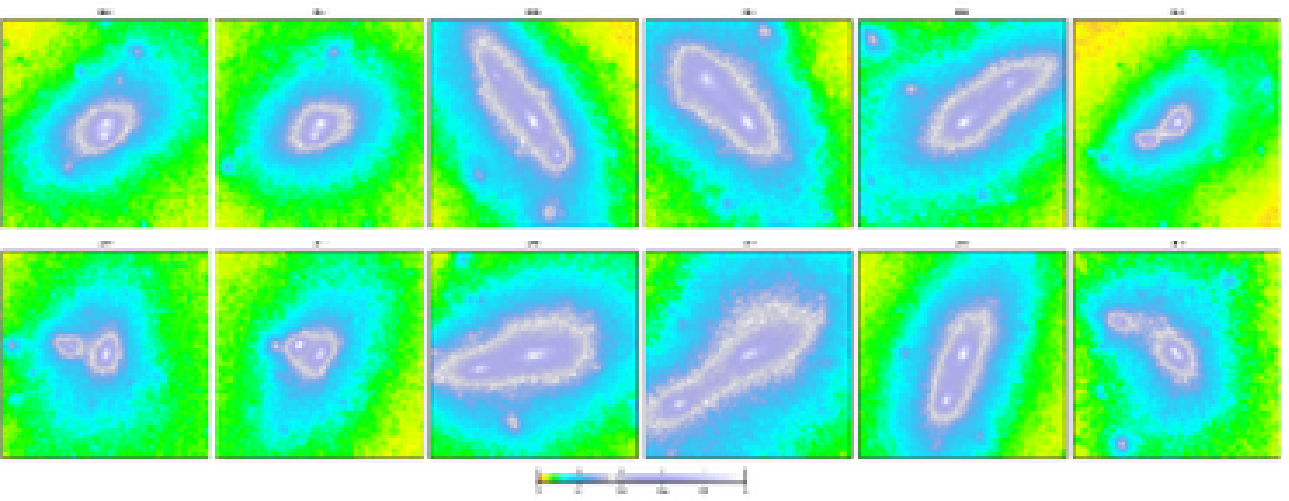}
\caption{Convergence maps of the clusters classified as ``peculiar''. The
  side-length of each panel is $226''$.}
\label{fig:kpeculiar}
\end{figure*}

The second subsample contains the clusters which we classified as
``peculiar''. The convergence maps of these halos are shown in
Fig.~\ref{fig:kpeculiar}. Visually it is clear that there is likely
to be some difficulty in applying the (pseudo-)elliptical lens models
to these clusters. The critical lines are very asymmetric, mainly as a
result of disturbances due to the passage of large mass concentrations
through the cluster core during merger events.

Applying the method to these perturbed clusters generally leads to
incorrect results for the determination of the parameters characterising
their density profiles. For such clusters a more detailed mass
modelling of the lens, including multiple mass components, is necessary
in order to reproduce the shape of the critical lines and the
positions where arcs form. The shear fields produced by secondary mass
concentrations cause the critical lines to be extended in the
direction connecting the cluster centre with the perturbing mass clump
and usually enhance a cluster's ability to produce both radial and
tangential gravitational arcs, as discussed by \cite{TO04.1}.

\begin{figure*}
  \centering
  \includegraphics[width=4.2cm]{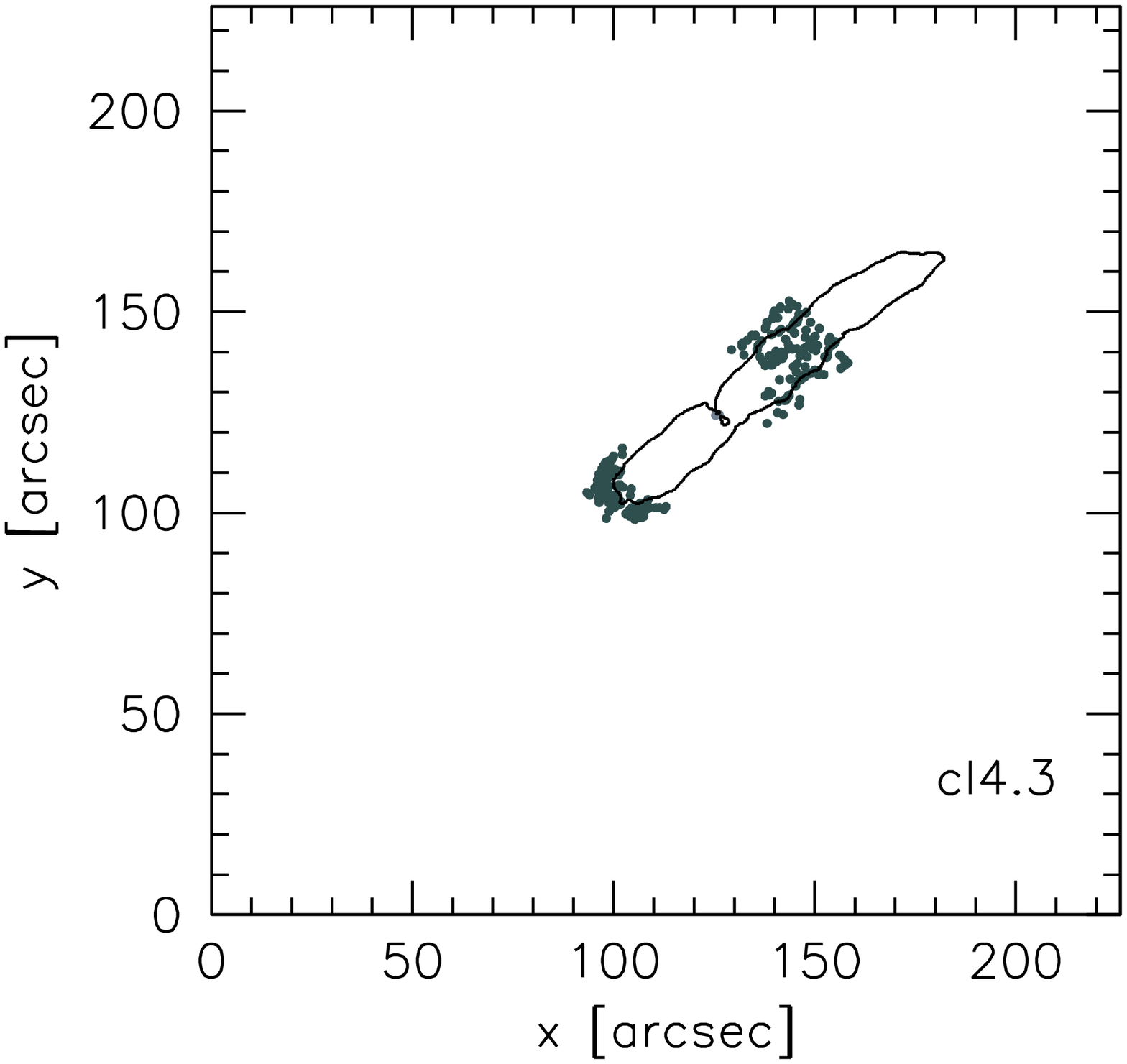}
  \includegraphics[width=4.2cm]{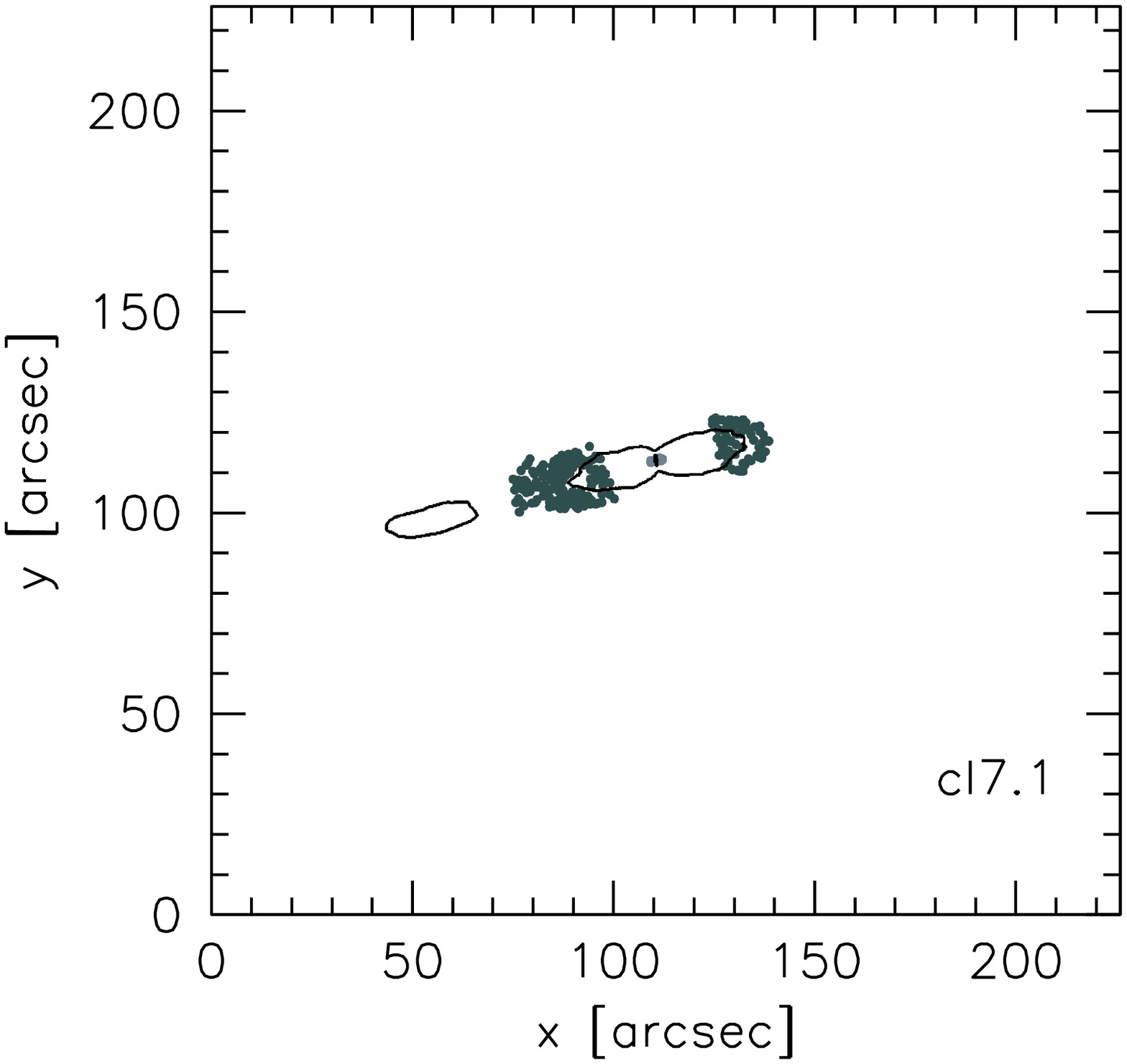} 
  \includegraphics[width=4.2cm]{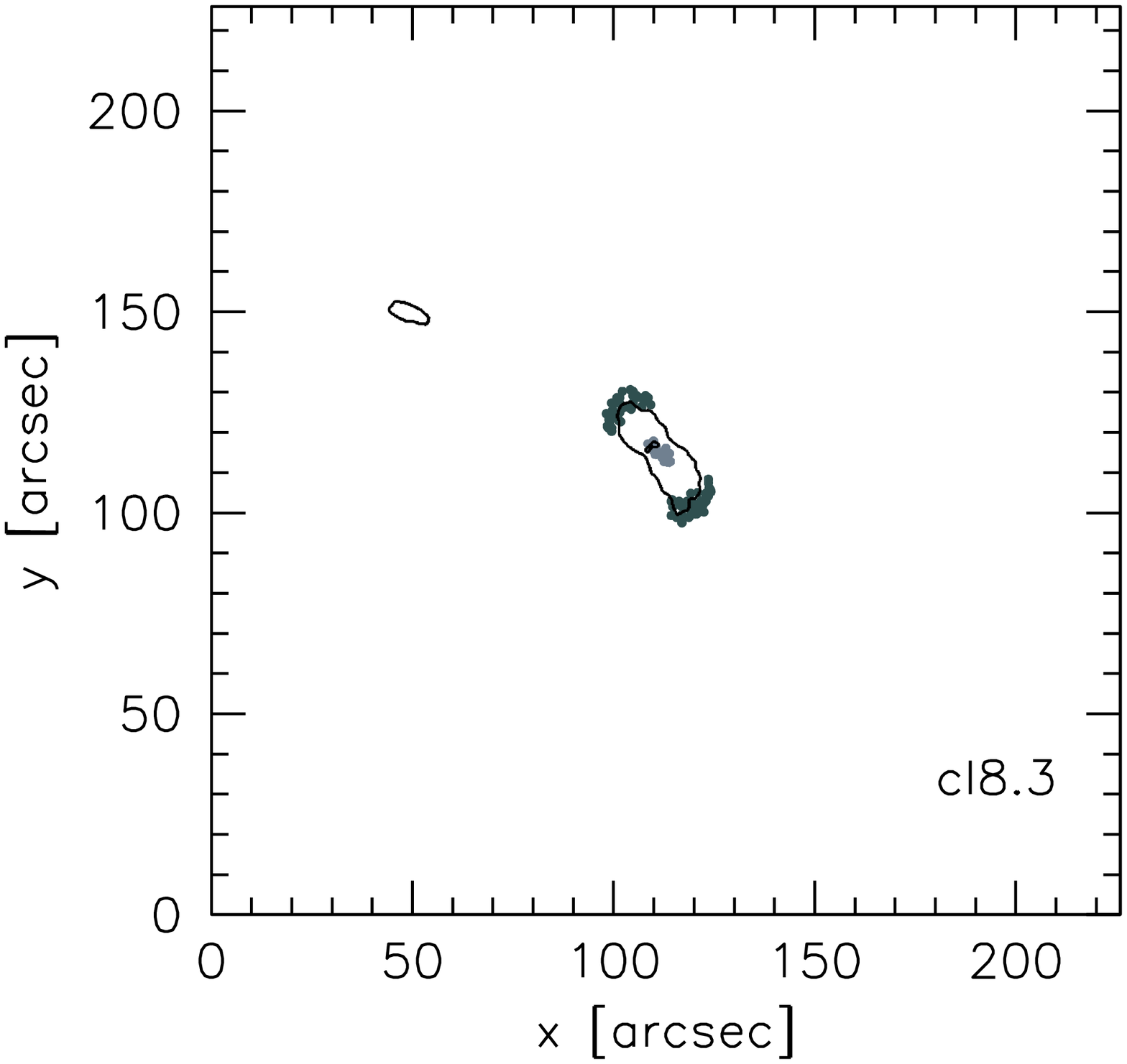}
  \includegraphics[width=4.2cm]{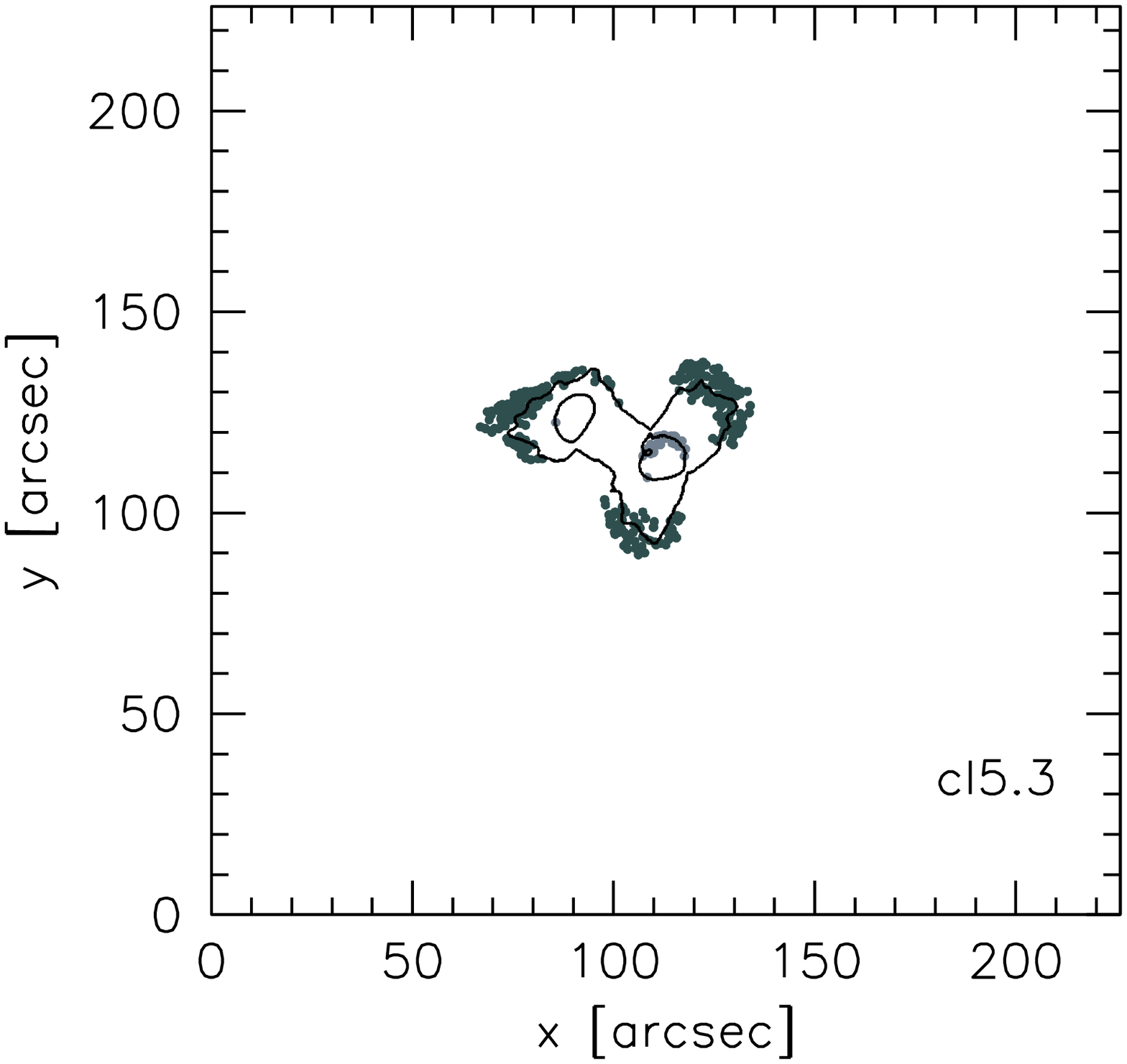} \\
  \includegraphics[width=4.2cm]{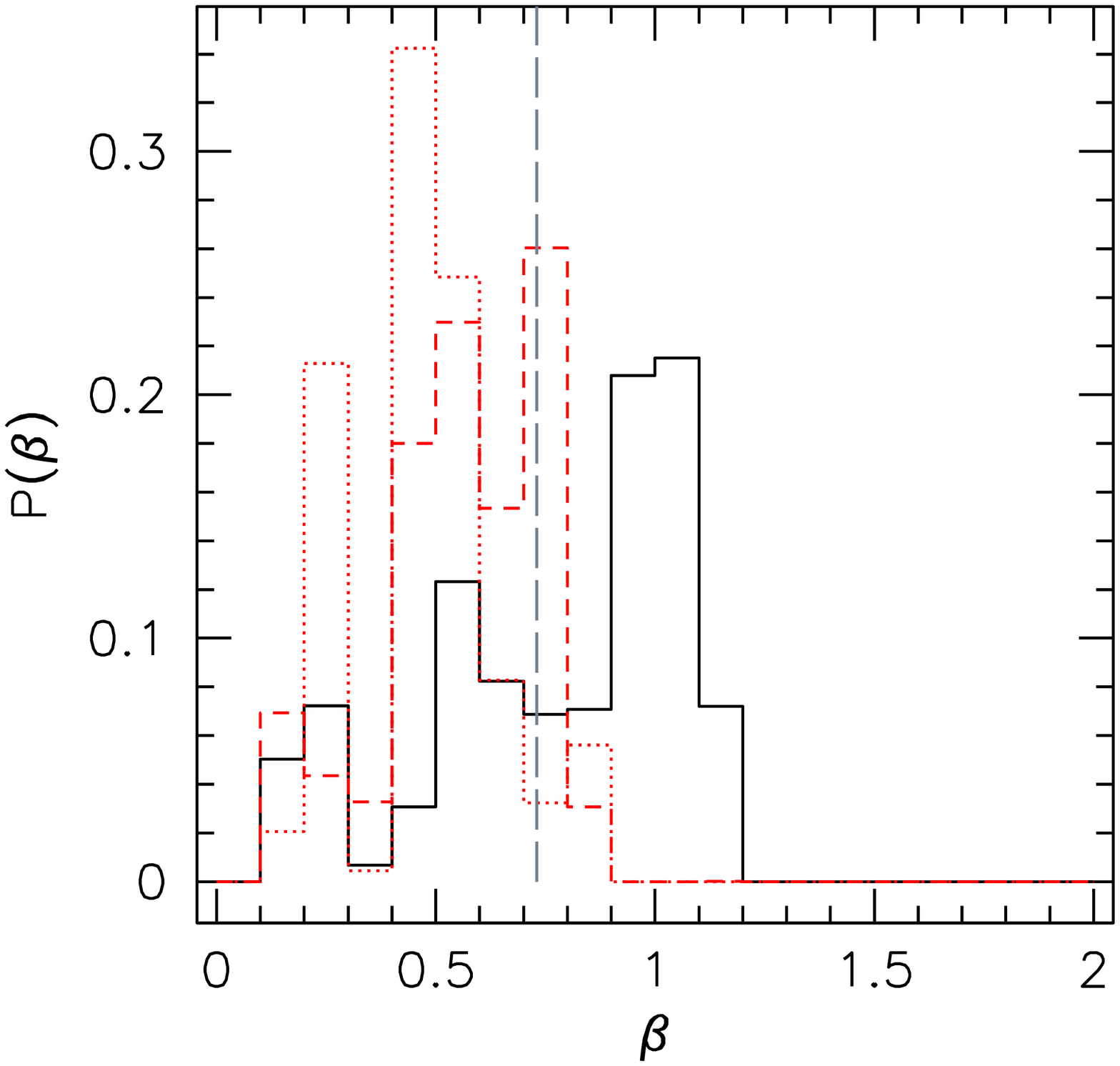} 
  \includegraphics[width=4.2cm]{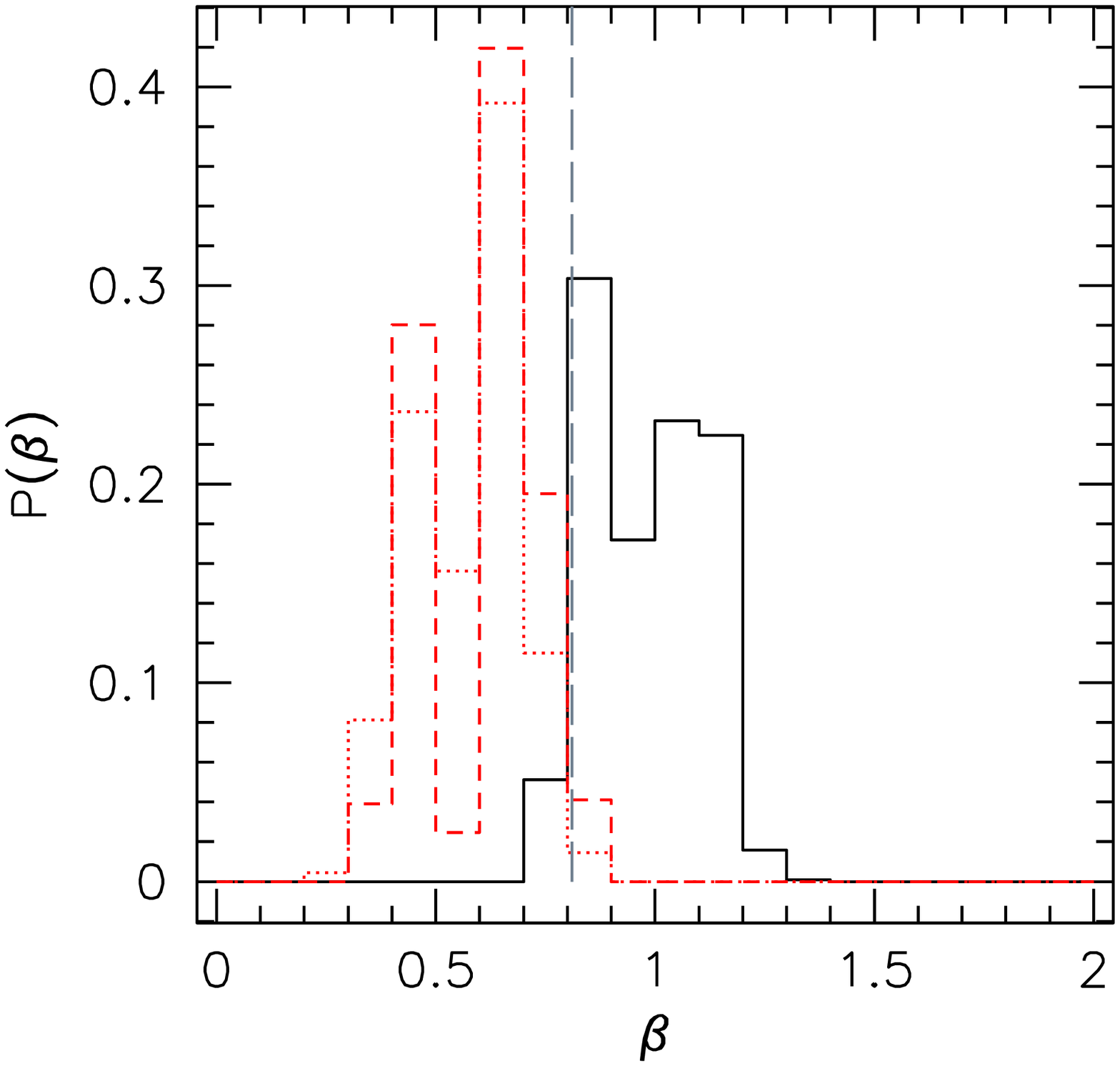}
  \includegraphics[width=4.2cm]{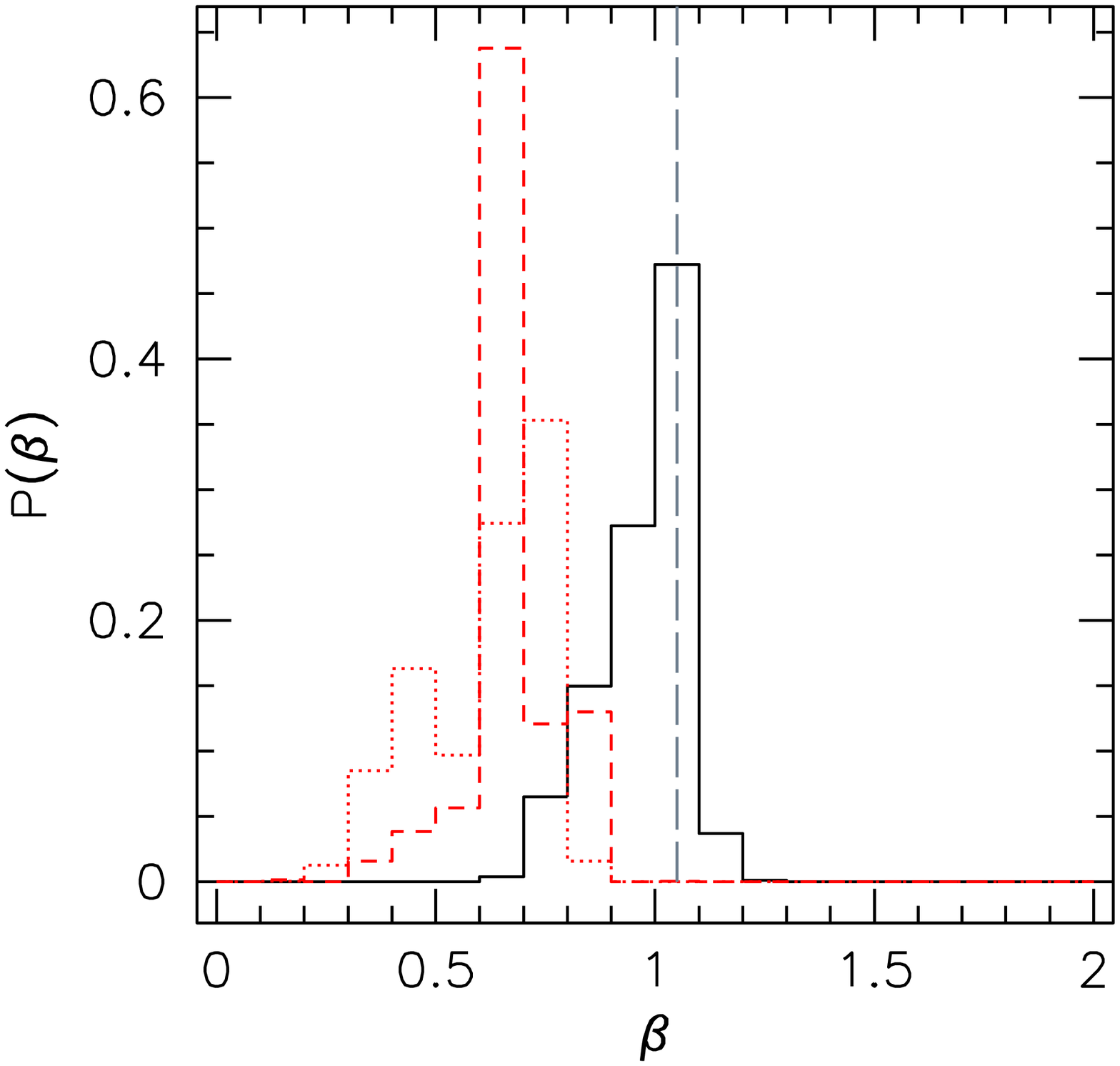}
  \includegraphics[width=4.2cm]{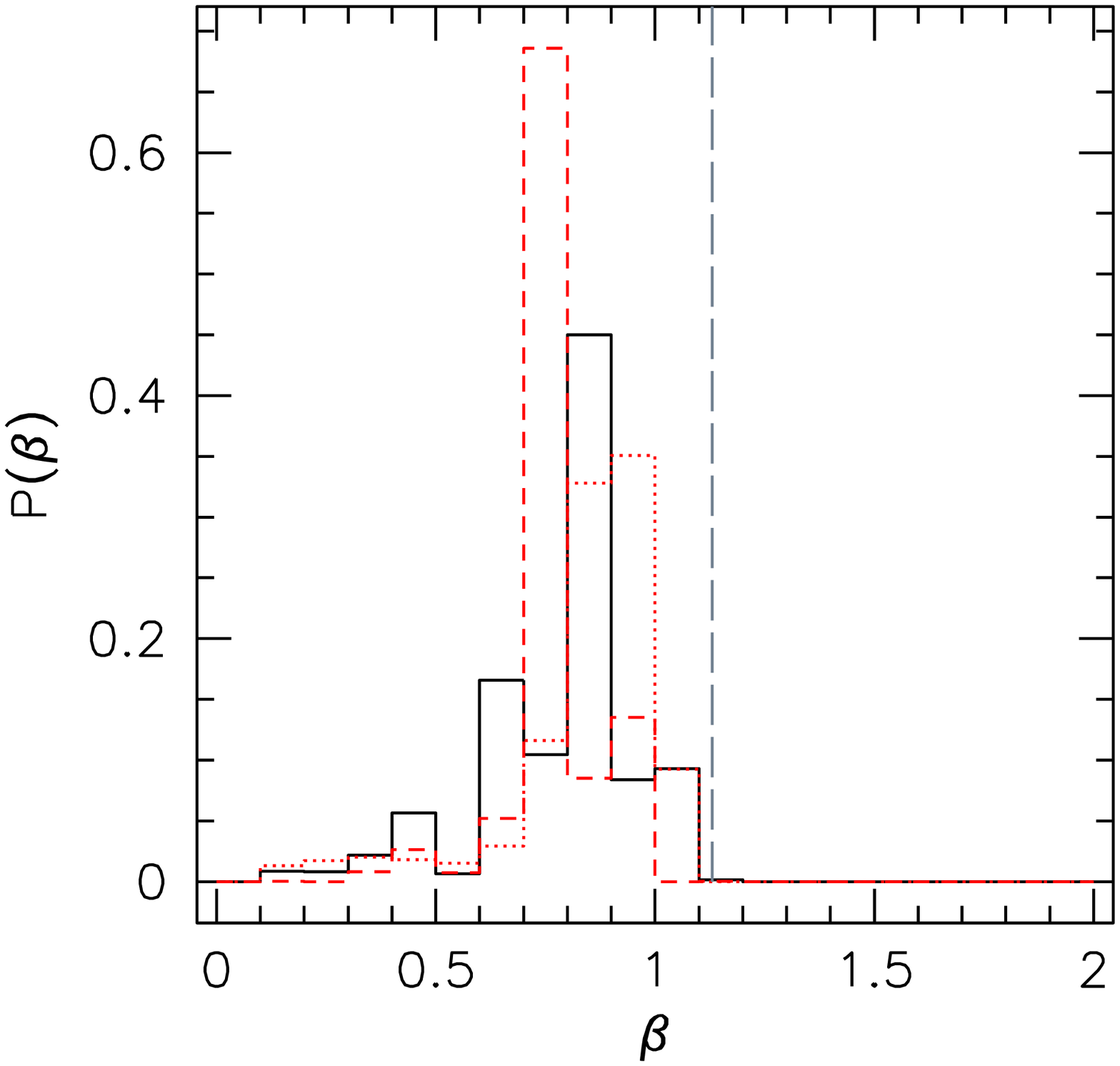}
\caption{Analysis of some peculiar clusters. Top panels: critical lines and
  arc positions for clusters cl4.3, cl7.1, cl8.3 and cl5.3. Bottom panels:
  probability distribution functions of the inner slope. The line types are
  the same as fig.~\ref{fig:reg_distrib}}
\label{fig:pec_distrib}
\end{figure*}

We show examples of the critical lines and distributions of the inner
slope for some peculiar clusters in Fig.~\ref{fig:pec_distrib}.
Massive substructures distort the shape of the iso-contours of the
lensing potential, stretching and elongating them along preferred
directions. If such substructures happen to be aligned with the major
axis of the main cluster clump, the ellipticity of their lensing
potential is generally overestimated, when using a single component
lens model. This occurs for example in cluster cl4.3. To compensate
for such an overestimate ellipticity, the probability distribution
function of the inner slope shifts towards high values of $\beta$. In
the previous section we argued that the inner slope is underestimated
if an intrinsically elliptical lens is fit with an axially symmetric
model.  Similarly, $\beta$ is overestimated when a model with too
large ellipticity is used to describe a cluster of moderate
ellipticity. Again, the central density is constrained by the velocity
dispersion data. Tangential arcs form where $\kappa+\gamma \sim 1$. If
the ellipticity of the lens is overestimated, the shear $\gamma$ will
be overestimated as well. As a result, the convergence $\kappa$ needs to be
reduced, implying that the cluster density has to be a steeper
function of distance from the centre.

Less massive or more distant substructures have a smaller impact on the
reliability of the method. For example, cluster cl7.1 is less sensitive
to the perturbation of a secondary mass clump whose presence is evidenced by
a secondary critical line. The resulting probability distribution function of
the inner slope is still shifted towards large $\beta$, but less
significantly so than for cluster cl4.3. Finally, the substructure in the cluster
cl5.3 is too small to affect the shape of the cluster critical lines. The
distribution of $\beta$ for this cluster peaks at its correct value. 

For some other clusters, however, the opposite results are found. For example
in cluster cl8.3, substructures close to the cluster core and not aligned
with the major axis of the cluster mass distribution mimic an
axially symmetric lensing potential. Consequently, the inner slope if the
cluster density profile results to be underestimated, when our method is
applied. 

\section{Comparison to previous observational work}

We now discuss the implications of our analysis for the results
obtained previously by \cite{SA03.1} for a sample of real clusters.
In the preceding sections, we have demonstrated the importance of
taking ellipticity and substructures into account in order to derive a
correct measurement of the inner slope of density profiles for a set
of clusters randomly chosen from a cosmological simulation. Since the
number of lenses available in the simulation is not large enough, we
were not able to sample uniformly the parameter space of clusters. In
particular, our clusters cover a range of ellipticities between
$\sim0.15$ and $\sim 0.4$. Only for one cluster projection, could we
measure an ellipticity below $0.2$. Interestingly, for this
particular lens, the ellipticity derived from fitting the simulated
lensing and velocity dispersion data is in good agreement with the
true ellipticity even under the assumption of axial symmetry.

In their work, \cite{SA03.1} carefully chose a set of clusters which
appear very round and relaxed, both in optical and X-ray images. For
the three clusters in their sample containing both radial and
tangential arcs, MS2137.3-2353, A383, and RXJ1133, the ellipticity is
presumably in the range $0.1-0.2$ \citep{GA03.1,MI02.1,SM01.1}, as
\cite{SA03.1} quote in their paper.

Unfortunately, as shown in Fig.~\ref{fig:ellipticity}, there is
substantial scatter in the inferred values of the inner slope even for
clusters classified as `regular'' and this makes it difficult to
establish the minimum value of the ellipticity for which fitting an
axially symmetric model, as \cite{SA03.1} did, would strongly bias the
results. A more quantitative comparison with the results of Sand et
al. would require a larger number of cluster projections with low
ellipticities. Furthermore, as we pointed out earlier, there are some
methodological differences between our analysis and that performed by
Sand et al. Our definition of $\chi^2_{lens}$ differs from theirs in
that they fit the location of the critical line whereas we fit the
eigenvalues of the Jacobian matrix at the positions of the arcs. In
one sense, the two approaches are equivalent, since they both require
that the lens critical lines should pass close to the radial and
tangential arcs. On the other hand, our estimate of the errors, based
on the Jacobian eigenvalues, would require calibration with analytical
models before it can used with observational data.

Finally, in our previous analysis, we have used all the radial arcs
found in the simulations, regardless of their position relative to the
BCG. As we have already mentioned, several of these arcs are probably
undetectable in real observations, because they are too close to the
cluster centre and thus would be completely embedded in the light of
the dominant galaxy. In the case of elliptical radial critical lines,
i.e. when the lenses are elliptical, this implies that those sections
of the critical line reaching farthest from the cluster centre would
be more easily traced by lensing observations. Assuming axial symmetry
in these cases would be quite dangerous and potentiall have a strong
effect on the determination of the inner slope. Indeed, steep density
profiles produce small radial critical lines due to the large
curvature of the time delay surface at the central maximum.

\begin{figure}
\plottwo{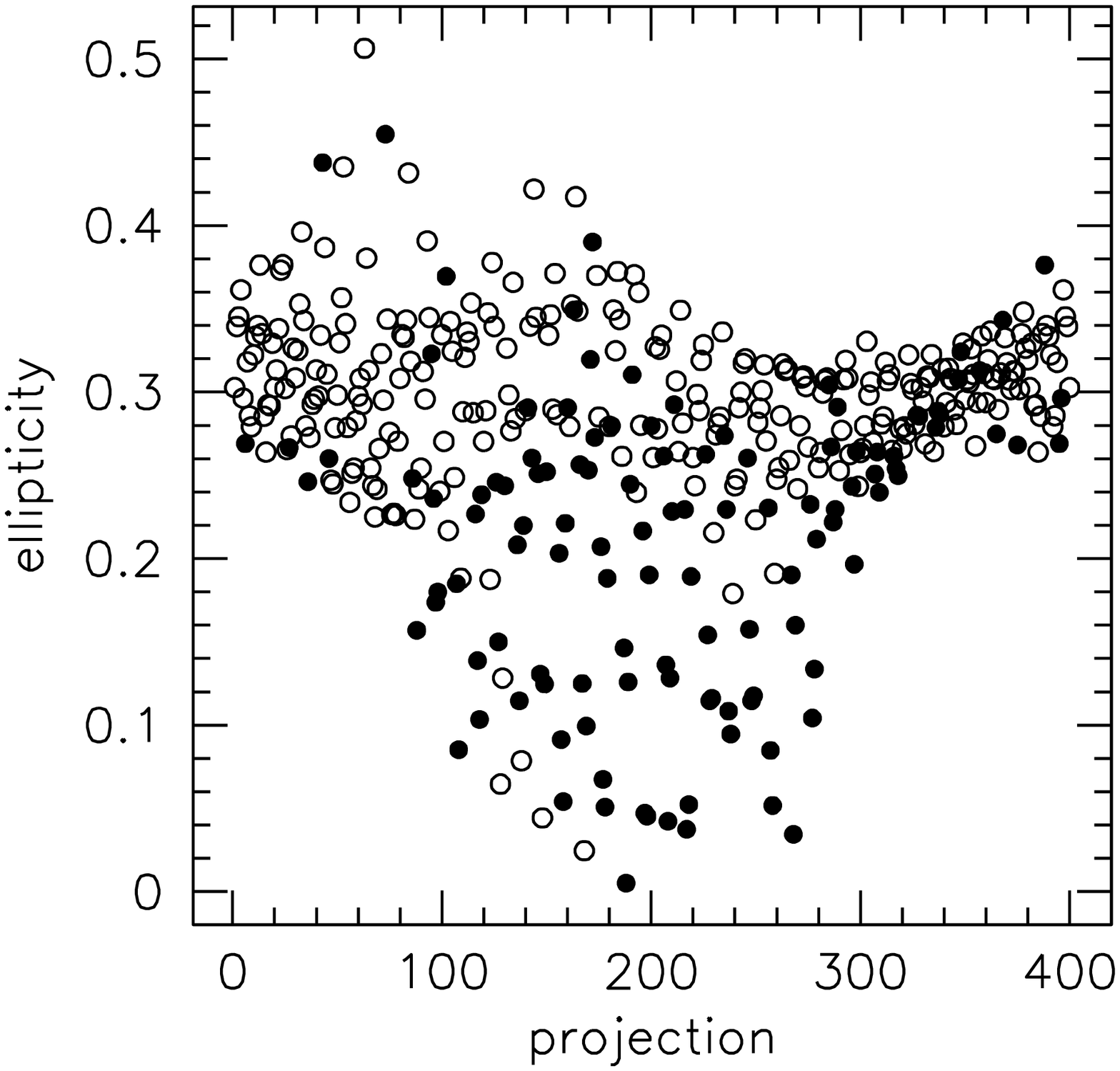}{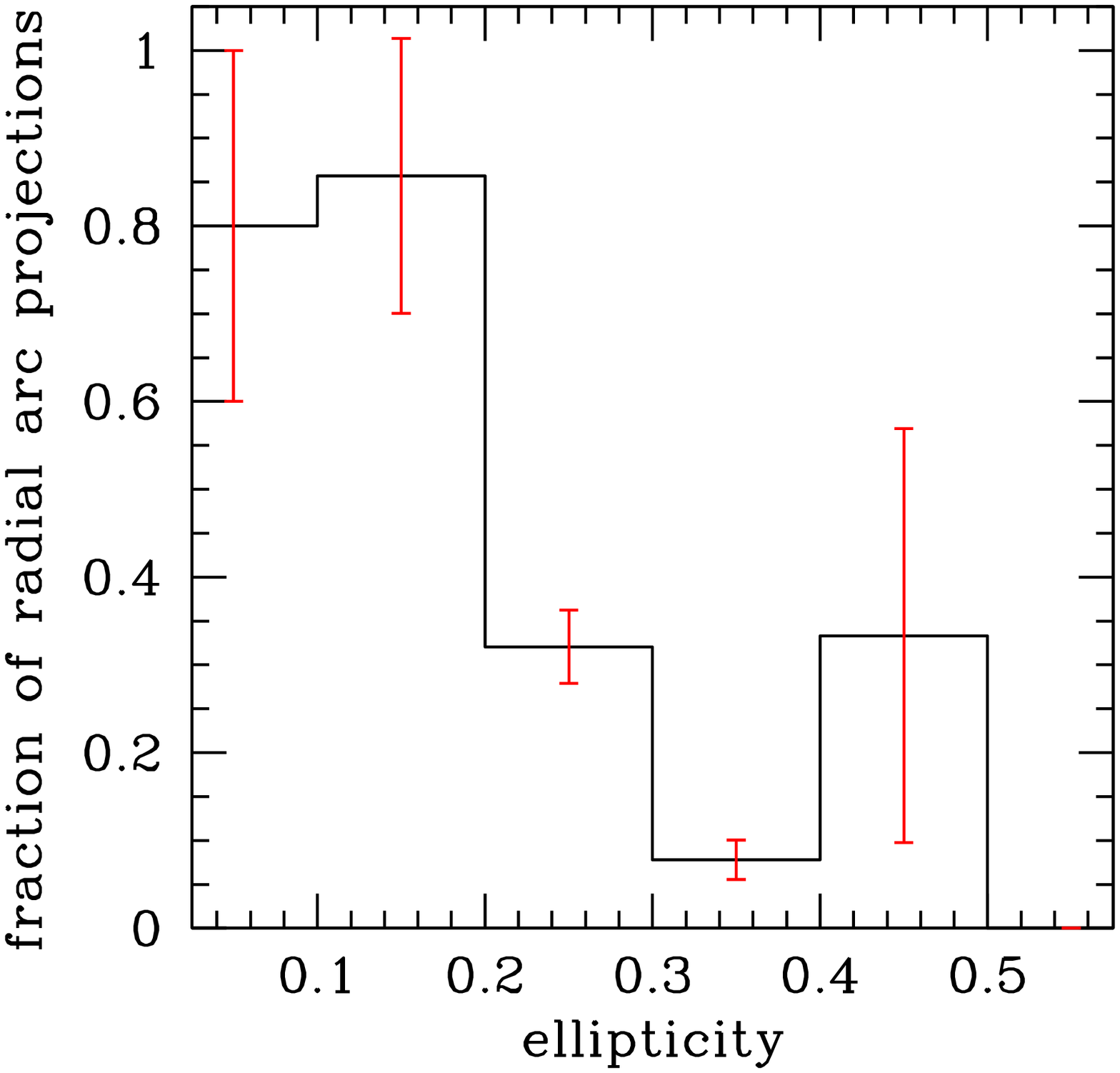}
\caption{Left panel: ellipticities of the lensing potential for 400 different projections
  of the cluster cl1. The projections which give rise to both radial
  and tangential arcs are indicated by filled circles. Right panel: fraction of radial arc projections per ellipticity bin.}
\label{fig:scattplot}
\end{figure}

In order to overcome some of these difficulties in comparing our
results to those of Sand et al. we have performed a different set of
simulations. We choose one of the clusters in our sample, cluster cl1,
which is the strongest lens, and project it along 400 different
lines-of-sight with direction uniformly distributed on the surface of
a sphere centred on the cluster BCG.  We carry out simulated lensing
measurements on each of the projections.  The ellipticity of the
lensing potential in the different projections varies between $\sim 0$
and $\sim 0.5$, indicating that the cluster is highly triaxial. The measured ellipticity for each of the 400
projections are plotted as a scatter diagram in the left panel of Fig.~\ref{fig:scattplot}.

An interesting property of this cluster is that most projections characterized by small ellipticities of the lensing potential are able to produce both radial and tangential arcs. We quantify this in the right panel of Fig.~\ref{fig:scattplot}, where we show the fraction of projections with radial and tangential arcs (and the corresponding errorbars) per ellipticity bin. We see that for ellipticities below 0.2 more than $80\%$ of the projections produce radial arcs in addition to tangential arcs. Such projections correspond to viewing the cluster nearly along its major axis.  

Compared to our previous, more general approach, we now tailor the criteria for
selecting radial and tangential arcs and for estimating position errors to the
procedures adopted by Sand et al. The aim is to mimic the observational process
as closely as possible. Firstly, we use only radial arcs lying at distances
larger than $4\,h^{-1}$kpc from the cluster centre, which is the minimal
distance of the radial arcs in the cluster sample used by Sand et al. Secondly,
we use only those arcs arising from mergers of multiple images of the same
source, i.e. arcs that in the observations would have multiple brightness
peaks. These are identified in the simulations by mapping the centre of each
source galaxy onto the lens plane. We use the multiple peaks along the same arc
to obtain realistic measurements of the positions, $x_c^{t,r}$, of the tangential
and of the radial critical lines and to estimate the errors $\Delta
x_c^{t,r}$. The former are determined by finding the intermediate points between
two close peaks with inverse image parity (opposite sign of the
magnification). The latter are estimated by measuring the distance between the
same brightness peaks. For fitting these data, we define a new $\chi^2_{lens}$
variable that is completely consistent with that used by Sand et al.,
\begin{eqnarray} \chi^2_{lens}(r_{\rm s},\rho_{\rm s},\beta) & = &
\left[\frac{x_c^t-\hat{x}_c^t(r_{\rm s},\rho_{\rm s},\beta)}{\Delta
x_c^t}\right]^2 \nonumber \\ & + & \left[\frac{x_c^r-\hat{x}_c^r(r_{\rm
s},\rho_{\rm s},\beta)}{\Delta x_c^r}\right]^2\;, \label{eq:newchi2}
\end{eqnarray} 
where $\hat{x}_c^{t,r}$ denote the expected positions of the tangential and the
radial critical lines, given a set of input parameters $(r_{\rm s},\rho_{\rm
s},\beta)$.

The model cluster used for this new set of simulations is not able to produce
arcs which meet the selection criteria in all its projections. As a result, we
can only use a subsample of them; these are identified by the filled circles in
Fig.~\ref{fig:scattplot}. The vast majority is characterised by ellipticities
below $0.3$. In particular, we now have $\sim 70$ projections whose lensing
potential has an ellipticity less than $0.2$.

For each projection, we fit all possible combinations of radial and tangential
arcs found in the simulations that meet the selection criteria. This is done 
by minimising the $\chi^2$ variables in Eqs.~\ref{eq:newchi2} and
\ref{eq:chi2vd} and by assuming axial symmetry, as was done by Sand et al. In
order to perform a direct comparison with the analysis of Sand et al., we also
centre the innermost bin of the simulated velocity dispersion profile on
$R=0.4\,h^{-1}$kpc, and extend the measurements to $R=5\,h^{-1}$kpc with a total
of four equidistant bins.

\begin{figure}
\plotone{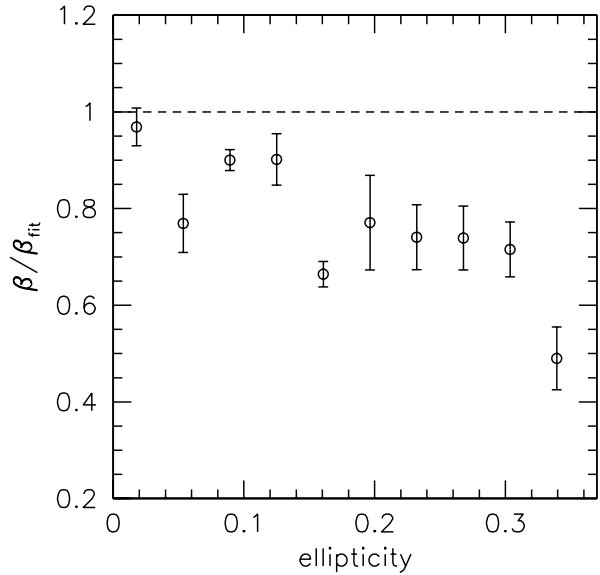}
\caption{Mean inner slopes in different bins of the ellipticity of the lensing
potential. These are estimated by fitting with axially symmetric models all the
possible pairs of radial and tangential arcs produced by many projections of
cluster cl1. The values of $\beta$ are normalised to the best-fit inner slope
$\beta_{fit}$. The errorbars show 3$-\sigma$ statistical errors.}
\label{fig:avbeta}
\end{figure}

Each determination of $\beta$ is normalised to the best-fit inner slope
$\beta_{fit}$ of cluster cl1. The estimates are divided into equidistant
ellipticity bins of size $\sim 0.03$. In each bin, we calculate the mean
value of the normalised inner slopes, weighting each measurement with
the corresponding $\chi^2$. The resulting mean normalised $\beta$ is
shown as a function of the ellipticity of the lensing potential in
Fig.~\ref{fig:avbeta}. The errorbars give the 3$-\sigma$ statistical
errors.

The results are in good agreement with those previously discussed in
Sect.~\ref{results}. Firstly, changing the definition of $\chi^2_{lens}$, in
line with that adopted by Sand et al. does not modify our conclusions either
qualitatively or quantitatively. The inner slope is underestimated when the
model used to fit the data is axially symmetric. Secondly, we find that it is
possible to underestimate the inner slope even in cases where the lensing
potential has very low ellipticity. Indeed, for ellipticities between $0$ and
$0.15$, the inner slopes inferred by assuming axial symmetry are typically
$10-35\%$ smaller than the true ones.

The joint probability distribution function of the inner slope of the clusters
in the sample studied by Sand et al. peaks at $\beta = 0.52_{-0.5}^{+0.5}$ (68\%
CL). Taking into account several possible systematic effects, such as anisotropy
of the stellar orbits, an incorrect choice of scale radius or of the luminous
mass model, this best-fit value may be shifted by $\sim 0.2$ (see the discussion
on systematics in Sect.~6 of Sand et al.). Thus, we conclude that our results
for axially symmetric fitting models are compatible with those of Sand et
al. Our results indicate that the small values of the inner slope that they
estimate from their cluster sample can be explained by the effects of the small
but non-zero ellipticity of the lenses.

We also note that the joint probability distribution function found by Sand et
al. is strongly conditioned by two clusters, MS2137.3-2353 and A383. Existing
two-dimensional mass reconstructions of these two lenses, based on the NFW
density profile, indicate that these clusters have very small scale radii ($\sim
60\,h^{-1}$kpc) \citep[see e.g.][]{CO06.1}. Thus, any flattenting of their
density profile occurs on scales where the BCG already dominates the mass
distribution. In such regions, the density profile of the dark matter mass
component is likely to be very poorly constrained. 

\section{Summary and discussion}
\label{summary}
In this paper we have {explored, using numerical simulations, the possibility of
constraining the density profiles of galaxy clusters.} The method, proposed by
Sand et al. (2002,2004), consists of using a two component lens model,
comprising a dark matter halo and a central dominant galaxy, to fit the
positions of radial and tangential gravitational arcs and the velocity
dispersion profile of the cluster BCG. While Sand et al.  used axially symmetric
models for describing the mass distribution of clusters, we have used a more
general pseudo-elliptical lens model, where ellipticity is included in the
lensing potential.

Using galaxy clusters obtained from cosmological N-body simulations with
high mass resolution and investigating their lensing properties with
ray-tracing techniques, we have created mock catalogues of radial and
tangential arcs and simulated velocity dispersion profiles. We have
fit the data with both the generalised pseudo-elliptical and the
axially symmetric lens models. This allowed us to evaluate the
reliability of the method and the impact that the ellipticity of the
cluster lensing potential has on the correct estimate of the profile
parameters. The projected clusters have been divided in two
sub-samples: those which do not contain substructure in the inner
region are classified as ``regular'', while those with perturbed
cores are classified as ``peculiar''. By comparing the results
obtained with the two sub-samples, we have studied the systematic
errors due to incorrect modelling of the cluster. In our fit, the
cluster density profile is assumed to be a generalised NFW profile,
depending on three free parameters, namely the inner logarithmic slope
$\beta$, the scale radius $r_{\rm s}$ and the characteristic density
$\rho_{\rm s}$.

Our main findings can be summarised as follows:
\begin{itemize}
  
\item If a model with the correct ellipticity and orientation of the
  iso-contours of the cluster lensing potential is used, the inner slope of
  the density profile is accurately recovered, at least for those clusters
  which we classify as ``regular''. On the other hand, the scale
  radius and the characteristic density are poorly constrained,
  because a strong degeneracy exists between these two parameters.
  
\item When using the axially symmetric lens models to fit the data,
the inner slope is generally significantly underestimated. The degree
to which $\beta$ is incorrectly determined depends on the ellipticity of
the cluster lensing potential, and is larger for larger
ellipticities. For our clusters with ellipticities in the range $0.2 -
0.4$, the inner slopes resulting from axially symmetric fits are $\sim
20\%$ to $\sim70\%$ smaller than the true values. When averaging over
all ``regular'' clusters , we find that axially symmetric fits
typically underestimate $\beta$ by $\sim 40\%$.  By contrast,
when the cluster ellipticities are properly taken into account, the
averaged probability distribution function of the inner slopes is in
good agreement with the distribution of the true inner slopes of the
density profiles of the clusters in our sample.
  
\item The fit produces incorrect results if applied to clusters whose
critical regions for strong lensing are perturbed by massive
substructures. i.e. for clusters which are undergoing major merger
events and which we classified as ``peculiar''. The shear field
produced by secondary mass clumps distorts the shape of the lens
critical lines and so the position of radial and tangential arcs are
badly reproduced by simple lens models.  In such cases a more detailed
lens model is required. The effect of the substructures depends on
their mass and on their location with respect to the main cluster
clump. For clusters where massive substructures are located close to
the critical regions, the inner slope can be both over- or
underestimated. However, in some cases we are able to obtain a good
measurement of the inner slope even for relatively perturbed clusters,
because the perturbing substructure is too small or too distant from
the cluster centre to  affect the shape of the lensing
potential iso-contours significantly in the region where arcs form.

\item A large number of lensing simulations, obtained by projecting the
same cluster in 400 different directions, have been performed with the
specific aim of determining the range of ellipticities for which the
assumption of axial symmetry is applicable. We verify that even for
ellipticities below $0.2$, the inner slopes can be underestimated by
$\sim10-35\%$.

\end{itemize}

We conclude that using strong lensing and velocity dispersion data is
potentially a very powerful method for constraining the mass
distribution in the inner parts of galaxy clusters, provided that the
lensing potential is accurately modelled.  In particular, the impact
of ellipticity cannot be neglected even for clusters which deviate
from axial symmetry only moderately. The effect of substructure in the
inner region of clusters also needs to be taken into account.

By neglecting the effects of ellipticity, Sand et al (2002,2004) could be led to
underestimate the slope of the inner profiles of the clusters they analysed and
to the conclusion that their data disagreed with the predictions of the cold
dark matter model. Our analysis shows such a conclusion may be unjustified. At
the same time however, it also demonstrates that the general approach pioneered
by Sand et al provides a powerful means to probe the central distribution of
mass in clusters.

\section*{Acknowledgements}
\label{acknowledgements}
We are grateful to Lauro Moscardini, Klaus Dolag, Elena Rasia, Stefano Ettori and Piero Rosati for helpful discussions. We thank Richard Ellis, Graham Smith and Tommaso Treu for useful comments. The simulations used in this paper was carried out by the Virgo Supercomputing Consortium.

\bibliographystyle{mn2e}
\bibliography{paper}

\label{lastpage}

\end{document}